\ifdefined\pdfoutput\pdfoutput=1\fi

\PassOptionsToPackage{dvipsnames,x11names,table}{xcolor}
\PassOptionsToPackage{hyphens}{url}

\documentclass[11pt,a4paper,british]{article} 

\usepackage[T1]{fontenc}
\usepackage[utf8]{inputenc}

\usepackage{jheppub}

\usepackage{amsmath}
\usepackage{amsthm}
\usepackage{amssymb}
\usepackage{epsfig}
\usepackage{graphicx}
\usepackage[numbers,sort&compress]{natbib}
\usepackage{color}
\usepackage[dvipsnames,x11names,table]{xcolor}

\usepackage[export]{adjustbox}
\usepackage[all]{xy}
\usepackage{tikz-cd}
\usetikzlibrary{matrix,intersections,arrows.meta,bending,positioning,decorations.markings}
\usepackage{float}
\usepackage{eso-pic}
\usepackage{subcaption}
\usepackage[font={small,singlespacing}]{caption}
\usepackage{setspace}
\usepackage{multirow}
\usepackage{longtable}
\usepackage{listings}
\usepackage[most]{tcolorbox}
\usepackage{framed}
\usepackage{quoting}

\usepackage{accents}
\usepackage{empheq}
\usepackage{bm}
\usepackage{soul}
\usepackage{mathrsfs}
\usepackage{bbm}
\usepackage{slashed}
\usepackage{stmaryrd}
\usepackage[normalem]{ulem}
\usepackage{cancel}

\usepackage[shortlabels]{enumitem}
\usepackage{lineno}
\usepackage[bottom,hang,flushmargin]{footmisc}
\usepackage[useregional,showdow]{datetime2}
\usepackage[hyphens]{url}

\newif\ifshowkeys
\showkeysfalse     

\newif\ifshowmarginalia
\showmarginaliafalse

\newif\ifshowlinenumbers
\showlinenumbersfalse

\makeatletter
\renewcommand\@makefnmark{%
  \hbox{\@textsuperscript{\normalfont\@thefnmark}}%
}
\makeatother

\definecolor{refkey}{rgb}{0,0,1}
\definecolor{labelkey}{rgb}{1,0,0}

\makeatletter
\newcommand{\xMapsto}[2][]{\ext@arrow 0599{\Mapstofill@}{#1}{#2}}
\def\Mapstofill@{\arrowfill@{\Mapstochar\Relbar}\Relbar\Rightarrow}
\makeatother

\def\be{ \begin{equation} }
\def\ee{ \end{equation} }

\numberwithin{equation}{section}
\numberwithin{figure}{section}

\makeatletter
\newcommand{\DeclareEquationTheorem}[3]{%
  \theoremstyle{#1}%
  \@ifundefined{c@#2}{%
    \newtheorem{#2}[equation]{#3}%
  }{%
    \expandafter\let\csname c@#2\endcsname\c@equation
    \expandafter\let\csname the#2\endcsname\theequation
  }%
}
\makeatother

\DeclareEquationTheorem{plain}{theorem}{Theorem}
\DeclareEquationTheorem{plain}{corollary}{Corollary}
\DeclareEquationTheorem{plain}{lemma}{Lemma}
\DeclareEquationTheorem{plain}{proposition}{Proposition}
\DeclareEquationTheorem{definition}{definition}{Definition}
\DeclareEquationTheorem{remark}{remark}{Remark}
\DeclareEquationTheorem{definition}{example}{Example}

\AddToHook{env/remark/begin}{\pushQED{\qed}}
\AddToHook{env/remark/end}{\popQED}

\tcolorboxenvironment{proposition}{%
  enhanced, breakable,
  colback=blue!4!white, colframe=blue!45!black,
  boxrule=0.6pt, arc=2pt,
  left=6pt, right=6pt, top=4pt, bottom=4pt,
  before skip=10pt, after skip=10pt}

\newcommand{\thmref}[1]{\hyperref[#1]{Theorem~\ref*{#1}}}
\newcommand{\corref}[1]{\hyperref[#1]{Corollary~\ref*{#1}}}
\newcommand{\lemref}[1]{\hyperref[#1]{Lemma~\ref*{#1}}}
\newcommand{\propref}[1]{\hyperref[#1]{Proposition~\ref*{#1}}}
\newcommand{\defref}[1]{\hyperref[#1]{Definition~\ref*{#1}}}
\newcommand{\remref}[1]{\hyperref[#1]{Remark~\ref*{#1}}}
\newcommand{\exref}[1]{\hyperref[#1]{Example~\ref*{#1}}}
\newcommand{\appref}[1]{\hyperref[#1]{Appendix~\ref*{#1}}}
\newcommand{\footrefnew}[1]{\hyperref[#1]{footnote~\ref*{#1}}}
\newcommand{\figref}[1]{\hyperref[#1]{Figure~\ref*{#1}}}

\AtBeginDocument{%
}

\makeatletter
\DeclareRobustCommand{\tightfootnote}[1]{\unskip\footnote{#1}}
\makeatother

\makeatletter
\newcommand*\bigbullet{\mathpalette\bigbullet@{1.8}}
\newcommand*\bigbullet@[2]{\mathbin{\vcenter{\hbox{\scalebox{#2}{$\m@th#1\bullet$}}}}}
\makeatother

\def\rem{$\clubsuit$}

\ifshowmarginalia

\makeatletter
\def\cvs@ubreak#1_#2\@nil{#1\_\allowbreak\cvs@ubreak#2\@nil}
\def\cvs@ubreak#1\@nil{#1}
\newcommand*\cvsbreak[1]{\cvs@ubreak#1\@nil}

\def\cm@ubreak#1_#2\@nil{#1\_\allowbreak\cm@ubreak#2\@nil}
\def\cm@ubreak#1\@nil{#1}
\newcommand*\cmbreak[1]{\cm@ubreak#1\@nil}

\makeatother

\newcommand{\cvs}[1]{%
  \marginpar{%
    \raggedright\tiny\color{blue}%
    \parbox{0.75\marginparwidth}{\rem VS: \cvsbreak{#1} \rem}%
  }%
}

\newcommand{\cm}[1]{%
  \marginpar{%
    \raggedright\tiny\color{black}%
    \parbox{0.75\marginparwidth}{\rem MDZ: \cmbreak{#1} \rem}%
  }%
}

\else
  \newcommand{\cvs}[1]{}
  \newcommand{\cm}[1]{}
 \fi

\def\det{{\mathsf{det}}}

\def\End{\mathsf{End}}
\def\exp{{\rm exp}}

\def\Tor{\mathsf{Tor}}

\def\Hom{\mathsf{Hom}}

\renewcommand{\Im}{{\rm Im }}

\def\log{\mathsf{log}}

\def\Tr{{\mathsf{Tr\,}}}

\def\vol{\mathsf{vol}}

\def\one{{\hbox{ 1\kern-.8mm l}}}

\def\CA{{\cal A}}

\def\CD {{\cal D}}

\def\CG {{\cal G}}

\def\CI {{\cal I}}
\def\CJ {{\cal J}}
\def\CK {{\cal K}}
\def\CL {{\cal L}}

\def\CN {{\cal N}}
\def\CO {{\cal O}}
\def\CP {{\cal P}}

\def\CV {{\cal V}}
\def\CW {{\cal W}}

\def\CI {{{\cal I}}}

\def\CT {{\cal T}}
\def\CU {{\cal U}}

\def\IC{\mathbb{C}}

\def\IP{\mathbb{P}}
\def\IQ{\mathbb{Q}}
\def\IR{{\mathbb{R}}}

\def\IZ{{\mathbb{Z}}}

\def\fo{\mathfrak{o}}

\def\fs{\mathfrak{s}}

\def\fu{\mathfrak{u}}

\def\fC{\mathfrak{C}}

\def\rmk#1{\bigskip\noindent{\bf Remarks} }

\DeclareUnicodeCharacter{0393}{\Gamma}
\DeclareUnicodeCharacter{0394}{\Delta}
\DeclareUnicodeCharacter{0398}{\Theta}
\DeclareUnicodeCharacter{039B}{\Lambda}
\DeclareUnicodeCharacter{039E}{\Xi}
\DeclareUnicodeCharacter{03A0}{\Pi}
\DeclareUnicodeCharacter{03A3}{\Sigma}
\DeclareUnicodeCharacter{03A5}{\Upsilon}
\DeclareUnicodeCharacter{03A6}{\Phi}
\DeclareUnicodeCharacter{03A8}{\Psi}
\DeclareUnicodeCharacter{03A9}{\Omega}
\DeclareUnicodeCharacter{03B1}{\alpha}
\DeclareUnicodeCharacter{03B2}{\beta}
\DeclareUnicodeCharacter{03B3}{\gamma}
\DeclareUnicodeCharacter{03B4}{\delta}
\DeclareUnicodeCharacter{03B5}{\epsilon}
\DeclareUnicodeCharacter{03B6}{\zeta}
\DeclareUnicodeCharacter{03B7}{\eta}
\DeclareUnicodeCharacter{03B8}{\theta}
\DeclareUnicodeCharacter{03D1}{\vartheta}
\DeclareUnicodeCharacter{03B9}{\iota}
\DeclareUnicodeCharacter{03BA}{\kappa}
\DeclareUnicodeCharacter{03BB}{\lambda}
\DeclareUnicodeCharacter{03BC}{\mu}
\DeclareUnicodeCharacter{03BD}{\nu}
\DeclareUnicodeCharacter{03BE}{\xi}
\DeclareUnicodeCharacter{03C0}{\pi}
\DeclareUnicodeCharacter{03C1}{\rho}
\DeclareUnicodeCharacter{03C3}{\sigma} 
\DeclareUnicodeCharacter{03C4}{\tau}
\DeclareUnicodeCharacter{03C5}{\upsilon}
\DeclareUnicodeCharacter{03C6}{\phi}
\DeclareUnicodeCharacter{03D5}{\varphi}
\DeclareUnicodeCharacter{03C7}{\chi}
\DeclareUnicodeCharacter{03C8}{\psi}
\DeclareUnicodeCharacter{03C9}{\omega}
\DeclareUnicodeCharacter{21D0}{\Leftarrow}
\DeclareUnicodeCharacter{0212B}{\AA}
\DeclareUnicodeCharacter{00B7}{\cdot}
\DeclareUnicodeCharacter{00B0}{^{\circ}}
\DeclareUnicodeCharacter{266A}{\eighthnote}
\DeclareUnicodeCharacter{266B}{\twonotes}
\DeclareUnicodeCharacter{2202}{\partial}

\newcommand{\n}{\bm{\nabla}}

\newcommand{\sfG}{\mathsf{G}}

\newcommand{\sfL}{\mathsf{L}}

\newcommand{\sfO}{\mathsf{O}}

\newcommand{\sfS}{\mathsf{S}}

\newcommand{\sfU}{\mathsf{U}}

\newcommand{\sfZ}{\mathsf{Z}}

\newcommand{\ov}[1]{\overline{#1}}

\tikzset{
    marrow/.style={decoration={markings,mark=at position 0.5 with {\arrow{#1}}}, postaction=decorate}
}

\def\imag{\mathsf{i}}
\newcommand{\wh}{\widehat }
\newcommand{\wt}{\widetilde }
\newcommand{\smod}{\mathsf{\,\,mod\,\,}}

\DeclareFontFamily{U}{mathx}{}
\DeclareFontShape{U}{mathx}{m}{n}{<-> mathx10}{}
\DeclareSymbolFont{mathx}{U}{mathx}{m}{n}
\DeclareMathAccent{\widecheck}{0}{mathx}{"71}

\colorlet{mylinkcolor}{Blue1}
\colorlet{mycitecolor}{Purple1}
\colorlet{myurlcolor}{Aquamarine}

\colorlet{linkcolormix}{mylinkcolor}
\colorlet{citecolormix}{mycitecolor}
\colorlet{urlcolormix}{myurlcolor!90!black}

\usepackage[
  pagebackref=true,
  colorlinks=true,
  linkcolor=linkcolormix,
  citecolor=citecolormix,
  urlcolor=urlcolormix,
  linktoc=all,
  pdfusetitle=true
]{hyperref}
\usepackage{orcidlink}

\renewcommand*{\backref}[1]{}
\renewcommand*{\backrefalt}[4]{%
  \ifcase #1\relax
  \or [Cited on: pg.~#2.]%
  \else [Cited on: pgs.~#2.]%
  \fi
}

\newcommand{\ssq}{\ensuremath \mathsf{Sq}}

\colorlet{shadecolor}{LavenderBlush2}

\ifshowkeys
  \usepackage[notref,notcite,color]{showkeys}

  \usepackage{seqsplit}

  \definecolor{refkey}{rgb}{0,0,1}
  \definecolor{labelkey}{rgb}{1,0,0}
\fi

\ifshowlinenumbers
  \AtBeginDocument{\linenumbers}
\fi

\title{\centering {\huge Generalized Symmetries of QED$_4$ on Non-Spin Manifolds}}

\author{Michele Del Zotto$^{α,β,γ}$\,\orcidlink{0000-0001-8988-0574} and}
\affiliation[α]{Department of Mathematics, Uppsala University, Box 480, SE-75106, Uppsala, Sweden}
\affiliation[β]{Department of Physics and Astronomy, Uppsala University, Box 516, SE-75120, Uppsala, Sweden}
\affiliation[γ]{Center for Geometry and Physics, Uppsala University, Box 480, SE-75106, Uppsala, Sweden}

\author{Vivek Saxena$^{δ,ε}$\,\orcidlink{0000-0001-9653-5912}}
\affiliation[δ]{Department of Physics, The Ohio State University, Columbus, OH 43210, USA}
\affiliation[ε]{New High Energy Theory Center and Department of Physics and Astronomy, Rutgers University, 126 Frelinghuysen Rd., Piscataway, NJ 08855, USA}
\hypersetup{pdfauthor={Michele Del Zotto, Vivek Saxena}}

\emailAdd{michele.delzotto@math.uu.se}
\emailAdd{vivek.hepth@gmail.com}

\abstract{
  Generalized symmetries are sensitive to the tangential structure of spacetime. We study this interplay for massless QED on oriented non-spin four-manifolds, where the photon becomes a spin$^c$ connection and the path integral sums over spin$^c$ structures. The shifted flux quantization extends the magnetic 1-form symmetry into a \emph{$w_2$-twisted 1-form symmetry}: its $2\pi$ rotation acts as a gravitational sign, fixed by the normal bundle of the surface the operator wraps. The obstruction to treating it as an ordinary magnetic 1-form symmetry is an order-two mixed magnetic--gravitational anomaly, which is classified by reduced oriented bordism and detected on the Wu manifold. In a fixed spin$^c$ sector, this anomaly is removable by a counterterm, so it is a property of the sum over spin$^c$ structures. We discuss its consequences for gapped, Coulomb, and Higgs phases. The chiral anomaly is governed by the index of the spin$^c$ Dirac operator, so rational chiral defects carry gravitational as well as electromagnetic responses. A spin$^c$ Hall topological field theory cancels the electromagnetic part but leaves a framing anomaly. At the level of anomaly inflow, this is canceled for every rational angle by stacking a neutral, generally noninvertible Witten--Reshetikhin--Turaev layer. The same framing subtlety affects the chiral defects of QED on curved spin manifolds. The complex projective plane illustrates each of these features explicitly.
  \\$\,$\\
  \DTMnow
}

\begin{document}

\maketitle

\section{Introduction}\label{sec:intro}

A generalization of the notion of symmetries is necessary to describe the conserved quantum numbers of extended operators within QFT. For continuum QFTs, generalized symmetries comprise their (quasi-)topological defects of various codimensions \cite{Gaiotto:2014kfa}. Topological defects can form intricate networks which organize into a \emph{higher structure} encoding many interesting features of generalized symmetries such as anomalies and charge fractionalization.
\tightfootnote{The idea that topological defects correspond to symmetries in QFT has a long history, as does their interpretation in terms of higher categories. We refer readers to recent reviews of this topic for further references and alternative approaches. See, for example, \cite{Petkova:2000ip,Frohlich:2004ef,Frohlich:2006ch,Chang:2018iay,Kapustin:2010ta,Kapustin:2014gua,Gaiotto:2014kfa,Cordova:2022ruw,McGreevy:2022oyu,Gomes:2023ahz,Schafer-Nameki:2023jdn,Brennan:2023mmt,Bhardwaj:2023kri,Luo:2023ive,Shao:2023gho,Carqueville:2023jhb,Copetti:2023mcq}.}
Interestingly, the higher structure of a symmetry is also sensitive to global features of spacetime, such as its tangential structure. In this paper, we begin an exploration of the interplay of the symmetry category with tangential structure, using massless quantum electrodynamics (QED) on oriented \emph{non-spin} four-manifolds as our main example.

\medskip

Ordinary QED is formulated on spin manifolds. Its unit-charge fermion obeys a \emph{spin-charge} relation -- odd-charge particles are fermions and even-charge particles are bosons -- which allows QED to be placed on oriented non-spin manifolds.
\tightfootnote{See, for example, the analogous discussion of the differential-topological conditions defining a ``physical theory'' on non-spin four-manifolds in \cite{Moore:2024vsd}.}
There, the Maxwell field of QED is promoted to a spin$^c$ connection, and compact QED sums over its spin$^c$ sectors, that is, over spin$^c$ structures \cite{Witten:1995gf,Seiberg:2016rsg,Wang:2018qoy}. This summed theory depends only on the orientation, unlike the theory in a fixed spin$^c$ sector, which depends on a chosen spin$^c$ structure. The main purpose of this paper is to describe how the resulting shifted flux quantization modifies the well-known topological defects of QED. In particular, we show that the magnetic 1-form symmetry, the Adler--Bell--Jackiw (ABJ) anomaly \cite{Adler:1969gk,Bell:1969ts}, and the topological defects associated with (rational) chiral rotations \cite{Cordova:2022ieu,Choi:2022jqy,Copetti:2023mcq} are affected in interesting ways. 

\medskip

The main results discussed in this paper are the following:
\begin{enumerate}[leftmargin=*]
  \item \emph{A $w_2$-twisted magnetic 1-form symmetry.} QED has a well-known magnetic 1-form generalized symmetry.
  \tightfootnote{The dynamical unit-charge fermion makes integral Wilson lines endable and hence completely breaks the electric 1-form symmetry. It does not break the magnetic 1-form symmetry, whose conservation follows from the Bianchi identity $dF=0$. The latter would instead be broken by dynamical magnetically charged matter. The magnetic surface operators measure the flux $\int_{\Sigma_2}F/(2\pi)$, which distinguishes the flux classes of the spin$^c$ sectors summed over in compact QED.}
  The shifted flux quantization makes the magnetic surface operator at rotation parameter $2\pi$ equal to the gravitational phase $(-1)^{\int w_2}$, so that the magnetic 1-form symmetry is a $4\pi$-periodic, $w_2$-twisted symmetry obeying
 \be
  U_m(\alpha+2\pi;\Sigma_2)=(-1)^{\int_{\Sigma_2} w_2}\,U_m(\alpha;\Sigma_2) ~,
  \ee
where $\Sigma_2$ is an oriented surface in the four-manifold, and $w_2$ is the second Stiefel--Whitney class of the four-manifold. On $\Sigma_2$, $w_2$ equals the second Stiefel--Whitney class of the normal bundle, so the sign is $(-1)^{\Sigma_2\cdot\Sigma_2}$. It is fixed by the normal bundle of the surface, through its self-intersection (see \propref{prop:w2-twist}). The obstruction to treating it as an ordinary $2\pi$-periodic symmetry is an order-two mixed magnetic--gravitational anomaly. For the summed theory, we show that it is classified by reduced oriented bordism and detected on the Wu manifold.
\tightfootnote{See also \cite{DavighiLohitsiriDebray2023} for related discussions about this anomaly and \cite{JiaWangZhang2026} for bordism group computations.}
  \item \emph{A gauge--gravitational ABJ anomaly.} Rational chiral rotations
  \tightfootnote{\label{foot:rationalangle}Throughout this paper, a \emph{rational chiral rotation} is an axial rotation by an angle $2\pi p/N$, with $p\in\IZ$, $N\in\IZ_{>0}$, and $\gcd(p,N)=1$. We call such angles \emph{rational}, and the corresponding defects \emph{rational chiral defects}. Since only $p/N$ modulo $1$ matters, rational chiral rotations are labeled by $\IQ/\IZ$.}
  are governed by the full spin$^c$ Dirac index density, so the axial anomaly carries both an electromagnetic and a gravitational (signature) part. The corresponding gauge--gravitational theta angle is exactly $2\pi$-periodic.
\tightfootnote{A $2\pi$ shift of the electromagnetic term alone produces only the field-independent gravitational phase $e^{\imag\pi\sigma(M_4)/4}$, so the familiar $4\pi$ periodicity of an ordinary $\sfU(1)$ connection on non-spin manifolds does not arise (see \appref{app:theta-periodicity}).}
  \item \emph{Framing completion of the chiral defect.} Stacking the wall with a spin$^c$ Hall TFT
\tightfootnote{\label{foot:tft-terminology}We use the term ``topological field theory'' (TFT) for both invertible topological theories defined by exponentiated classical actions and topological theories obtained by quantization. V.S. thanks G. Moore for related discussions.}
 cancels the electromagnetic anomaly but leaves a residual framing anomaly controlled by $(c_- - \sigma_H)$, where $\sigma_H$ is the Hall conductivity of the 3d TFT and $c_-$ is its chiral central charge. For every rational angle, we complete the wall at the level of anomaly inflow by stacking it with a neutral bosonic Witten--Reshetikhin--Turaev (WRT) layer \cite{Witten:1989Jones,Reshetikhin:1990,Reshetikhin:1991pm,Turaev:2016} (see \propref{prop:unframed-completion}). The completed defect is generally noninvertible, and the choice of framing completion is noncanonical. This generalizes the construction by Putrov and Wang \cite{Putrov:2023jqi} to the present context. Complementary aspects of this matching, from the perspective of monodromy defects, appear in \cite{Copetti:2026monodromy}.
  \item \emph{Scope of the spin constructions.} On a general curved spin manifold, the topological defects for spin QED of \cite{Choi:2022jqy,Cordova:2022ieu} depend on a choice of $2$-framing of the wall, so they are framed or relative (see \autoref{sec:contextandmotivation}). Making them genuine topological defects requires canceling a residual gravitational anomaly. An invertible completion suffices when $(c_- - \sigma_H) \in \frac{1}{2}\IZ$, which for the minimal Abelian TFT $\CA^{N,p}$ \cite{Hsin:2018vcg} happens only when $N\in\{1,2\}$ (see \remref{rem:spin-versus-spinc-completion}). For every other $N$, the completed defect contains a noninvertible neutral TFT, which enters its fusion rules. The electromagnetic data, such as the action on 't Hooft lines and the selection rules, are unchanged.
  \item \emph{The example of $\IC\IP^2$.} As the simplest simply connected non-spin four-manifold, $\IC\IP^2$ makes each of these features explicit: the half-integral flux, the $w_2$-twist, the full spin$^c$ index, and the residual framing phase. The boundary of a tubular neighborhood of $\IC\IP^1 \subset \IC\IP^2$ is a three-sphere, which splits $\IC\IP^2$ into this neighborhood and a four-ball. These two extensions of the same $S^3$ realize the extension dependence of the wall geometrically.
\end{enumerate}

In each case, the tangential structure enters: through $w_2(TM_4)$ it twists the magnetic 1-form symmetry, through $p_1(TM_4)$ it contributes the signature term to the ABJ anomaly, and through the $2$-framing of the wall it leaves a framing anomaly on the chiral defect. We construct these chiral defects and show how their anomalies cancel. The fusion interfaces and coherence data of the symmetry category are left to \cite{DelZottoSaxena:symmetry-category}.

This paper is organized as follows. In \autoref{sec:contextandmotivation}, which can serve as an extended introduction, we review the rational chiral defects of spin QED$_4$, explain why they are framed or relative on curved backgrounds, and outline their spin$^c$ refinement. In \autoref{sec:spinc}, we review spin$^c$ structures, determinant line bundles, and the shifted flux lattice, including the role of 2-torsion. We also recall the four variants of compact $\sfU(1)$ gauge theory and isolate the spin$^c$ theory among them. In \autoref{sec:spinc-QED4}, we distinguish a fixed spin$^c$ sector from the compact QED path integral that sums over all spin$^c$ sectors, and we characterize spin$^c$ QED$_4$ in terms of the latter. We then analyze the magnetic 1-form symmetry: its $w_2$-twist, the local magnetic charge lattice, and the mixed magnetic--gravitational anomaly, which is classified by reduced oriented bordism. In \autoref{sec:chiral-defect-revisited}, we derive the ABJ anomaly from the spin$^c$ Dirac index and we reformulate the chiral rotation in terms of a gauge--gravitational theta angle interface. We then construct the gauge sector Hall TFT, determine its residual framing anomaly and its WRT completion, and compare its spin limit with the earlier spin QED construction. We also describe its action on 't Hooft lines. \autoref{sec:CP2-example} works the construction out explicitly on $\IC\IP^2$. In \autoref{sec:dynamics}, we use the mixed anomaly to constrain gapped, Coulomb, and Higgs phases, discuss $\sfU(N)$, $\mathsf{SU}(2)$, and $\mathsf{Spin}(10)$ gauge theories whose infrared photon is a spin$^c$ connection, and show how the rational chiral defects constrain renormalization group flows. \autoref{sec:conclusion} summarizes the status of the proposed defect. \appref{app:symbol-list} lists our notation, and \appref{app:QEDconventions} fixes our differential-form and QED$_4$ conventions. In \appref{app:theta-periodicity}, we derive the periodicities of the electromagnetic and gauge--gravitational theta angles. \appref{app:bordism-anomaly-proof} gives the bordism computation behind \propref{prop:bordism-anomaly}. In \appref{app:minimal-TFT}, we review the minimal TFTs $\CA^{N,p}$ and $\CV^{N,p}$. In \appref{app:spin-QED4-defect}, we derive the spin QED defect from a finite Gauss sum. Finally, in \appref{app:magnetic-sheet}, we distinguish intrinsic half-integral spin$^c$ flux from externally imposed electromagnetic sheets and formulate their fusion and wall-transport questions.

\section*{Acknowledgments}
We thank T. Banks, C. Copetti, G. Festuccia, A. Grigoletto, G. Moore, K. Ohmori, and Y. Wang for interesting discussions. We especially thank G. Moore for helpful comments and suggestions. The work of MDZ is supported by the European Union (ERC, HIGH, 101171852). Views and opinions expressed are, however, those of the author(s) only and do not necessarily reflect those of the European Union or the European Research Council. Neither the European Union nor the granting authority can be held responsible for them. MDZ also acknowledges support by the VR project grant No. 2023-05590, the VR Centre for Geometry and Physics (VR grant No. 2022-06593), and the Simons Foundation International Grant for the Simons Collaboration on Global Categorical Symmetries. The work of VS was supported by the US Department of Energy under grant DE-SC0010008 to Rutgers. We acknowledge the use of OpenAI's Codex and Anthropic's Claude Code for editorial assistance in the concluding stages of preparing this manuscript, but we take full responsibility for any errors or omissions that remain.

\section{Rational Chiral Defects: From Spin To \texorpdfstring{Spin$^c$}{Spin-c} \texorpdfstring{QED$_4$}{QED4}}\label{sec:contextandmotivation}
Let us begin with the familiar construction of rational chiral defects in massless QED on a spin four-manifold $M_4$ -- we will sometimes refer to this as spin QED$_4$.
\tightfootnote{\label{foot:spin-QED-background}For the moment, $M_4$ is a smooth, compact, oriented, spin manifold endowed with a Riemannian metric $g$. Later, we will relax the spin assumption. Also, adhering to tradition, we will not be careful to distinguish between operators and defects.}
\phantomsection\label{par:axial-current}
Let $M_3 \subset M_4$ be a smooth, closed, oriented, three-dimensional embedded submanifold. Let $j^A$ denote the conventional axial current 1-form and $\CJ_A^{(3)} := \frac{1}{2}\star_{4}j^{A}$ its Hodge dual.
\tightfootnote{The normalization $\frac{1}{2}$ is a matter of convention to adapt to \eqref{eq:axial-rotation}.} 
The codimension-one operator
\be
 \exp\left(\imag \alpha \int_{M_3}\CJ_{A}^{(3)} \right) ~,
\ee
implements a chiral rotation through an angle $\alpha \in \IR/2\pi\IZ$, acting as
\be
  \label{eq:axial-rotation}
  \Psi\longmapsto e^{\imag\alpha\gamma^5/2}\Psi ~,
  \qquad
  \ov\Psi\longmapsto\ov\Psi e^{\imag\alpha\gamma^5/2} ~.
\ee

By general principles \cite{Gaiotto:2014kfa}, a quantum 0-form symmetry corresponds to a topological defect of codimension one. However, the ABJ anomaly \cite{Adler:1969gk,Bell:1969ts} prevents the bare axial symmetry operator from being topological. For rational values of the rotation angle, $\alpha = 2\pi p/N$, with $\gcd(p,N)=1$, the electromagnetic contribution to the ABJ anomaly can be canceled by coupling QED$_4$ to the minimal Abelian TFT $\CA^{N,p}$ \cite{Hsin:2018vcg} supported on $M_3$. 
\tightfootnote{For a brief review of this theory, see \appref{app:minimal-TFT}.}
The TFT realizes this cancellation by coupling to a $\IZ_{N}^{(1)}$ subgroup of the $\sfU(1)^{(1)}$ magnetic 1-form symmetry, whose current is $F/(2\pi)$, with $F$ the electromagnetic curvature. Before its framing anomaly is included, the resulting \emph{gauge sector} defect, i.e., the wall before its gravitational completion, is
\be\label{eq:defect-schematic}
  \wh{\CD}_{p/N}(M_3)
  = \CA^{N,p}\left[\frac{F}{2\pi}\right]
    \otimes \exp\left(\imag\frac{2\pi p}{N}\int_{M_3}\CJ_A^{(3)}\right) ~.
\ee
Here and below, this formula is schematic: the first factor denotes the three-dimensional TFT together with its coupling to the bulk magnetic current. On flat spacetime, or with appropriate framing or relative data, this is the familiar spin QED gauge sector defect of \cite{Choi:2022jqy,Cordova:2022ieu}, derived from a finite Gauss sum in \appref{app:spin-QED4-defect}. However, on a general curved background, this codimension-one wall is only framed or relative (to be defined momentarily) until its residual gravitational response and framing anomaly are canceled. Therefore, it is not yet an unframed defect in the sense defined below. We will construct the required completion in \autoref{sec:coupling-tft-to-spinc-qed4}. After normalizing the physical period of the axial rotation to one, the rational rotation parameters form $\IQ/\IZ$.
\tightfootnote{Proposals to go full circle have appeared in the literature \cite{GarciaEtxebarria:2022jky,Arbalestrier:2024oqg}. In our analysis, we restrict the rotation parameter to $\IQ/\IZ$ to guarantee the finiteness of the corresponding topological defect free energies without the need for regularization.}
The relation of this parameter group to the noninvertible symmetry of four-dimensional QED and to the categorification of the Chern--Simons dressing was discussed in \cite{Putrov:2022qz}. Here $\IQ/\IZ$ initially describes only the rational angle defect objects. Constructing a fusion-coherent $\IQ/\IZ$ symmetry category requires the morphism and coherence data.
\tightfootnote{For a first step in this direction in the cases of QED and Axion-Maxwell theories, see \cite{Copetti:2023mcq, DelZotto:2024ngj}.}

\medskip
\medskip
\noindent \emph{Framing, Framed, Unframed, and Relative Defects.}
\phantomsection\label{par:framing-data}

A \emph{framed defect} depends on framing data for its three-dimensional worldvolume $M_3$, more precisely on a \emph{$2$-framing}, which is a homotopy class of trivializations of $TM_3 \oplus TM_3$ \cite{Atiyah:1990framing}.
\tightfootnote{See also \cite{Witten:1989Jones,Freed:2009CS}, \cite[Sec. 3.5.3]{Wernli:2019CS}, \cite[Sec. 26.2]{Moore:2025tmt}, and \cite{FreedTeleman:2026p1} for discussions of framing and \emph{$p_1$-structures}.}
 For connected $M_3$, the $2$-framings form a $\IZ$-torsor. Taken relative to the canonical $2$-framing of \cite{Atiyah:1990framing},
\tightfootnote{The canonical $2$-framing $f_{\mathrm{can}}$ is the unique one with $\sigma(Y_4)=\frac{1}{6}\,p_1(2TY_4,f_{\mathrm{can}})$ for any oriented extension $Y_4$ of $M_3$, where $\sigma(Y_4)$ is the signature of the intersection form on $H^2(Y_4,M_3;\IR)$, with $p_1(2TY_4,f)$ as in \eqref{eq:gravitational-CS-functional} below. For it, the gravitational Chern--Simons functional \eqref{eq:gravitational-CS-functional} equals $3\,\eta(M_3,g_3)$, where $\eta$ is the eta invariant of the signature operator \cite{Atiyah:1990framing}.}
a $2$-framing is therefore labeled by an integer $f \in \IZ$, and $f \mapsto f+1$ denotes a unit shift.

The \emph{framing anomaly} is the phase by which the defect partition function changes when the $2$-framing is changed. Under a unit shift $f\mapsto f+1$, the defect partition function acquires the phase $\exp(2\pi\imag\,c_-/24)$, which depends only on the chiral central charge $c_-$ modulo $24$. An \emph{unframed defect} is independent of this auxiliary choice. A \emph{relative defect} is defined only together with a four-dimensional anomaly inflow theory whose variation cancels that of the defect.

Let $g_3$ be the induced metric on $M_3$, with Levi--Civita connection $\omega$.
\tightfootnote{\label{foot:ambient-connection}The ambient Levi--Civita connection $\omega_4$, which enters the axial anomaly, restricts to a connection on $TM_4|_{M_3}\simeq TM_3\oplus\mathbb R$. It generally differs from the direct-sum connection $\omega\oplus0$ by terms involving the second fundamental form.}
We define the gravitational Chern--Simons term
\be\label{eq:gravitational-CS-form}
  \mathsf{GCS}(\omega)
  :=-\frac{1}{8\pi^2}\Tr\left(
    \omega\wedge d\omega
    +\frac{2}{3}\omega\wedge\omega\wedge\omega
  \right) ~.
\ee
Here $\Tr$ is the trace in the real tangent representation. On a four-dimensional extension $Y_4$ of $M_3$, i.e., a compact oriented four-manifold with $\partial Y_4=M_3$ (also called a filling), with Levi--Civita curvature $R=d\omega+\omega\wedge\omega$, the Pontryagin form is
\be\label{eq:Pontryagin-CS-conventions}
  p_1(R)=-\frac{1}{8\pi^2}\Tr(R\wedge R) ~,
  \qquad \text{ so } \qquad 
  d\mathsf{GCS}(\omega)=p_1(R) ~.
\ee
The local expression \eqref{eq:gravitational-CS-form} depends on the choice of orthonormal frame. 
To make the dependence on a $2$-framing $f$ explicit, choose an oriented extension $Y_4$ of $M_3$ and extend the metric with a product collar near its boundary. The real gravitational Chern--Simons functional is then given by \cite[Sec.~5]{Atiyah:1990framing}
\be\label{eq:gravitational-CS-functional}
  \mathsf{CS}_{\mathrm{grav}}[M_3,g_3;f]
  :=\int_{Y_4}p_1(R)-\frac{1}{2}p_1(2TY_4,f) ~.
\ee
Here $2TY_4:=TY_4\oplus TY_4$, and $p_1(2TY_4,f)\in2\IZ$ is its relative Pontryagin number, using $f$ and the outward normal to trivialize the bundle along the boundary. Changing the extension changes both terms on the right by the same amount, so the functional depends only on $M_3$, $g_3$, and $f$. It is the global, framing-dependent meaning of the notation $\int_{M_3}\mathsf{GCS}(\omega)$.

Let $\CT$ be a three-dimensional TFT whose gravitational response is specified by the chiral central charge $c_-(\CT)$, and denote by $\sfZ_{\CT}^{\mathrm{fr}}[M_3,f]$ its framed partition function. Under a unit change $f\mapsto f+1$ of the $2$-framing, the relative Pontryagin number changes by two \cite[Sec.~2]{Atiyah:1990framing}:
\be\label{eq:relative-Pontryagin-framing-shift}
  p_1(2TY_4,f+1)=p_1(2TY_4,f)+2 ~.
\ee
So the framed partition function transforms as \cite[Sec.~26.2]{Moore:2025tmt}:
\be\label{eq:framing-conventions}
  \begin{aligned}
    \mathsf{CS}_{\mathrm{grav}}[M_3,g_3;f+1]
    &=\mathsf{CS}_{\mathrm{grav}}[M_3,g_3;f]-1 ~,\\
    \sfZ_{\CT}^{\mathrm{fr}}[M_3,f+1]
    &=\exp\left[\frac{2\pi\imag c_-(\CT)}{24}\right]
      \sfZ_{\CT}^{\mathrm{fr}}[M_3,f] ~.
  \end{aligned}
\ee
We define the corresponding partition function on the metric background by
\be\label{eq:metric-framed-TFT}
  \sfZ_{\CT}^{\mathrm{met}}[M_3,g_3]
  =
  \exp\left[
    \frac{2\pi\imag c_-(\CT)}{24}
    \mathsf{CS}_{\mathrm{grav}}[M_3,g_3;f]
  \right]
  \sfZ_{\CT}^{\mathrm{fr}}[M_3,f] ~.
\ee
Under $f\mapsto f+1$, the first factor in \eqref{eq:metric-framed-TFT} acquires $\exp[-2\pi\imag c_-(\CT)/24]$, which cancels the phase of the second factor. The product is therefore independent of the auxiliary framing. It generally changes when $g_3$ changes, through the Chern--Simons factor. This is what we mean by a \emph{metric-dependent realization}: the TFT partition function including its specified gravitational response on a Riemannian background. We call the metric dependence of this phase the \emph{local gravitational response}. For a spin TFT, the spin structure is also held fixed and is suppressed in the notation.

\phantomsection\label{par:defect-category}
We recall the categorical description of topological defects as developed in \cite{Copetti:2023mcq,Carqueville:2023jhb}. We can think of QFTs in $d+1$ spacetime dimensions as objects in a $d+1$ category with morphisms defined by topological interfaces. In this context, the symmetries of a given $d+1$-dimensional QFT $\mathcal T$ are encoded by the monoidal $d$-category of its endomorphisms, $\End(\mathcal T)$. In this paper, we consider the bulk QED theory as an object and a codimension-one topological defect as a $1$-endomorphism of that object. After fixing the bulk theory, we write $\End(\mathsf{QED}_4)$ 
%
for the resulting monoidal $3$-category. Its objects are the three-dimensional defects, its $1$-morphisms are two-dimensional topological interfaces between them, its $2$-morphisms are one-dimensional junctions of interfaces, and its $3$-morphisms are point junctions. The monoidal product is the fusion of defects. Anticipating the extension to spin$^c$ structures, we write 
\be\nonumber
  \fC_{\chi}^{\mathrm{spin}^c}\subset \End(\mathsf{QED}_4)  ~,
\ee
for the putative monoidal symmetry $3$-category generated by the rational chiral defects, the magnetic surface operators, and their junctions on spin$^c$ backgrounds. The word \emph{putative} is important: this paper constructs the relevant codimension-one objects and describes anomaly cancellation, but it does not construct all fusion interfaces, associators, and higher coherence data required to specify $\fC_{\chi}^{\mathrm{spin}^c}$ completely. A systematic construction of these categorical data is the subject of \cite{DelZottoSaxena:symmetry-category}.

The earlier spin QED constructions \cite{Choi:2022jqy,Cordova:2022ieu} correctly cancel the electromagnetic part of the ABJ anomaly and thereby determine the gauge sector defect, its fusion rules, its action on line operators, and its local operator selection rules. What those analyses do not track is the signature phase on a curved spin manifold. The axial Jacobian
\tightfootnote{\label{foot:axial-jacobian}By the \emph{axial Jacobian}, we mean the Jacobian of the fermionic functional measure under a chiral rotation \cite{Fujikawa:1979ay,Fujikawa:1980eg}, which is a phase. In our conventions, a rotation through an angle $\alpha$ contributes $\exp[\imag\alpha\,\mathsf{Ind}(\slashed{D}_{\fs})]$, and we also refer to it as the ABJ phase.}
is obtained by exponentiating the full Dirac index and therefore includes its gravitational contribution, while the three-dimensional TFT $\CA^{N,p}$ has its own framing anomaly. After the electromagnetic terms cancel, the combined gravitational response of the axial Jacobian and $\CA^{N,p}$ is governed by the difference
\be\nonumber
  c_-(\CA^{N,p})-\frac{p}{N} ~.
\ee
Thus the defects described above are topological in flat spacetime, with the framing choice understood. On curved spin backgrounds, however, they are topological only as framed or relative defects. Promoting them to unframed defects requires cancellation of the residual gravitational response and full framing anomaly. This distinction is invisible to flat-space analyses. In \appref{app:spin-QED4-defect}, we reformulate the discrete gauging construction of \cite{Choi:2022jqy} as a finite sum over cohomology classes. On a spin four-manifold with torsion-free cohomology, this path integral is the Gauss sum of a quadratic function on $H^2(M_4;\IZ_N)$. Completing the square separates the flux-dependent factor that shifts the theta angle from a flux-independent normalization.

On spin backgrounds, both the gauge sector and its gravitational completion simplify. First, on a spin four-manifold the electromagnetic flux is integrally quantized, the magnetic surface operator at parameter $2\pi$ is the identity, and the ordinary minimal TFT $\CA^{N,p}$ may be used without an additional spin-charge constraint. Second, the possible dependence on the four-dimensional extension is weaker on spin backgrounds. 

\phantomsection\label{par:spin-wall-filling}
Concretely, the gravitational part of the wall is defined relative to a four-dimensional extension, i.e., an auxiliary spin four-manifold $Y_4$ with $\partial Y_4=M_3$, whose spin structure restricts to that of $M_3$.
\tightfootnote{\label{foot:spin-wall-filling}Since $\Omega_3^{\mathsf{Spin}}=0$ \cite{Milnor1963Spin,Kirby1989Topology}, every closed 3-manifold $M_3$ with a spin structure is the boundary of some compact four-manifold with a spin structure, such that the spin structures agree on $M_3$. Therefore, such a $Y_4$
 always exists. It need not be a region of the spacetime $M_4$.}
Given two extensions $Y_4$ and $Y_4'$, gluing along $M_3$ gives the closed spin four-manifold $X_4=Y_4\cup_{M_3}\overline{Y_4'}$, where the bar denotes orientation reversal. In the conventions of \eqref{eq:Pontryagin-CS-conventions}, the residual gravitational inflow has coefficient $\frac{2\pi\imag}{24}(c_-(\CA^{N,p})-p/N)$ multiplying $\int_{Y_4}p_1(R)$. Since $\int_{X_4}p_1(R)=3\sigma(X_4)$, the wall partition functions defined using the two extensions differ by a phase:
\be\label{eq:phase-ratio}
   \frac{\sfZ[Y_4]}{\sfZ[Y_4']}
   =\exp\left[\frac{2\pi\imag}{8}\left(c_-(\CA^{N,p})-\frac{p}{N}\right)\sigma(X_4)\right] ~.
\ee
By Rokhlin's theorem \cite{Rokhlin1952}, the signature $\sigma(X_4)$ is divisible by $16$. Equation \eqref{eq:phase-ratio} therefore shows that the defect is independent of the chosen four-dimensional extension precisely when 
\be
   c_-(\CA^{N,p})-\frac{p}{N}\in\frac{1}{2}\IZ ~.
\ee 
This condition is necessary but not sufficient for an unframed defect. In the framed description, the local Pontryagin response of the axial factor is canceled by the gravitational transgression $\exp[2\pi\imag p\,\mathsf{CS}_{\mathrm{grav}}[M_3,g_3;f]/(24N)]$. By \eqref{eq:framing-conventions}, a positive unit change of $2$-framing gives this factor the phase $\exp[-2\pi\imag p/(24N)]$, whereas $\CA^{N,p}$ contributes $\exp[2\pi\imag c_-(\CA^{N,p})/24]$. The combined phase is
\be\label{eq:2-framing-phase}
   \exp\left[\frac{2\pi\imag}{24}\left(c_-(\CA^{N,p})-\frac{p}{N}\right)\right] ~.
\ee
So while extension independence requires the mismatch to vanish modulo $\frac{1}{2}$, the 2-framing independence requires it to vanish modulo $24$. We choose a completion that cancels the mismatch exactly. When $c_{-}(\CA^{N,p}) - p/N \in \frac{1}{2}\IZ$, this completion can be obtained by stacking neutral $p+\imag p$ spin TFT layers with suitably chosen chiralities.
\tightfootnote{\label{foot:pip-superconductor}The $p+\imag p$ (or $p_x+\imag p_y$) spin TFT is invertible and has chiral central charge $c_- = \frac{1}{2}$ \cite{ReadGreen2000}. Reversing its chirality yields $c_- = -\frac{1}{2}$. These layers use the spin structure induced on $M_3$ by the chosen bulk spin structure. Being decoupled from the electromagnetic gauge field, they shift the chiral central charge without changing the Hall conductivity.}
If $c_{-}(\CA^{N,p}) - p/N \notin \frac{1}{2}\IZ$, no neutral invertible spin TFT suffices. Since $\CA^{N,p}$ is Abelian, $c_-(\CA^{N,p})\in\frac{1}{2}\IZ$, so the first case occurs only for $N\in\{1,2\}$ and the second for every $N>2$ (see \remref{rem:spin-versus-spinc-completion}). A neutral, generically noninvertible completion can instead be constructed using Witten--Reshetikhin--Turaev (WRT) theories \cite{Putrov:2023jqi}. We describe this construction in \autoref{sec:coupling-tft-to-spinc-qed4}.

\medskip
\medskip
\noindent \emph{Spin$^c$ Refinement.}
\phantomsection\label{par:spinc}

The spin$^c$ theory modifies this picture in several related ways. For a fixed spin$^c$ structure $\fs$, the de Rham class of the electromagnetic curvature satisfies $[F/(2\pi)]_{\mathrm{dR}}=\frac{1}{2}c_1(\CL_\fs)_{\IR}\in H^2(M_4;\IR)$, where $\CL_\fs$ is the determinant line bundle of the spin$^c$ structure and the subscript $\IR$ denotes the image in real cohomology of its integral Chern class modulo torsion (see \autoref{sec:spinc} for details). After choosing an auxiliary basepoint spin$^c$ structure $\fs_0$, the set of spin$^c$ structures is the $H^2(M_4;\IZ)$-torsor $\{\fs_0+\varepsilon\mid\varepsilon\in H^2(M_4;\IZ)\}$. Accordingly, the full compact-QED path integral sums over these spin$^c$ sectors, including flat torsion data not detected by the de Rham curvature class. Writing $U_m(\alpha;\Sigma_2)$ for the magnetic surface operator of parameter $\alpha$ supported on $\Sigma_2$, spin$^c$ flux quantization gives
\be
  U_m(2\pi;\Sigma_2)
  =(-1)^{\int_{\Sigma_2}w_2(TM_4)} ~.
\ee
This value at $\alpha=2\pi$ is the magnetic manifestation of the Lorentz symmetry fractionalization encoded by the spin-charge relation \cite[Sec.~2.5.1]{Brennan:2022tyl}. A central contribution of this paper is the operator-level formulation of this identity as a $4\pi$-periodic, \emph{$w_2$-twisted magnetic 1-form symmetry} and its incorporation into the same symmetry category as the rational chiral defects, as $1$-endomorphisms of the trivial defect (see \autoref{subsubsec:w2-twist}). 

At the same time, the ABJ anomaly is governed by the full spin$^c$ Dirac index density
\be
  \left[\wh A(TM_4)\exp\left(\frac{F}{2\pi}\right)\right]^{(4)}
  = \frac{F\wedge F}{8\pi^2}-\frac{p_1(TM_4)}{24} ~.
\ee
Thus, on general curved backgrounds, a chiral wall carries both electromagnetic and gravitational response data. The same index density defines the natural gauge--gravitational spin$^c$ theta coupling, which is invariant under $\theta\mapsto\theta+2\pi$ on every closed oriented four-manifold. \appref{app:theta-periodicity} contrasts this coupling with the isolated electromagnetic theta term.

Motivated by the fractional quantum Hall interpretation of the topological defects of spin QED, we choose a spin$^c$ Hall TFT for the electromagnetic part of the anomaly. The Hall TFT cancels the flux-dependent electromagnetic anomaly of the axial wall, but the resulting wall remains framed or relative until its residual gravitational response and full 2-framing dependence are canceled. 

Recall that a Hall TFT carries two distinct response coefficients. Its Hall conductivity $\sigma_H$ controls the electromagnetic Chern--Simons response, and canceling the electromagnetic anomaly fixes $\sigma_H=p/N$. Its chiral central charge $c_-$ controls the gravitational Chern--Simons response and hence the framing anomaly. These coefficients need not agree. For example, the Abelian theory $\sfU(1)_N$ with positive odd $N$ has $\sigma_H = 1/N$ and $c_- = 1$. With the sign conventions of \eqref{eq:2-framing-phase}, the residual gravitational mismatch is $c_- - p/N$. The spin$^c$ Gauss--Milgram relation \cite{Kobayashi:2021jsc,Cheng:2025ube} implies that a neutral invertible spin$^c$ layer shifts $c_-$ only by a multiple of $8$, while leaving $\sigma_H$ unchanged. An invertible completion is therefore possible precisely when $c_- - p/N \in 8\IZ$. In that case, neutral $E_8$ layers with suitable chiralities cancel the mismatch exactly. Otherwise, a noninvertible neutral TFT is required. For an Abelian Hall TFT with $N>1$, this is always the case, and the neutral TFT must be non-Abelian (see \autoref{sec:coupling-tft-to-spinc-qed4}). At $p/N=\frac{1}{2}$, by contrast, the particle--hole symmetric Pfaffian (PH-Pfaffian) \cite{Son:2015xqa} has $c_-=\sigma_H$ and needs no neutral completion (see \remref{rem:half-wall}).

The mismatch is nevertheless rational, since $\sigma_H=p/N$ and the chiral central charge of a unitary TFT is rational \cite{Anderson:1987ge,Vafa:1988ag}. A neutral bosonic completion can therefore be constructed using WRT theories \cite{Putrov:2023jqi}. We choose the neutral sector, including its gravitational response, so that the Hall central charge minus the neutral central charge is exactly $p/N$ (see \eqref{eq:full-framing-completion}). This cancels both the local gravitational response and the full 2-framing dependence, while leaving the electromagnetic response and the spin-charge relation unchanged. Constructing a fusion-closed monoidal subcategory additionally requires suitable topological interfaces and coherent junction and associator data, of the type studied in \cite{Copetti:2023mcq}. These fusion and coherence data will be developed in \cite{DelZottoSaxena:symmetry-category}. Extending the discrete gauging argument of \appref{app:spin-QED4-defect} to manifolds with torsion also requires a differential cohomology formulation.

The recent work \cite{Copetti:2026monodromy} has some overlap with our findings, though it addresses a different defect problem. It also uses the spin$^c$ gauge--gravitational theta term, finds that rational chiral anomaly inflow requires matching both $\sigma_H$ and $c_-$, analyzes the spin$^c$ compatibility of the minimal Abelian sector at odd and even denominator, and uses the partition function on $\IC\IP^2$ with half-integral flux to detect $c_- - \sigma_H$. Its approach starts from monodromy defects and focuses on the codimension-two termination of the wall.

Our approach begins instead with the global spin$^c$ geometry of compact QED on a general oriented four-manifold. We determine the shifted flux lattice and the sum over dynamical spin$^c$ sectors, derive the magnetic surface operator algebra, interpret its bordism anomaly \cite{DavighiLohitsiriDebray2023} for the summed theory, and use the spin$^c$ index theorem to compute the full ABJ phase. The closed codimension-one wall is then studied by comparing its four-dimensional extensions and canceling both its Hall and framing responses. These steps yield the $4\pi$-periodic, $w_2$-twisted magnetic 1-form symmetry and, for every rational angle, an anomaly-level unframed completion by a neutral bosonic WRT factor. We retain this completion as part of the object label in the symmetry category and realize the extension problem geometrically using the two extensions of $S^3$ in $\IC\IP^2$. Constructing the monodromy terminations and their protected gapless modes is left for future work.

\section{\texorpdfstring{Spin$^c$}{Spin-c} Structures, Flux Quantization, And Variants Of Maxwell Theory}\label{sec:spinc}

In this section, to fix our notation and conventions, we review in detail the notion of a spin$^c$ structure on a four-manifold $M_4$ and its associated spin$^c$ connection.
\tightfootnote{\label{foot:donaldson-witten}A supersymmetric extension of spin$^c$ QED, with suitable background couplings, appears in the low-energy description of Donaldson--Witten theory~\cite{Witten:1994cg,Moore:1997pc}
near its monopole and dyon singularities. The corresponding twisted Abelian theories localize on the Seiberg--Witten monopole equations, and the contributions from these singularities involve sums over spin$^c$ structures weighted by their Seiberg--Witten invariants, together with factors determined by the observables. Extending the present symmetry analysis to these theories and their weighted sums is an interesting direction for future work. We thank G. Moore for this suggestion. For a recent discussion of generalized spin$^c$ structures in the context of four-dimensional gauge theories -- and particularly $d=4$ $\CN=2$ supersymmetric field theories --  see \cite{Moore:2024vsd}. For additional discussions of spin$^c$ structures, see \cite{Witten:1995gf,Witten:1998cd,Freed:1999vc,Scorpan:2005,Seiberg:2016gmd,Seiberg:2016rsg,Witten:2016cio,Wang:2018qoy,Korpas:2019cwg,Davighi:2019rcd,Manschot:2021qqe} and references therein. Supergravity provides another setting in which tangential structures and gauge--gravitational anomalies interact. Lambert and Moore~\cite{Lambert:2005dx} showed that the auxiliary chiral gauge field of four-dimensional $\CN=1$ new minimal supergravity permits an off-shell formulation on spin$^c$ manifolds. Relatedly, family Donaldson invariants~\cite{Cushing:2023rha} admit a
formulation using twisted $\CN=2$ Yang--Mills theory coupled
to a twisted truncation of background conformal supergravity. It would be interesting to extend the present analysis of generalized symmetries and anomalies to these settings, including the additional fields and the conditions required for anomaly cancellation.}
Readers familiar with these concepts can safely skip to \autoref{sec:u1-variants}.

\subsection{\texorpdfstring{Spin$^c$}{Spin-c} Structures}
Let $M_4$ denote an oriented, smooth, closed (i.e., compact and without boundary) four-manifold. These assumptions will be in force throughout unless specified otherwise. 

\smallskip

\noindent A \emph{spin structure} on $M_4$ is an equivariant lift of the $\sfS\sfO(4)$ structure group of the frame bundle on $M_4$ to $\mathsf{Spin}(4)$. In practice, let $\{\CU_\alpha\}_{\alpha \in I}$ be a good \v{C}ech cover of $M_4$ with transition functions $ g_{\alpha\beta}: \CU_\alpha \cap \CU_\beta \to \sfS\sfO(4)$ such that
\[
  g_{\alpha\beta} = g_{\beta\alpha}^{-1} \quad\text{on}\quad \CU_{\alpha}\cap \CU_{\beta} \qquad\text{and}\qquad g_{\alpha\beta}g_{\beta\gamma}g_{\gamma\alpha} = 1 \quad \text{on} \quad \CU_{\alpha}\cap \CU_{\beta}\cap \CU_{\gamma} ~.
\]
A spin structure on $M_4$ exists provided one can lift the $\sfS\sfO(4)$-valued transition functions to $\mathsf{Spin}(4)$-valued transition functions $\wt{g}_{\alpha\beta}: \CU_{\alpha}\cap \CU_{\beta} \to \mathsf{Spin}(4)$ satisfying the cocycle condition on triple overlaps. This is possible if and only if the second Stiefel--Whitney class of the tangent bundle $TM_4$ vanishes, i.e., iff $w_2(TM_4) = 0 \in H^{2}(M_4;\IZ_2)$. If $w_2(TM_4) \neq 0$, the cocycle condition is violated due to some inevitable factors of $-1$. %
%
For example, $w_2(T\IC\IP^2)\neq0$, so $\IC\IP^2$ is non-spin. Naively, one cannot have spinorial fields on non-spin manifolds, but this obstruction can be overcome when the spinorial fields carry odd $\sfU(1)$ charge. Indeed, it turns out that lifts of $\{g_{\alpha\beta}\}$ to the group $\mathsf{Spin}^{c}(4)$, defined as
\be \label{eq:spin-c-group}
 \mathsf{Spin}^{c}(4) =\big(\mathsf{Spin}(4)\times \sfU(1)\big)/\langle (-1, -1)\rangle ~,
\ee
\emph{always} exist \cite{TeichnerVogt2000}, i.e., every such $M_4$ admits a \emph{spin$^c$ structure}  $\fs = \{ \wt{g}_{\alpha\beta}^{\,c}(x)\}$, given by a collection of maps $\wt{g}^{\,c}_{\alpha\beta}(x) : \CU_{\alpha}\cap \CU_{\beta} \to \mathsf{Spin}^{c}(4)$ (considered up to cocycle isomorphisms) satisfying the cocycle conditions. Essentially, the errant $-1$ signs in the cocycle relations for $\wt{g}_{\alpha\beta}\wt{g}_{\beta\gamma}\wt{g}_{\gamma\alpha}$ on triple overlaps can always be absorbed by $-1$ signs from the cocycle relations of the transition functions of a locally defined line bundle (which need not be globally defined).
\tightfootnote{A complex line bundle has structure group $\sfG\sfL(1,\IC) \cong \IC^\times$. By choosing a Hermitian metric on the fibers of the line bundle, one can reduce the structure group to $\sfU(1)$. Conversely, given a principal $\sfU(1)$ bundle, one can construct an associated complex line bundle.}
The proof of the existence of a spin$^c$ structure on any such $M_4$ relies on the important fact that $w_2(TM_4) \in H^{2}(M_4;\IZ_2)$ always admits an integral lift $\ov{w}_2(TM_4) \in H^{2}(M_4;\IZ)$, due to the existence of a characteristic element \cite{Hopf1958}.
\tightfootnote{This is equivalent to the vanishing of the integral class $W_3 := \mathsf{Bock}(w_2(TM_4)) \in H^{3}(M_4;\IZ)$. Here $\mathsf{Bock}: H^2(M_4;\IZ_2) \to H^3(M_4;\IZ)$ denotes the Bockstein map at degree 2 in the long exact cohomology sequence of $M_4$ induced by the short exact sequence $0 \rightarrow \IZ \xrightarrow{\times 2} \IZ \xrightarrow{\rho_2} \IZ_2 \rightarrow 0$ of coefficients, where $\times 2$ denotes multiplication by $2$ and $\rho_2$ denotes reduction modulo $2$.}

\medskip

The quotient in \eqref{eq:spin-c-group} identifies $(g,z)\sim(-g,-z)$. The map $\det:(g,z)\mapsto z^2$ is invariant under this identification and descends to $\mathsf{Spin}^c(4)$. Therefore, given a spin$^c$ structure $\fs= \{\wt{g}_{\alpha\beta}^{\,c}\}$, one can construct a $\sfU(1)$-valued cocycle $\{ \det\,\wt{g}^{\,c}_{\alpha\beta}\}$ which has a natural action on $\IC$ and thus can be assembled into a complex line bundle $\CL_\fs \to M_4$ called the \emph{determinant line bundle of the spin$^c$ structure $\fs$}. By definition, the Chern class of $\CL_\fs$ is the Chern class of the spin$^c$ structure $\fs$, which also happens to coincide with the chosen integral lift $\ov{w}_2(TM_4)$:
\be
 c_1(\fs) := c_1(\CL_\fs) = \ov{w}_2(TM_4) \in H^{2}(M_4;\IZ) ~.
\ee
Equivalently, a spin$^c$ structure can be described by locally defined chiral spinor bundles $S_\pm$ and a locally defined line bundle $\CW$ whose tensor products $\Sigma_\pm:=S_\pm\otimes\CW$ glue to globally defined bundles. A chiral spin$^c$ spinor, then, is a section of $\Sigma_{\pm}$. Furthermore, $\CW^2$ exists as a globally defined line bundle over $M_4$. The aptly named determinant line bundle of the spin$^c$ structure $\fs$ is simply given as $\CL_{\fs} = \det_{\IC}\,\Sigma_{\pm}$.
\tightfootnote{Note that $\det_{\IC}\,\Sigma_\pm = (\Lambda^{2}S_{\pm}) \otimes \CW^2 \cong \underline{\IC} \otimes \CL_{\fs} \cong \CL_{\fs}$: locally $S_\pm$ is the fundamental representation of an $\sfS\sfU(2)$ factor of $\mathsf{Spin}(4)=\sfS\sfU(2)\times\sfS\sfU(2)$, so its structure group has unit determinant and $\Lambda^2 S_\pm\cong\underline{\IC}$ canonically. Here $\underline{\IC}$ is the trivial complex line bundle over $M_4$.}
Therefore, $c_1(\Sigma_{\pm}) = c_1(\fs)$.

For every spin$^c$ structure $\fs$, the determinant line bundle $\CL_\fs$ is a globally defined complex line bundle on both spin and non-spin manifolds.

The set of isomorphism classes of spin$^c$ structures is a torsor for $H^2(M_4;\IZ)$: given a spin$^c$ structure $\fs = \{\wt{g}^{\,c}_{\alpha\beta}\}$, twisting it by a complex line bundle $L_{\varepsilon} \to M_4$ with cocycle $\{\lambda_{\alpha\beta}\}$ and Chern class $\varepsilon := c_1(L_{\varepsilon})$ yields another spin$^c$ structure: $\fs \mapsto \fs + \varepsilon$, with a spin$^c$ cocycle $\{ \wt{g}_{\alpha\beta}^{\,c} \cdot \lambda_{\alpha\beta} \}$. Recalling the map $\det$ above, the cocycle for the corresponding determinant bundle is twisted by $\lambda_{\alpha\beta}^2$. Therefore, 
\be\label{eq:sbluorg}
  \CL_{\fs + \varepsilon} = \CL_{\fs} \otimes L_{\varepsilon}^{\otimes 2} ~,
\ee
which in turn implies that
\be\label{eq:sbluorg2}
   c_1(\fs + \varepsilon) = c_1(\fs) + 2\varepsilon ~.
\ee
So if $H^2(M_4;\IZ)$ has 2-torsion and $\varepsilon$ happens to be a nonzero 2-torsion element (so that $2\varepsilon = 0$), then $\fs + \varepsilon$ is a spin$^c$ structure distinct from $\fs$ even though their Chern classes coincide. Hence, the Chern class $c_1(\fs)$ of a spin$^c$ structure $\fs$ uniquely labels the spin$^c$ structure only if $H^2(M_4;\IZ)$ has no 2-torsion, in which case there is a canonical bijection
\begin{align}\begin{split}
 \{ \text{spin$^c$ structures on $M_4$} \} &\cong \{ \text{integral lifts $\ov{w}_2(TM_4)$ of $w_2(TM_4)$} \}\\ 
 &\qquad = \{ x \in H^2(M_4;\IZ) ~|~ \rho_2(x) = w_2(TM_4) \} ~.
\end{split}
\end{align}
Choosing a basepoint identifies this torsor with $H^2(M_4;\IZ)$. The fibers of the map $\fs\mapsto c_1(\fs)$ are torsors for the 2-torsion subgroup $H^2(M_4;\IZ)[2]$, so 2-torsion can distinguish structures with the same determinant class. Thus the set of spin$^c$ structures is non-canonically isomorphic to $H^2(M_4;\IZ)$, whereas the set of integral lifts of $w_2(TM_4)$ forgets this 2-torsion data.

\subsection{Connections On \texorpdfstring{Spin$^c$}{Spin-c} Bundles And Determinant Line Bundles}\label{subsec:spinc-connections}
A spin$^c$ bundle with connection on a Riemannian four-manifold $(M_4,g)$ is a principal bundle with connection for the structure group \eqref{eq:spin-c-group}, which projects to the oriented orthonormal frame bundle of $TM_4$, with its connection projecting to the Levi--Civita connection of $g$. Since the spin$^c$ group is \emph{locally} isomorphic to a product of $\mathsf{Spin}(4)$ and $\sfU(1)$, one can represent a connection on a spin$^c$ bundle as a pair $(\omega_4, A)$, where $\omega_4$ is the spin connection and $A$ is the real gauge potential representing the $\sfU(1)$ component. The corresponding Lie algebra-valued connection is formally $\omega_4\oplus\imag A$, and both components transform inhomogeneously.

We use the physics convention in which the locally defined gauge potential $A$ is a real 1-form and its real fieldstrength is
\be
  F:=dA ~.
\ee
All Maxwell, theta angle, and anomaly formulas below use this convention.
\tightfootnote{In the corresponding anti-Hermitian convention, the connection form is $\imag A\in\Omega^1(M_4;\fu(1))$ and its curvature is $\imag F$.}

The potential $A$ is defined locally as the connection on $\CW$. Its global geometric meaning comes from pairing it with the spin connection to form a spin$^c$ connection. The real curvature $F$ obeys shifted flux quantization and can have non-integral periods.
\tightfootnote{\label{foot:spinc-surface-torsion}Caution: a non-spin $M_4$ need not possess \emph{any} 2-cycle supporting a non-integral flux. By the Wu formula, $\langle w_2(TM_4),\rho_2[\Sigma_2]\rangle=[\Sigma_2]\cdot[\Sigma_2]\smod2$, so if the intersection form of $M_4$ is even, the fluxes of $F$ are integral on every integral 2-cycle. The Enriques surface is an example: it is non-spin, but $w_2(TM_4)$ is the mod 2 reduction of a 2-torsion class in $H^2(M_4;\IZ)$, and the intersection lattice on the free quotient, $\mathrm{E}_8(-1)\oplus H$, is even. Its $w_2(TM_4)$ pairs nontrivially only with classes in $H_2(M_4;\IZ_2)$ that come from $\Tor(H_1(M_4;\IZ),\IZ_2)=\IZ_2$ and are not reductions of integral 2-cycles. Curvature periods therefore cannot detect the spin$^c$ shift there, or more generally the flat torsion data distinguishing spin$^c$ structures. A complete treatment of the spin$^c$ sectors must retain this information.}

A smooth homomorphism $\varphi: G_1 \to G_2$ between Lie groups induces a functor $\varphi_*$ between groupoids of principal bundles with connection over $M_4$ for the domain and codomain structure groups $G_1$ and $G_2$, respectively. In particular, if $(P_1 \to M_4,\n_1)$ is a principal $G_1$-bundle with connection $\n_1$, then $\varphi_{*}(P_1,\n_1)$ is a principal $G_2$-bundle over $M_4$ with induced connection $\varphi_{*}\n_1$. See \cite[App.~B]{Moore:2024vsd} for a detailed discussion. This is called the \emph{transfer of structure group associated with a group homomorphism}. In the present case, there \emph{is} a well-defined group homomorphism $\det:\mathsf{Spin}^{c}(4) \to \sfU(1)$ given by $(g,z) \mapsto z^2$. Hence a principal $\mathsf{Spin}^{c}(4)$ bundle $P_{\fs}\to M_4$ determines the transferred principal $\sfU(1)$ bundle $\det_*P_{\fs}:=P_{\fs}\times_{\det}\sfU(1)\to M_4$. Its associated complex line bundle
$\CL_{\fs}:=(\det_*P_{\fs})\times_{\sfU(1)}\IC\cong P_{\fs}\times_{\det}\IC$
is precisely the determinant line bundle of the spin$^c$ structure. A connection $\n^{\fs}$ on $P_{\fs}$ induces a connection on $\CL_{\fs}$, represented in our real convention by a (locally defined) gauge potential $A_{{\rm det}}$ with fieldstrength $F_{{\rm det}}=dA_{{\rm det}}$ satisfying the integral quantization condition
\be\label{eq:integer-flux-det}
  \int_{\Sigma_2}\frac{F_{{\rm det}}}{2\pi} = \left\langle c_1(\CL_\fs), [\Sigma_2]\right\rangle  \in \IZ ~, \quad \text{ for any 2-cycle }\Sigma_2 \text{ in } M_4 ~.
\ee
The determinant representation gives $A_{\mathrm{det}}=2A$ and hence $F_{\mathrm{det}}=2F$. For a fixed spin$^c$ structure $\fs$, the normalized curvature therefore satisfies
\be\label{eq:spin-c-physics-v3}
  \int_{\Sigma_2}\frac{F}{2\pi}
  =\frac{1}{2}
    \left\langle c_1(\CL_\fs),[\Sigma_2]\right\rangle ~.
\ee
This fixes the de Rham flux class, while allowing the connection to vary within that class. Since $c_1(\CL_\fs)\equiv w_2(TM_4)\smod2$, it implies the following weaker parity constraint:
%
\tightfootnote{This holds for every integral 2-cycle $\Sigma_2$. If $[\Sigma_2]$ is torsion in $H_2(M_4;\IZ)$, both sides vanish: $\int_{\Sigma_2}F=0$, and $\langle w_2(TM_4),\rho_2[\Sigma_2]\rangle=\rho_2\langle c_1(\CL_\fs),[\Sigma_2]\rangle=0$. A basis of $H_2(M_4;\IZ)/\mathsf{torsion}$ therefore describes the de Rham flux lattice, and the remaining torsion data are discussed in \footrefnew{foot:spinc-surface-torsion}.}
\be\label{eq:spin-c-math}
  \int_{\Sigma_2}\frac{F}{2\pi} \in \left\{ \begin{array}{ll} 
    \IZ ~, & \text{if } \langle w_2(TM_4), \rho_2[\Sigma_2]\rangle = 0 ~,\\
    \IZ + \frac{1}{2} ~, & \text{if } \langle w_2(TM_4), \rho_2[\Sigma_2]\rangle = 1
  \end{array}\right. ~. 
\ee
Since $w_2(TM_4)$ has $\IZ_2$ coefficients, the integral class $[\Sigma_2]$ must first be reduced modulo $2$ by $\rho_2:H_2(M_4;\IZ)\to H_2(M_4;\IZ_2)$. Whenever we write $\langle w_2(TM_4),[\Sigma_2]\rangle$ or $\int_{\Sigma_2}w_2(TM_4)$ elsewhere, this reduction is implicit.
It is conventional in physics to denote the $\smod 2$ pairing $\langle w_2(TM_4), \rho_2[\Sigma_2]\rangle$ by an integral
\be
 \int_{\Sigma_2}w_2(TM_4) := \langle w_2(TM_4), \rho_2[\Sigma_2]\rangle \in \IZ/2\IZ ~.
\ee
With this notation, the quantization condition can be written as
\be\label{eq:spin-c-physics-v1}
  2\int_{\Sigma_2}\frac{F}{2\pi} \equiv \int_{\Sigma_2}w_2(TM_4) \smod 2 ~, 
\ee
or equivalently,
\be \label{eq:spin-c-physics-v2}
 \int_{\Sigma_2}\frac{F}{2\pi} \in \IZ + \frac{1}{2} \int_{\Sigma_2}w_2(TM_4) \subset \IR ~. 
\ee
Here the right hand side is the inverse image in $\IR$ of the canonical class $\frac{1}{2}\int_{\Sigma_2}w_2(TM_4)\in\IR/\IZ$. Equivalently, after representing the $\IZ/2\IZ$-valued integral by $0$ or $1$, it is the corresponding coset $\IZ$ or $\IZ+\frac{1}{2}$. Thus \eqref{eq:spin-c-physics-v2} is precisely the parity constraint \eqref{eq:spin-c-math}.

We now describe the flux classes obtained by varying the spin$^c$ structure. Choose a basepoint $\fs_0$ and set $\ov{w}_2(TM_4)=c_1(\CL_{\fs_0})$. By \eqref{eq:sbluorg2}, twisting $\fs_0$ by $L_\varepsilon$ changes the determinant Chern class by $2\varepsilon$, and therefore changes the normalized de Rham flux class by the image of $\varepsilon$ in real cohomology.
Every integral lift of $w_2(TM_4)$ occurs as the first Chern class of a spin$^c$ structure, although in the presence of 2-torsion it need not determine that structure uniquely. We write $a_{\IQ}$ for the image of $a\in H^2(M_4;\IZ)$ under the map that forgets torsion,
  \be\label{eq:embedding}
    H^2(M_4;\IZ) \twoheadrightarrow H^2(M_4;\IZ)/\mathsf{torsion} \subset H^2(M_4;\IQ) ~, \qquad a\longmapsto a_{\IQ} ~.
  \ee
and $\Lambda$ for the image of $H^2(M_4;\IZ)$. We define
\be
   \bm{\mu} := \ov{w}_2(TM_4)_{\IQ} \in H^2(M_4;\IQ) ~.
\ee
The quantization condition can be stated more invariantly as
\be\label{eq:spin-c-moore-witten}
  \left[\frac{F}{2\pi}\right] \in \Lambda + \frac{1}{2} \bm{\mu} ~. 
\ee
Although the representative $\bm\mu$ is obtained from a chosen basepoint spin$^c$ structure, the affine coset $\Lambda+\bm\mu/2$ is independent of that choice: twisting the basepoint by $\varepsilon$ sends $\bm\mu/2$ to $\bm\mu/2+\varepsilon_{\IQ}$ and hence translates it by an element of $\Lambda$. The coset describes the union of the flux classes obtained by twisting any basepoint.

\begin{remark}\label{rem:spinc-terminology}\
  \begin{enumerate}
    \item The discussion in this section has been limited to four-manifolds. More generally, every oriented smooth manifold of dimension at most $4$ admits a spin$^c$ structure. 
    \item Every oriented 3-manifold is in fact parallelizable, and hence spin, but it need not carry a preferred spin structure. Sometimes the term `non-spin three-manifold' is used to mean an oriented 3-manifold on which no spin structure has been chosen \cite{Seiberg:2016gmd}. We instead reserve \emph{non-spin manifold} for a manifold that does not admit a spin structure. Correspondingly, a \emph{non-spin theory} is a theory whose definition does not require a choice of spin structure.
    \item A spin$^c$ structure geometrizes the spin-charge relation. For fields or local operators of integer $\sfU(1)$ charge $Q$, the quotient by $\langle(-1,-1)\rangle$ in \eqref{eq:spin-c-group} requires
    \be\label{eq:spin-charge-relation}
      (-1)^F=e^{\imag\pi Q} ~.
    \ee
    Thus odd-charge local particles are fermions and even-charge local particles are bosons \cite{Seiberg:2016gmd}. In this paper, a spin$^c$ theory is one that is defined using spin$^c$ structures. The spin$^c$ structure may be fixed as background data, or the path integral may sum over spin$^c$ structures when the associated electromagnetic gauge field is dynamical, as explained in \remref{rem:sum-spinc-structures}. 
    
    Note, however, that charged matter alone does not impose the spin-charge relation: a charge-one scalar, for instance, violates it. Conversely, when the spin and charge assignments obey \eqref{eq:spin-charge-relation}, the theory can be defined using spin$^c$ structures alone.
  \end{enumerate}
\end{remark}

\subsection{Different Variants Of Maxwell Theory}\label{sec:u1-variants}
We first consider Maxwell theory with an ordinary $\sfU(1)$ connection on a complex line bundle $L \to M_4$. The action can be written as
\be\label{eq:Maxwell}
  \begin{aligned}
    S_{\text{Maxwell}} &= \int_{M_4} d^{4}x\left(\frac{1}{4e^2}\sqrt{g} F_{\mu\nu}F^{\mu\nu} - \frac{\imag \theta}{32\pi^2}\varepsilon^{\mu\nu\rho\sigma}F_{\mu\nu}F_{\rho\sigma} \right) \\
    &= \int_{M_4}\left(\frac{1}{2e^2}F \wedge \star F - \frac{\imag \theta}{8\pi^2}F \wedge F \right) ~,
  \end{aligned}
\ee
where $A$ is the local gauge potential of a unitary connection on a complex line bundle $L\to M_4$, with curvature $F$ (so locally, $F=dA$, i.e., $F_{\mu\nu}=\partial_{\mu}A_{\nu}-\partial_{\nu}A_{\mu}$). Thus $[F/(2\pi)]_{\mathrm{dR}}$ is the image of $c_1(L)$ modulo torsion, i.e., its class in $\Lambda\subset H^2(M_4;\IQ)\subset H^2(M_4;\IR)$ (cf. \autoref{sec:spinc}). For this ordinary $\sfU(1)$ flux lattice, the electromagnetic theta angle is $2\pi$-periodic on spin manifolds and $4\pi$-periodic for a theory defined uniformly on arbitrary oriented manifolds.
\tightfootnote{\label{foot:ordinary-u1-theta-enriques}The period on a fixed non-spin manifold can nevertheless be $2\pi$ when its free intersection form is even. The Enriques surface is an example. See \footrefnew{foot:spinc-surface-torsion}.}

The $4\pi$ periodicity described above concerns an ordinary integral $\sfU(1)$ flux lattice. It does not apply to the spin$^c$ gauge potential used in the rest of this paper, for which the natural theta coupling contains both electromagnetic and gravitational terms and is $2\pi$-periodic. These distinctions are discussed in \appref{app:theta-periodicity} and used again in \autoref{sec:chiral-defect-revisited}.

Different choices for the statistics of the fundamental electric and magnetic line operators give distinct versions of Maxwell theory on non-spin manifolds \cite{Metlitski:2015yqa,Wang:2018qoy,Ang:2019txy}. For a related discussion of Abelian gauge theory on non-spin four-manifolds, including the behavior of Wilson--'t Hooft lines under electromagnetic duality and the appearance of a spin$^c$ gauge field in a duality frame, see \cite[§4]{Cordova:2018acb}.
The electromagnetic charge lattice $\Gamma_{\mathrm{em}}$ of the theory is a rank-2 lattice equipped with a natural symplectic pairing, the quantization of which is the Dirac--Schwinger--Zwanziger (DSZ) condition,
\be
  \langle \gamma_1, \gamma_2 \rangle := Q_{e,1} Q_{m,2} - Q_{e,2} Q_{m,1} \in 2\pi \IZ ~, \quad \text{for all} \quad \gamma_{i} = (Q_{e,i}, Q_{m,i}) \in \Gamma_{\mathrm{em}} ~,
\ee
which reflects the quantization of the angular momentum of the electromagnetic field produced by a pair of dyons with charges $\gamma_1=(Q_{e,1},Q_{m,1})$ and $\gamma_2 =(Q_{e,2}, Q_{m,2})$, respectively.
\tightfootnote{More precisely, the angular momentum of the electromagnetic field configuration is an integer multiple of $1/2$ (in appropriately chosen units), which corresponds to integrality of the Dirac pairing (given by $\frac{1}{2\pi}$ times the symplectic pairing on $\Gamma_{\mathrm{em}}$).}
The charge lattice contains the fundamental Wilson line even without a dynamical minimally charged particle: the line need not admit a genuine local endpoint.

The magnetic charge in Dirac units is defined locally by
\be\label{eq:local-magnetic-charge}
  m:=\int_{S^2_{\mathrm{link}}}\frac{F}{2\pi} ~,
\ee
where $S^2_{\mathrm{link}}$ is a small positively oriented sphere linking the line once. For an ordinary $\sfU(1)$ connection, this flux is the first Chern number on the linking sphere, so $m\in\IZ$. The same local integrality holds for a spin$^c$ connection, as shown in \autoref{sec:fractionalization}. The flux definition requires no choice of spin structure. Spin or spin$^c$ data determine the allowed line operators and their statistics.

After choosing electric and magnetic generators, the lattice points are labeled by $(n,m)\in\IZ^2$. It is convenient to combine the physical charges into
\be
  Q_e+\imag Q_m=e(n+m\tau) ~,
  \qquad
  \tau:=\frac{\theta}{2\pi}+\frac{2\pi\imag}{e^2} ~.
\ee
This normalization follows from the Maxwell action \eqref{eq:Maxwell} together with the DSZ pairing above. In particular, $\Im\,\tau=\frac{2\pi}{e^2}>0$, and the line labeled by $(n,m)$ has charges
\be\label{eq:dyon-charges}
  Q_e=\left(n+\frac{\theta}{2\pi}m\right)e ~,
  \qquad
  Q_m=\frac{2\pi}{e}m ~.
\ee
It corresponds to the Wilson--'t Hooft line $W^nT^m$.

Magnetic monopoles require the compact gauge group $\sfU(1)$.
For a theory defined without an independent spin structure, the statistics of the electric and magnetic probe lines are part of the specification of the theory.  In particular, we must specify whether the unit electric probe line, labeled by $(n,m) = (1,0)$, is bosonic or fermionic. 
The question can also be posed for the monopole $(n,m)=(0,1)$ or the dyon $(n,m)=(1,1)$. In writing charge assignments, we are implicitly assuming that $\theta = 0$, as otherwise a magnetically charged line has a $\theta$-dependent shift in its physical electric charge via the Witten effect \cite{Witten:1979ey}, already evident in \eqref{eq:dyon-charges}. As outlined in \cite{Wang:2018qoy}, if any two of these lines are bosonic or any two are fermionic, the third is necessarily fermionic, whereas if one is bosonic and another is fermionic, the third is necessarily bosonic. These statements follow from the relation
\be\label{eq:spin-inevitable}
 \text{spin}(1,1) = \text{spin}(1,0) + \text{spin}(0,1) + \frac{1}{2} \smod 1 ~,
\ee
where the last term is the contribution from the angular momentum of the electromagnetic field configuration itself. These considerations lead to four possible variants
of $\sfU(1)$ gauge theories, listed in \autoref{tbl:u1-variants}, where we have adopted in column 1 the naming conventions of \cite{Ang:2019txy}. The variants differ by the fractionalization of Lorentz symmetry on their line operators, and the gauging maps and duality interfaces between them are constructed in \cite{Kan:2024fbn}.
\begin{table}[h]
  \centering
  \begin{tabular}{|c|c|c|c|c|}\hline
    Label & $(n,m)=(1,0)$ & $(n,m)=(0,1)$ & $(n,m)=(1,1)$ & generating lines\\ \hline
    $\sfU(1)_{\text{b}}$ & boson & boson & fermion & $W_{\text{b}}, T_{\text{b}}$ \\ \hline
    $\sfU(1)_{\text{f}}$ & boson & fermion & boson & $W_{\text{b}}, T_{\text{f}}$ \\ \hline
    $\mathsf{Spin}^{c}_{\text{b}}$ & fermion & boson & boson & $W_{\text{f}}, T_{\text{b}}$ \\ \hline
    $\mathsf{Spin}^{c}_{\text{f}}$ & fermion & fermion & fermion & $W_{\text{f}}, T_{\text{f}}$ \\ \hline
  \end{tabular}
  \caption{The four versions of $\sfU(1)$ gauge theory at $\theta = 0$. Columns 2--4 give the statistics assigned to the indicated probe lines (or to heavy particles supported on them). Column 5 lists generating lines.}
  \label{tbl:u1-variants}
\end{table}
The last two rows describe theories with a fermionic electric generator, whose gauge field is naturally a spin$^c$ connection as reviewed in \autoref{sec:spinc}. These correspond to  \emph{fermionic matter} in the parlance of \cite{Metlitski:2015yqa}
\tightfootnote{In the condensed-matter literature, these are also called \emph{electronic phases} \cite{Cheng:2025ube}. Although every oriented 3-manifold admits a spin structure, the choice is not canonical: a spin TFT may depend on that choice, whereas a bosonic TFT does not require one. A spin$^c$ TFT is a fermionic TFT with $\sfU(1)$ symmetry obeying the spin-charge relation. It is naturally formulated on spin$^c$ backgrounds and requires no spin structure. See \remref{rem:spinc-terminology}.}
which means that all excitations with an odd electric charge (and zero magnetic charge) are fermions, whereas all excitations with even electric charge (and zero magnetic charge) are bosons. In this work, we instead refer to them as spin$^c$ theories. The key feature of these theories is this spin-charge relation, which notably implies that the electron in this theory is a fermion (or equivalently, the endpoint of the Wilson line, when such an endpoint exists, is a fermion).
\tightfootnote{As discussed in \autoref{sec:spinc}, the local $\fu(1)$ potential $A$ combines with the spin connection to define a global spin$^c$ connection. Parallel transport of an odd-charge spin$^c$ field is defined using the combined spin$^c$ connection.}
In particular, the last entry in the table corresponds to ``all-fermion electrodynamics,'' which has a gravitational anomaly \cite{Wang:2013zja,Thorngren:2014pza,Kravec:2014aza}. See \cite[§6]{Thorngren:2014pza}, \cite[§3.2]{Wang:2018qoy} and \cite[§2]{Cordova:2018acb} for nice topological interpretations of the anomaly.

As emphasized in \cite{Metlitski:2015yqa,Wang:2018qoy}, electromagnetic duality corresponds to a relabeling of dyons that preserves their statistics and interactions. In the conventions of \cite[§2.3]{Ang:2019txy}, the generators act on a line of charge $(n,m)$ as
\be\label{eq:ST-lines}
  S:(n,m)\longmapsto(m,-n) ~,
  \qquad
  T:(n,m)\longmapsto(n-m,m) ~,
\ee
and on the coupling as $S:\tau\mapsto-1/\tau$ and $T:\tau\mapsto\tau+1$. With the coupling transformed simultaneously, these maps define dualities up to purely gravitational invertible phases. At the level of the four variants in \autoref{tbl:u1-variants}, $T$ exchanges $\sfU(1)_{\text{b}}$ and $\sfU(1)_{\text{f}}$, while $S$ exchanges $\sfU(1)_{\text{f}}$ and $\mathsf{Spin}^{c}_{\text{b}}$. Moreover, $S$ fixes $\sfU(1)_{\text{b}}$ and $T$ fixes $\mathsf{Spin}^{c}_{\text{b}}$. Thus the three non-anomalous variants form a single $\sfS\sfL(2,\IZ)$ orbit. By contrast, the anomalous all-fermion theory $\mathsf{Spin}^{c}_{\text{f}}$ is fixed by both $S$ and $T$. Equivalently, the three non-anomalous spin assignments are the three even quadratic refinements of the mod-2 DSZ pairing, whereas the all-fermion assignment is the unique odd quadratic refinement.
\tightfootnote{The interpretation of duality as an equivalence between Maxwell theories at different couplings also appears in \cite{Cordova:2023ent}.}

In the bulk of this paper, the Maxwell sector is the $\mathsf{Spin}^{c}_{\text{b}}$ theory (row 3 of \autoref{tbl:u1-variants}). We couple this Maxwell sector to a dynamical massless unit-charge Dirac fermion and refer to the resulting theory as spin$^c$ QED$_4$.

\section{Quantum Electrodynamics On Non-Spin Manifolds (\texorpdfstring{Spin$^c$}{Spin-c} \texorpdfstring{QED$_4$}{QED4})}\label{sec:spinc-QED4}

\subsection{The Theory In A Fixed \texorpdfstring{Spin$^c$}{Spin-c} Sector}
\label{subsec:fixed-spinc-sector}

Let $(M_4,g,\fo)$ be a smooth, closed, oriented Riemannian four-manifold. We consider the $\mathsf{Spin}^c_{\mathrm b}$ Maxwell sector described in \autoref{sec:u1-variants}, coupled to a dynamical massless unit-charge Dirac fermion. We call the resulting theory spin$^c$ QED$_4$ (made precise in \autoref{subsec:summed-spinc-theory}). The spin-charge relation allows its charged fermions to be defined using a spin$^c$ structure, without choosing an independent spin structure. This formulation applies in particular when $M_4$ is non-spin.

\medskip

\noindent We first fix a spin$^c$ structure $\fs$ and integrate over connections compatible with it. The action in this sector is
\be\label{eq:L-QED-4}
 S_{\text{QED}}[A,\Psi,\ov{\Psi};g;\fo;\fs;(e^2,\theta)] = \int_{M_4}\left( \frac{1}{2e^2}F \wedge \star F - \frac{\imag \theta}{8\pi^2} F \wedge F +  \imag \ov{\Psi}\slashed{D}\Psi \,\vol(g)\right) ~,
\ee
where $\fo$ is the orientation of $M_4$, $\star$ is the Hodge star with respect to the metric $g$, $\Psi$ is a Dirac spin$^c$ spinor (i.e., a section of 
$\Sigma_{+} \oplus \Sigma_{-} \to M_4 $),
\tightfootnote{Recall that $\Sigma_{\pm}=S_{\pm}\otimes\CW$ already includes the locally defined factor $\CW$.}
$\slashed{D}$ is the Dirac operator associated with the spin$^c$ connection, and $\vol(g)$ is the Riemannian volume form. In our conventions $A=A_\mu dx^\mu$ is real and $F=dA$ locally. Thus $D_{\mu}\Psi = \n_{\mu}\Psi + \imag A_{\mu}\Psi$. A gauge transformation $A\mapsto A+d\lambda$ acts by $\Psi\mapsto e^{-\imag\lambda}\Psi$ and $\ov\Psi\mapsto\ov\Psi e^{\imag\lambda}$. The parameters $(e^2,\theta)$ are the coupling and the coefficient of the purely electromagnetic theta term in \eqref{eq:L-QED-4}. On a spin$^c$ background, it is often more natural to combine this term with its gravitational counterpart into a $2\pi$-periodic gauge--gravitational theta coupling (see \cite[App.~A.2]{Seiberg:2016rsg}). In \appref{app:theta-periodicity} we discuss these conventions and the corresponding periodicities of the theta couplings -- see also \autoref{sec:chiral-defect-revisited}. As explained in \autoref{sec:ABJ-spinc}, the ABJ anomaly gives an interface shifting both the electromagnetic and gravitational theta couplings. In the chiral wall discussion, $\mathsf{QED}_\theta$ denotes this gauge--gravitational family. It differs from the electromagnetic-only family of \eqref{eq:L-QED-4} by the gravitational term. The existence of this topological interface is the spin$^c$ analog of the well-known fact that one can, at no energy cost, change the parameter $\theta$ by performing a chiral rotation in (massless) QED.

Let $\CA(\fs)$ be the space of spin$^c$ connections compatible with $\fs$, and let $\CG$ be the group of gauge transformations. All connections in $\CA(\fs)$ have the same de Rham flux class
\be\label{eq:fixed_spin_struct}
    \left[\frac{F}{2\pi}\right]_{\mathrm{dR}} =\frac{1}{2}c_1(\CL_{\fs})_{\IR} \quad\text{in }H^2(M_4;\IR) ~.
\ee
The determinant connection $A_{\rm det}=2A$ is a genuine $\sfU(1)$ connection on $\CL_{\fs}$, so $F_{\rm det}/2\pi$ has integral periods. The unit-charge fermion couples to the full spin$^c$ connection, whose electromagnetic part $A=A_{\rm det}/2$ therefore obeys the shifted quantization \eqref{eq:fixed_spin_struct}.

The partition function in the fixed spin$^c$ sector is
\be\label{eq:fixed-spinc-partition}
  \sfZ_{M_4}^{\fs}[g;\fo;(e^2,\theta)] := \int_{\CA(\fs)/\CG}[\CD A][\CD\Psi][\CD\ov{\Psi}] e^{-S_{\text{QED}}[A,\Psi,\ov{\Psi};g;\fo;\fs;(e^2,\theta)]} ~.
\ee
In the Euclidean path integral, $\ov{\Psi}$ is an independent Grassmann-valued spinorial field. In \eqref{eq:fixed-spinc-partition}, the connection is dynamical, but the spin$^c$ structure $\fs$ and hence the de Rham flux class \eqref{eq:fixed_spin_struct} remain fixed. This defines the theory in a fixed spin$^c$ sector. Compact QED requires a sum over all spin$^c$ sectors.

\subsection{The Sum Over \texorpdfstring{Spin$^c$}{Spin-c} Structures}
\label{subsec:summed-spinc-theory}

We define the fully summed compact theory by including all spin$^c$ sectors compatible with the orientation of $M_4$. 

To characterize the sum over spin$^c$ sectors, we proceed as follows. Choose an auxiliary basepoint spin$^c$ structure $\fs_0$. Since the set of spin$^c$ structures is a torsor for $H^2(M_4;\IZ)$, the map
\be
  H^2(M_4;\IZ)\longrightarrow \{\text{spin$^c$ structures on }M_4\} ~,
  \qquad \varepsilon\longmapsto\fs_0+\varepsilon ~,
\ee
is a bijection. If $L_{\varepsilon}$ is a line bundle with $c_1(L_{\varepsilon})=\varepsilon$, then by \eqref{eq:sbluorg} and \eqref{eq:sbluorg2},
\be
  \CL_{\fs_0+\varepsilon}=\CL_{\fs_0}\otimes L_{\varepsilon}^{\otimes2} ~,
  \qquad
  c_1(\CL_{\fs_0+\varepsilon})=c_1(\CL_{\fs_0})+2\varepsilon ~.
\ee
Thus, after mapping integral cohomology to real cohomology, the flux in the sector $\fs_0+\varepsilon$ is
\be\label{eq:spinc-sector-flux}
  \left[\frac{F_{\varepsilon}}{2\pi}\right]_{\mathrm{dR}} =\frac{1}{2}c_1(\CL_{\fs_0})_{\IR}
   +\varepsilon_{\IR} \quad\text{in }H^2(M_4;\IR) ~.
\ee
Equation \eqref{eq:spinc-sector-flux} describes the curvature flux in each sector. The fully summed compact QED partition function is
\be\label{eq:full-spinc-partition}
\begin{aligned}
  \sfZ_{M_4}[g;\fo;(e^2,\theta)]
  &:= \sum_{\fs\in\operatorname{Spin}^c(M_4)} \sfZ_{M_4}^{\fs}[g;\fo;(e^2,\theta)] \\
  &= \sum_{\varepsilon\in H^2(M_4;\IZ)}
     \int_{\CA(\fs_0+\varepsilon)/\CG}[\CD A][\CD\Psi][\CD\ov{\Psi}] 
  e^{-S_{\text{QED}}[A,\Psi,\ov{\Psi};g;\fo;\fs_0+\varepsilon;(e^2,\theta)]} ~.
\end{aligned}
\ee
The basepoint $\fs_0$ is only a device for coordinatizing the torsor and is not background data on which the fully summed partition function depends. Replacing $\fs_0$ by $\fs_0+\delta$ simply reindexes the sum by $\varepsilon\mapsto\varepsilon-\delta$. This is why $\fs_0$ does not appear among the arguments on the left-hand side of \eqref{eq:full-spinc-partition}.

\begin{remark}[To Sum Or Not To Sum\ldots{} Over Spin$^c$ Structures]\label{rem:sum-spinc-structures}
The fixed-sector partition function \eqref{eq:fixed-spinc-partition} and the summed partition function \eqref{eq:full-spinc-partition} retain different background data. The former depends on a prescribed spin$^c$ structure $\fs$, while the latter sums over these structures and does not depend on any chosen basepoint $\fs_0$. Throughout the following discussion, spin$^c$ QED$_4$ denotes the summed theory unless a fixed sector is specified explicitly. The sum runs over spin$^c$ structures, which are finer than determinant classes. Twisting $\fs$ by a nonzero $\varepsilon\in H^2(M_4;\IZ)$ with $2\varepsilon=0$ gives a distinct spin$^c$ structure with the same $c_1(\CL_{\fs})$, and more generally torsion twists are invisible to the de Rham flux. Such sectors are distinct terms in \eqref{eq:full-spinc-partition}.

The monopole and dyon contributions to Donaldson--Witten theory~\cite{Witten:1994cg,Moore:1997pc} involve a related sum over spin$^c$ structures weighted by Seiberg--Witten invariants. See also \footrefnew{foot:donaldson-witten}. 
\end{remark}

The sum over spin$^c$ sectors does not introduce dynamical magnetically charged matter. In the theory considered here, $dF = 0$ away from magnetic defect insertions. The dynamical unit-charge fermion screens integral-charge Wilson lines, while the magnetic 1-form symmetry remains. We next study its global realization in \autoref{sec:magnetic-spinc}.

\subsection{The Magnetic 1-Form Symmetry Of \texorpdfstring{Spin$^c$}{Spin-c} \texorpdfstring{QED$_4$}{QED4}}\label{sec:magnetic-spinc}
\subsubsection{The \texorpdfstring{$w_2$}{w2}-Twist}\label{subsubsec:w2-twist}
In the absence of dynamical magnetic charges, the Bianchi identity $dF = 0$ implies that the theory has a magnetic 1-form symmetry generated by the family of topological operators
\be\label{eq:Um}
  U_{m}(\alpha;\Sigma_2) = \exp\left(\imag \alpha \int_{\Sigma_2} \frac{F}{2\pi}\right) ~, \qquad \alpha\in\IR ~, \quad \Sigma_2 \subset M_4 \text{ closed} ~.
\ee
Spin$^c$ flux quantization implies that this family is not
$2\pi$-periodic:
\be\label{eq:Gee-sigma}
  G(\Sigma_2) := U_{m}(\alpha=2\pi; \Sigma_2) = (-1)^{\int_{\Sigma_2}w_2(TM_4)} ~.
\ee
Equivalently, the composition law is
\be\label{eq:twisted-identification}
  U_m(\alpha+2\pi;\Sigma_2)
  =G(\Sigma_2)\,U_m(\alpha;\Sigma_2) ~.
\ee
It follows that $U_m(\alpha+4\pi;\Sigma_2)=U_m(\alpha;\Sigma_2)$. The factor $G(\Sigma_2)$ is a c-number, fixed by $w_2(TM_4)$ and the mod-$2$ class of $\Sigma_2$.

For a closed oriented embedded surface $\Sigma_2$, the splitting $TM_4|_{\Sigma_2}\cong T\Sigma_2\oplus N_{\Sigma_2/M_4}$ and the Whitney sum formula give $w_2(TM_4)|_{\Sigma_2}=w_2(N_{\Sigma_2/M_4})$, since every closed oriented surface is spin. The latter class is the Euler class of the normal bundle modulo two \cite{MilnorStasheff1974}, and $\langle e(N_{\Sigma_2/M_4}),[\Sigma_2]\rangle=\Sigma_2\cdot\Sigma_2$. Hence $\langle w_2(TM_4),[\Sigma_2]\rangle\equiv\Sigma_2\cdot\Sigma_2\smod2$, which also follows from the Wu formula, and we obtain the following.
\begin{proposition}[The $w_2$-twisted composition law]\label{prop:w2-twist}
For every $\alpha\in\IR$ and every closed oriented embedded surface $\Sigma_2\subset M_4$, the magnetic surface operators of spin$^c$ QED$_4$ obey $U_m(\alpha+2\pi;\Sigma_2)=G(\Sigma_2)\,U_m(\alpha;\Sigma_2)$, with
\begin{equation}\label{eq:gravitational-surface-self-intersection}
  G(\Sigma_2)=(-1)^{\Sigma_2\cdot\Sigma_2}=(-1)^{\int_{\Sigma_2}e(N_{\Sigma_2/M_4})} ~.
\end{equation}
In particular, the family $U_m(\alpha;\Sigma_2)$ is $4\pi$-periodic, and it is $2\pi$-periodic on $\Sigma_2$ precisely when $\Sigma_2\cdot\Sigma_2$ is even.
\end{proposition}
For example, $\IC\IP^1\subset\IC\IP^2$ has self-intersection one, so $U_m(2\pi;\IC\IP^1)=-1$ without any additional insertion.

Thus $G(\Sigma_2)=(-1)^{\int_{\Sigma_2}e(N_{\Sigma_2/M_4})}$ is an invertible theory of the surface together with its normal bundle. It is not determined by $\Sigma_2$ alone: the projective line $\IC\IP^1\subset\IC\IP^2$ and a 2-sphere bounding a 3-ball embedded in $\IC\IP^2$ are both 2-spheres, yet $G=-1$ for the first and $G=+1$ for the second. On closed surfaces, its value depends only on the class of $\Sigma_2$ in $H_2(M_4;\IZ_2)$. This suggests a connection with the generalized-cohomological description of symmetry-protected topological phases \cite{Freed:2016rqq,Gaiotto:2017zba}, in which invertible field theories \cite{Freed:2004yc} of different dimensions decorate symmetry defects and their junctions. On surfaces with boundary, the bare expression is relative and requires compensating boundary data. See \appref{app:sheet-fusion-transport}.

\begin{figure}[htbp]
  \centering
  \includegraphics{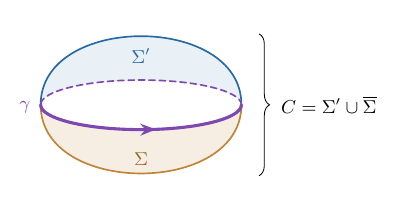}
  \caption{\label{fig:wilson-surface-change}Two surfaces $\Sigma$ and $\Sigma'$ with the same boundary $\gamma$. Together they form the closed 2-cycle $C=\Sigma'\cup\ov{\Sigma}$, and the two charge-$n$ Wilson surfaces differ by $G(C)^n$.}
\end{figure}

\begin{remark}[Surface dependence of Wilson surfaces]\label{rem:wilson-dressing}
For an open surface $\Sigma$ with $\gamma\subset\partial\Sigma$, define $W^q(\gamma;\Sigma):=\exp\left(\imag q\int_{\Sigma}F\right)$ for the charge-$q$ Wilson surface, the extension of $U_m(2\pi q;\,\cdot\,)$ in \eqref{eq:Um} to open surfaces. If $qA$ is a connection on a line bundle, as for $q\in2\IZ$, where $qA=\frac{q}{2}A_{\mathrm{det}}$, or for $q\in\IZ$ on a spin manifold, its holonomy is a differential character $\chi_{qA}$ on $1$-cycles with $\chi_{qA}(\partial\Sigma)=\exp\left(\imag q\int_\Sigma F\right)$ \cite[Def.~34.11 and Sec.~36]{Moore:2025tmt}. For $\partial\Sigma=\gamma$, $W^q(\gamma;\Sigma)=\chi_{qA}(\gamma)$ is then independent of $\Sigma$, which is why it is labeled by $\gamma$. Otherwise, the surface is part of the data, as we now show. Let $\Sigma'$ be another surface with $\partial\Sigma'=\partial\Sigma$, so that $C:=\Sigma'\cup\ov{\Sigma}$ is a closed 2-cycle disjoint from singular insertions (see \autoref{fig:wilson-surface-change}). For integral $n$, \eqref{eq:Gee-sigma} gives
\begin{equation}\label{eq:wilson-dressing-surface-change}
  \frac{W^n(\gamma;\Sigma')}{W^n(\gamma;\Sigma)}
  =G(C)^n
  =(-1)^{n\langle w_2(TM_4),[C]\rangle} ~.
\end{equation}
Even an integral Wilson surface therefore retains a sign when $n$ is odd and $M_4$ is non-spin (see also \cite[Eqs.~(1.10) and (B.36)]{BergCappelliVilla2026}). The genuine odd-charge Wilson line is the worldline of a charged fermion. Its holonomy is defined by the parallel transport of a spin$^c$ spinor along $\gamma$, using additional normal data along $\gamma$, and the resulting tangential factor compensates this sign. For fractional $q$, the ratio is $\exp(\imag q\int_C F)$, so the Wilson surface requires a specified surface and endpoint data. Such Wilson surfaces arise when the chiral wall acts on 't Hooft lines, as described in \autoref{sec:coupling-tft-to-spinc-qed4}.
\end{remark}

Uniformly over oriented backgrounds, this $4\pi$-periodic family is described by $\wt{\sfU(1)}_{m}^{(1)}\cong\IR/4\pi\IZ$, which fits into the central extension
\be\label{eq:extension-sequence}
  1 \to \IZ_{2}^{\text{grav}} \to \wt{\sfU(1)}_{m}^{(1)}
  \to \sfU(1)_{m}^{(1)} \to 1 ~.
\ee
The projection reduces $\alpha$ modulo $2\pi$, and its nontrivial kernel element is represented by $G(\Sigma_2)$. As an extension of continuous parameter groups, this is the non-split double cover of $\IR/2\pi\IZ$. The superscript ``grav'' emphasizes that the extension involves combined internal--tangential data: the kernel is represented by the gravitational surface factor $G$, determined by the ambient tangent bundle.

The kernel is invisible to genuine local lines. A small sphere linking a smooth magnetic line has trivial normal bundle, so $G=1$ there, and the magnetic charge of every genuine line is integral (see \autoref{sec:fractionalization}). The linking action therefore factors through the $2\pi$-periodic $\sfU(1)_{m}^{(1)}$, and the kernel is detected only by surfaces of odd self-intersection. On a spin background, or on an individual non-spin background on which $w_2(TM_4)$ pairs trivially with every integral 2-cycle, $G$ is invisible altogether. The latter happens, for example, on the Enriques surface of \footrefnew{foot:spinc-surface-torsion}.

\begin{remark}[Redefining the surface operators]\label{rem:surface-counterterms}
Since $G(\Sigma_2)$ is a c-number, the twist can be moved by $\alpha$-dependent c-number redefinitions of $U_m$. There are two natural candidates.
\begin{enumerate}
  \item Let $c_0\in H^2(M_4;\IZ)$ be a characteristic element, $\rho_2(c_0)=w_2(TM_4)$. Then
  \be
    U_m^{c_0}(\alpha;\Sigma_2):=U_m(\alpha;\Sigma_2)\,e^{-\imag\alpha\langle c_0,[\Sigma_2]\rangle/2} ~,
  \ee
  is $2\pi$-periodic, and its phase is linear in $[\Sigma_2]$. In the summed theory, it measures the image of the twisting class $\varepsilon$ in $H^2(M_4;\IZ)/\mathsf{torsion}$, relative to a reference structure with $c_1(\CL_{\fs_0})=c_0$. It therefore depends on a choice that does not enter the summed theory. In general, no choice is invariant under orientation-preserving diffeomorphisms.
  \tightfootnote{On $\IC\IP^2$, for instance, the involution $[z_0:z_1:z_2]\mapsto[\bar z_0:\bar z_1:\bar z_2]$ is an orientation-preserving diffeomorphism that acts on $H^2(\IC\IP^2;\IZ)$ by $H\mapsto-H$. It therefore maps every choice $c_0=(2r+1)H$ to $-c_0\neq c_0$.}
  \item For embedded $\Sigma_2$, the normal bundle supplies a canonical integral lift of $w_2(TM_4)|_{\Sigma_2}$, namely its Euler class. Unlike $c_0$, this lift involves no choice, since the embedding determines it. The operator
  \be
    U_m^{\mathrm{Euler}}(\alpha;\Sigma_2):=U_m(\alpha;\Sigma_2)\exp\left(-\frac{\imag\alpha}{2}\int_{\Sigma_2}e(N_{\Sigma_2/M_4})\right) ~,
  \ee
  is natural, topological, and $2\pi$-periodic by \eqref{eq:gravitational-surface-self-intersection}. It acts on genuine lines as $U_m$ does, since linking spheres have vanishing self-intersection. However, its phase is quadratic in $[\Sigma_2]$: resolving a transverse intersection point of sign $\pm1$ between two surfaces carrying $U_m^{\mathrm{Euler}}(\alpha)$ multiplies the result by $e^{\mp\imag\alpha}$.
\end{enumerate}
For embedded $\Sigma_2$, the two redefinitions differ by a phase,
\be
  \frac{U_m^{\mathrm{Euler}}(\alpha;\Sigma_2)}{U_m^{c_0}(\alpha;\Sigma_2)}
  =\exp\left(-\frac{\imag\alpha}{2}\left(\Sigma_2\cdot\Sigma_2-\langle c_0,[\Sigma_2]\rangle\right)\right) ~,
\ee
whose exponent is an integer multiple of $\alpha$ by the Wu formula. This rephasing depends on $\Sigma_2$: on $\IC\IP^2$ with $c_0=H$, a smooth complex curve of degree $d$, which is embedded in the class $dH$, gives $\Sigma_2\cdot\Sigma_2-\langle c_0,[\Sigma_2]\rangle=d(d-1)$. This vanishes for the line $\IC\IP^1$ but not for a smooth conic. A $2\pi$-periodic redefinition whose phase is linear in $[\Sigma_2]$ is of the first type, so in general no redefinition is both natural and linear. The invariant content of the twist is the mixed anomaly of \propref{prop:bordism-anomaly}, which is not removed by any counterterm.
\end{remark}

The obstruction to taking the $2\pi$-periodic quotient universally appears when the naive $\sfU(1)_{m}^{(1)}$ is coupled to an ordinary two-form background $B$: the partition function is not invariant under all large gauge transformations of $B$. This mixed magnetic--gravitational 't Hooft anomaly, derived in \autoref{sec:magnetic-gravitational-anomaly}, is the background manifestation of the same obstruction encoded operatorially by \eqref{eq:Gee-sigma}. Keeping $G$ as part of the composition law \eqref{eq:twisted-identification} instead gives what we call the \emph{$w_2$-twisted magnetic 1-form symmetry}.
\tightfootnote{We derive the $w_2$-twist and its mixed anomaly from the shifted flux quantization of the summed theory. Beyond the composition law \eqref{eq:Gee-sigma}, this yields its geometric form $G(\Sigma_2)=(-1)^{\Sigma_2\cdot\Sigma_2}$ (see \propref{prop:w2-twist}), the $4\pi$-periodic extension \eqref{eq:extension-sequence}, the obstruction to removing the twist by a redefinition that is both natural and linear (see \remref{rem:surface-counterterms}), the classification of the anomaly by reduced oriented bordism with its detection on the Wu manifold (see \propref{prop:bordism-anomaly}), and the role of the twist as $1$-endomorphisms of the trivial defect in the symmetry category of rational chiral defects. The anomaly also appears, as $w_2\smile\tau_m$, in \cite[Sec.~4.2]{DavighiLohitsiriDebray2023}, which computes the oriented bordism group \cite[App.~B.1]{DavighiLohitsiriDebray2023}. Here $\tau_m\in H^3(B^2\sfU(1)_m;\IZ_2)$ is the universal degree-three class, which pulls back to $\rho_2[dB/(2\pi)]$. The same anomaly, together with Wilson lines attached to surfaces carrying $w_2$ flux, appears in the analysis of superconductors with a dynamical spin$^c$ connection in \cite{BergCappelliVilla2026}.}
The relation between the extended composition law and the ordinary background anomaly is made precise in \propref{prop:bordism-anomaly} and \remref{rem:anomaly-vs-extension} below.

The identity codimension-one defect is the tensor unit $\mathbf 1\in\fC_{\chi}^{\mathrm{spin}^c}$ of the symmetry category, and the magnetic surface operators are $1$-endomorphisms of $\mathbf 1$. Equation \eqref{eq:Gee-sigma} states that their composition law in $\End_{\fC_{\chi}^{\mathrm{spin}^c}}(\mathbf 1)$ is twisted by the invertible tangential $1$-endomorphism $G$. Thus the $w_2$-twisted magnetic 1-form symmetry is lower morphism data inside the same symmetry $3$-category as the chiral codimension-one defects.

\begin{remark}[Condensation defects]\label{rem:condensation} A higher gauging construction would proliferate the magnetic surface operators into a foam on a chosen submanifold.
\tightfootnote{Such defects are commonly called condensation defects, although, as stressed in \cite[App.~A]{Cheng:2026qax}, the underlying gauging operation does not constitute condensation in a dynamical sense, since it involves no spontaneous breaking of a global symmetry. Here ``condensation'' means a weighted sum over all insertions of the symmetry operators on the cycles of the support.}
 Whether the $w_2$-twist is visible on that support depends on its dimension. 

\emph{Three-dimensional supports.} Let $M_3\subset M_4$ be closed and oriented. Since $M_4$ is oriented, the normal bundle of $M_3$ is trivial, so $TM_4|_{M_3}\cong TM_3\oplus\CO$ and
\be
w_2(TM_4)\big|_{M_3}=w_2(TM_3)=0 ~.
\ee
Equivalently, a surface $\Sigma_2\subset M_3$ can be pushed off itself along the normal direction of $M_3$, so $\Sigma_2\cdot\Sigma_2=0$ and the Wu formula gives $\langle w_2(TM_4),\Sigma_2\rangle=0$. Hence $G(\Sigma_2)=1$ for every surface in the support, $\rho_2\big(c_1(\CL_\fs)\big)\big|_{M_3}=w_2(TM_4)\big|_{M_3}=0$ for every spin$^c$ structure $\fs$, and $F/(2\pi)$ has integral periods on $M_3$. In particular, $U_m(\tfrac{2\pi}{K};\Sigma_2)$ has order $K$ on such surfaces. The $1$-gauging of the $\IZ_K$ it generates sums over $a\in H^1(M_3;\IZ_K)$, with $U_m(\tfrac{2\pi}{K})$ inserted on the Poincar\'e dual of $a$, as in spin QED. By contrast, for even $K$ the cyclic subgroup of the $4\pi$-periodic parameter group generated by $U_m(\tfrac{4\pi}{K})$ contains $U_m(2\pi)=G$, which acts trivially on the support. Condensing that subgroup on $M_3$ would add a decoupled $\IZ_2$ gauge theory.

This observation also has a direct interpretation in canonical quantization. If $M_3=Y_3$ is a cooriented spatial hypersurface, every closed surface $\Sigma_2\subset Y_3$ supporting a magnetic surface operator satisfies $G(\Sigma_2)=1$. Consequently, the action of magnetic surface operators supported on an ordinary spatial slice has the usual $2\pi$ periodicity. The additional gravitational kernel is instead detected by global four-dimensional configurations, such as surfaces of odd self-intersection in $\IC\IP^2$.

One must also account for discrete torsion factors obtained by stacking a suitable $\IZ_K$ SPT on $M_3$. For these discrete-symmetry-dependent factors, we remove the contributions present with trivial $\IZ_K$ background and use characters of reduced spin$^c$ bordism:
\be
\Hom\big(\wt\Omega_3^{\mathsf{Spin}^c}(B\IZ_K), \sfU(1)\big) ~.
\ee
Since $\Omega_3^{\mathsf{Spin}^c}(\mathrm{pt})=0$, the reduced and unreduced groups agree. The Atiyah--Hirzebruch computation of \cite[App.~C.1]{Garcia-Etxebarria:2018ajm} shows that this bordism group is an extension
\be
0 \to \IZ_K \to \Omega_3^{\mathsf{Spin}^c}(B\IZ_K) \to \IZ_K \to 0 ~,
\ee
whose subgroup comes from $H_1(B\IZ_K;\Omega_2^{\mathsf{Spin}^c})$ and whose quotient comes from $H_3(B\IZ_K;\IZ)$. The extension does not split in general: for example, $\Omega_3^{\mathsf{Spin}^c}(B\IZ_4)\cong\IZ_8\oplus\IZ_2$ \cite[Eq.~(C.9)]{Garcia-Etxebarria:2018ajm}. The discrete torsion choices are the characters of this group. The characters that are trivial on the subgroup form $\Hom(H_3(B\IZ_K;\IZ),\sfU(1))\cong H^3(B\IZ_K;\sfU(1))\cong\IZ_K$, the bosonic Dijkgraaf--Witten classes.

 \emph{Two-dimensional supports.} On a closed surface $\Sigma_2\subset M_4$, the twist can be visible: for example, $G(\IC\IP^1)=-1$ in $\IC\IP^2$. The gauging must therefore retain the relation $U_m(2\pi;\Sigma_2)=G(\Sigma_2)$ and use a finite subgroup of the $4\pi$-periodic parameter group, such as the $\IZ_K$ generated by $U_m(\tfrac{4\pi}{K};\Sigma_2)$. Here the symmetry surfaces fill the support, so for connected $\Sigma_2$ the $2$-gauging sums over $H_2(\Sigma_2;\IZ_K)\cong\IZ_K$. The condensation defect is a weighted sum of the operators $U_m(\tfrac{4\pi j}{K};\Sigma_2)$.

 Defining a condensation defect also requires compatible tangential and junction data. We do not carry out this construction here. The structure of the resulting condensation defects will be revisited in more detail in \cite{DelZottoSaxena:symmetry-category}.
\end{remark}

\subsubsection{Global Fluxes And The Local Magnetic Charge Lattice}\label{sec:fractionalization}
The half-integral flux on a global cycle such as $\IC\IP^1\subset\IC\IP^2$ reflects the spin-charge relation. In $\mathsf{Spin}^c$, the central $-1$ in the spin group is identified with the gauge transformation $-1$, allowing an odd-charge fermion to be defined without a separate spin structure.
This is the Lorentz symmetry fractionalization of the unit electric line \cite[Sec.~2.5.1]{Brennan:2022tyl}. See also \cite[Sec.~2.1.2]{Kan:2024fbn}. Its magnetic manifestation is \eqref{eq:Gee-sigma}. The fractionalization concerns the transformation properties of an odd-charge electric line under spacetime rotations. 

We now verify that the magnetic charge $m$ defined in \eqref{eq:local-magnetic-charge} remains integral for a spin$^c$ connection. Take a small sphere $S^2_{\mathrm{link}}$ linking a magnetic line $\gamma$ once, as in \eqref{eq:local-magnetic-charge}. Its normal bundle is trivial, so
\begin{equation}
  TM_4\big|_{S^2_{\mathrm{link}}}
  \cong TS^2_{\mathrm{link}}\oplus\CO\oplus\CO ~,
\end{equation}
where $\CO$ denotes the trivial real line bundle. Since the sphere is spin, $w_2(TM_4)$ restricts to zero there. The spin$^c$ quantization condition therefore gives
\begin{equation}\label{eq:local-spinc-magnetic-quantization}
  m=\int_{S^2_{\mathrm{link}}}\frac{F}{2\pi}\in\IZ ~,
  \qquad G(S^2_{\mathrm{link}})=1 ~.
\end{equation}
A standalone line $H_{1/2}(\gamma)$ with flux $\frac{1}{2}$ violates this condition. The full spin$^c$ quantization law therefore excludes a local half-monopole on both spin and non-spin backgrounds.

\begin{figure}[htbp]
  \centering
  \includegraphics[width=0.85\textwidth]{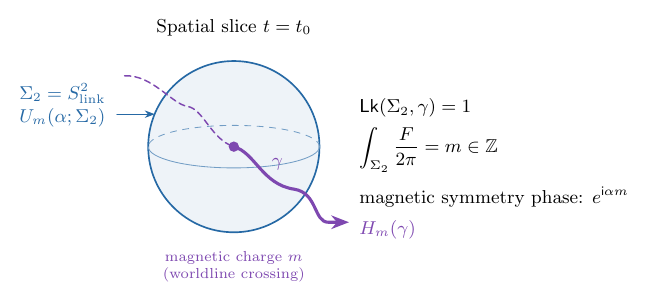}
  \caption{\label{fig:magnetic-surface-line-linking}Local surface--line linking. In a transverse spatial slice, the sphere surrounds a magnetic line of integral charge $m$. With linking number one, the magnetic surface contributes $e^{\imag\alpha m}$. The apparent intersection of the worldline with the sphere is an artifact of the projection.}
\end{figure}

For an allowed 't Hooft line $H_m(\gamma)$, the linking action illustrated
in \figref{fig:magnetic-surface-line-linking} is
\begin{equation}
  U_m(\alpha;\Sigma_2)H_m(\gamma) =e^{\imag\alpha m\,\mathsf{Lk}(\Sigma_2,\gamma)}H_m(\gamma) ~,
  \qquad m\in\IZ ~.
\end{equation}
At $\alpha=2\pi$ the phase is trivial, so every genuine
\tightfootnote{Genuineness means independence from a physical attached
surface \cite{Gaiotto:2014kfa}.}
magnetic line is neutral under $G$.

More generally, the magnetic 1-form symmetry acts on genuine Wilson--'t Hooft lines through their
magnetic charge $m\in\IZ$, independently of their electric charge. Purely electric Wilson lines have
$m=0$ and are therefore neutral under this symmetry.

Thus the local magnetic charge lattice is integral, while the global surface-operator algebra retains the kernel $G$ (see \autoref{subsubsec:w2-twist}).

Half-integral flux can also be imposed externally by an electromagnetic sheet. Such a sheet does not produce a genuine half-magnetically charged line. The sign it induces on intersecting surfaces occurs already in flat spacetime, where $w_2(TM_4)$ vanishes, so it does not diagnose the $w_2$-twist (see \appref{app:magnetic-sheet}).

\subsubsection{The Mixed Magnetic--Gravitational Anomaly}\label{sec:magnetic-gravitational-anomaly}
The $w_2$-twist appears as a mixed magnetic--gravitational anomaly of the naive $2\pi$-periodic symmetry. It follows from the global spin$^c$ flux quantization and can be formulated using bordism theory. For this discussion, spin$^c$ QED$_4$ means the fully summed compact theory of \eqref{eq:full-spinc-partition}, which is defined on oriented four-manifolds without choosing a background spin$^c$ structure. Couple its naive  $2\pi$-periodic $\sfU(1)_{m}^{(1)}$ to a background two-form gauge field $B$ by adding $-\frac{\imag}{2\pi}\int_{M_4}B\wedge F$ to the Euclidean action. Under a large background gauge transformation $B \mapsto B + \lambda$, with $\lambda$ closed with $2\pi\IZ$-valued periods, the path integral weight shifts by the phase
\be
  \exp\left(
    2\pi\imag\int_{M_4}
    \frac{\lambda}{2\pi}\smile
    \left[\frac{F}{2\pi}\right]
  \right)
  =
  (-1)^{\left\langle
    \rho_2\left[\frac{\lambda}{2\pi}\right]
    \smile w_2(TM_4),[M_4]
  \right\rangle} ~,
\ee
by the shifted quantization condition \eqref{eq:spin-c-physics-v3}. This non-invariance is a mixed 't Hooft anomaly between the naive magnetic 1-form symmetry and gravity, and it admits the following characterization.
\begin{proposition}\label{prop:bordism-anomaly}
The 't Hooft anomaly of the naive $2\pi$-periodic $\sfU(1)_{m}^{(1)}$ of spin$^c$ QED$_4$ on oriented four-manifolds is the order-two 5d invertible field theory
\be\label{eq:anomaly-theory-5d}
  \alpha_5[B;X_5] = \exp\left(\imag\pi\int_{X_5}\rho_2\left[\frac{dB}{2\pi}\right]\smile w_2(TX_5)\right) = \exp\left(\imag\pi\int_{X_5}\ssq^2\,\rho_2\left[\frac{dB}{2\pi}\right]\right) ~,
\ee
where the second equality follows from the Wu formula on a closed oriented $X_5$, on which the Wu class is $v_2 = w_2 + w_1^2 = w_2$. Its deformation class generates
\be
  \Hom\left(\wt{\Omega}^{\mathsf{SO}}_5\big(K(\IZ,3)\big),\, \sfU(1)\right) \cong \IZ_2 ~,
\ee
which is the full group of 't Hooft anomalies available to a $\sfU(1)$ 1-form symmetry of a 4d theory defined on oriented manifolds, modulo purely gravitational anomaly theories. The group $\wt{\Omega}^{\mathsf{SO}}_5\big(K(\IZ,3)\big)\cong\IZ_2$ is generated by the difference $[W_5,W_3]-[W_5,0]$. Here $W_5=\mathsf{SU}(3)/\mathsf{SO}(3)$ is the Wu manifold, which admits no spin$^c$ structure, and $W_3=W_3(TW_5)\in H^3(W_5;\IZ)$ is the background class $[dB/2\pi]$. Subtracting the trivial background class removes the purely gravitational component of $[W_5,W_3]$. Since $\alpha_5$ is trivial on $[W_5,0]$, it is detected by
\be\label{eq:anomaly-Wu-manifold}
  \alpha_5[B;W_5]=(-1)^{\int_{W_5}w_2w_3}=-1 ~.
\ee
\end{proposition}

To prove \eqref{eq:anomaly-Wu-manifold}, note that $W_3(TW_5)=\mathsf{Bock}(w_2(TW_5))$. Then $\rho_2\left[\frac{dB}{2\pi}\right] = \ssq^1 w_2 = w_3$ on an orientable manifold. The mod-$2$ cohomology of $W_5$ is $\IZ_2$ in degrees $0,2,3,5$, with $H^2(W_5;\IZ_2)$ and $H^3(W_5;\IZ_2)$ generated by $w_2$ and $w_3$ \cite{Stong1968}. Since $W_5$ is closed and oriented, Poincar\'e duality forces the cup product pairing $H^2(W_5;\IZ_2)\otimes H^3(W_5;\IZ_2)\to H^5(W_5;\IZ_2)\cong\IZ_2$ to be nondegenerate, so $\langle w_2w_3,[W_5]\rangle=1$ and $\alpha_5 = (-1)^{\int_{W_5}w_2 w_3} = -1$. Since $W_3(TW_5)\neq0$, the Wu manifold admits no spin$^c$ structure. This test therefore detects the anomaly of the fully summed oriented theory, and it would not be an admissible test background for a fixed spin$^c$ sector. The anomaly is nonetheless visible on spin$^c$ backgrounds. On $\IC\IP^2\times S^1$ with $[dB/2\pi]=H\smile u$, where $H$ is the hyperplane class and $u$ generates $H^1(S^1;\IZ)$, one finds $\alpha_5=(-1)^{\int H^2 u}=-1$. This is the mapping torus of the large gauge transformation above with $[\lambda/2\pi]=H$, as in \cite[Sec.~4.2]{DavighiLohitsiriDebray2023}.
\tightfootnote{The group $\wt{\Omega}^{\mathsf{SO}}_5(K(\IZ,3))$ is a summand of the group $\Omega^{\mathsf{SO}}_5(K(\IZ,3)^2)\cong\IZ_2^3$ computed in \cite[App.~B.1]{DavighiLohitsiriDebray2023} for the electric and magnetic 1-form symmetries of Maxwell theory, in which $\alpha_5$ is one of the three $\IZ_2$ factors. The unreduced group $\Omega_5^{\mathsf{SO}}(K(\IZ,3))\cong\IZ_2\oplus\IZ_2$ is also obtained in \cite{JiaWangZhang2026}.}
A self-contained derivation of the bordism group is given in \appref{app:bordism-anomaly-proof}.
\tightfootnote{\label{foot:AHSS-computation}The Wu classes $v_i\in H^i(X_5;\IZ_2)$ used in \propref{prop:bordism-anomaly} are defined by $\langle v_i\smile x,[X_5]\rangle=\langle\ssq^ix,[X_5]\rangle$ for all $x\in H^{5-i}(X_5;\IZ_2)$, and $w(TX_5)=\ssq(v)$ gives $v_2=w_2+w_1^2$ \cite{MilnorStasheff1974}. There is no perturbative part, since $\left[\frac{dB}{2\pi}\right]^2$ vanishes rationally. Retaining the purely gravitational factor $\Omega_5^{\mathsf{SO}}\cong\IZ_2$, detected by $w_2w_3$, gives the unreduced $\Omega^{\mathsf{SO}}_5(K(\IZ,3))\cong\IZ_2\oplus\IZ_2$.}

The use of oriented bordism here relies essentially on the fully summed formulation: the dynamical spin$^c$ structure is part of the gauge field sum, so no spin$^c$ structure remains as fixed background data. A fixed sector partition function $\sfZ^{\fs}$ is instead a theory relative to a chosen spin$^c$ structure, and its anomaly must be restricted to spin$^c$ backgrounds.
\tightfootnote{\label{foot:fixed-spinc-trivial}With a fixed spin$^c$ structure $\fs$, $w_2$ has the integral lift $c_1(\fs)$, so $\alpha_5=(-1)^n$ with $n=\int_{X_5}c_1(\fs)\smile[dB/2\pi]\in\IZ$ (see \remref{rem:anomaly-vs-extension}). It is the value at $\theta=\pi$ of the family $\exp(\imag\theta n)$, and is therefore deformation-trivial. Indeed, the reduced group $\wt\Omega_5^{\mathsf{Spin}^c}(K(\IZ,3))\cong\IZ$, generated by $\IC\IP^2\times S^1$ with $c_1(\fs)=H$ and $[dB/2\pi]=H\smile u$, has no torsion (see \appref{app:bordism-anomaly-proof}). The corresponding counterterm uses a connection on $\CL_\fs$, which is background data only in a fixed sector (see also \cite{BergCappelliVilla2026}). In the summed theory $c_1(\fs)$ is summed over, no natural integral lift of $w_2(TM_4)$ is available (see \remref{rem:surface-counterterms}), and the anomaly is nontrivial. This differs from the assignment to the symmetry type $\mathsf{Spin}^c\times G_m$ in \cite[Sec.~4.2]{DavighiLohitsiriDebray2023}.}

As a consistency check, consider the pullback of the anomaly along the double-cover homomorphism in \eqref{eq:extension-sequence}. In terms of conventionally $2\pi$-periodic coordinates on both parameter circles, the covering homomorphism has degree two. The induced map on the integral degree-three characteristic class of a magnetic two-form background is therefore multiplication by two. A background obtained by this pullback has characteristic class $[dB/2\pi]=2\wt h$, and hence
$\rho_2[dB/2\pi]=0$. The anomaly \eqref{eq:anomaly-theory-5d} is consequently trivial on such pulled-back backgrounds. This cohomological observation is consistent with the extended magnetic symmetry resolving the obstruction. It does not, however, specify the joint background. Changing the coefficient of $B\wedge F$ changes the symmetry to which $B$ couples and leaves the variation in place. No counterterm built from an integral lift of $w_2(TM_4)$ is natural (see \remref{rem:surface-counterterms}). Instead, the two-form background and the spin$^c$ flux must obey a joint cocycle condition that reproduces \eqref{eq:twisted-identification}. Its differential-cohomological formulation, including torsion and flat data, remains to be constructed.

\begin{remark}[Restriction to spin and spin$^c$ backgrounds]\label{rem:anomaly-vs-extension}
Pulled back to spin bordism, $\alpha_5$ is trivial: on a spin 5-manifold $X_5$, $v_2 = 0$, so
\begin{equation*}
  \int_{X_5} \ssq^2\,\rho_2\left[\frac{dB}{2\pi}\right] = 0 ~.
\end{equation*}
This is consistent with the magnetic 1-form symmetry of spin QED$_4$ being anomaly-free \cite{Choi:2022jqy}.

On a spin$^c$ five-manifold, $w_2=\rho_2(c_1(\fs))$, so
\begin{equation*}
  \alpha_5[B;X_5] = (-1)^{\int_{X_5} \left[\frac{dB}{2\pi}\right]\smile c_1(\fs)} ~.
\end{equation*}
Its boundary manifestation is the operator relation \eqref{eq:Gee-sigma}. Thus the bordism anomaly and the extension \eqref{eq:extension-sequence} encode the same $w_2$-twisted composition law after restriction and descent.
\end{remark}

In \autoref{sec:dynamics}, we use this anomaly to constrain gapped phases and the flux quantization of an infrared Coulomb photon.

\section{Chiral Symmetry Defects In \texorpdfstring{Spin$^c$}{Spin-c} \texorpdfstring{QED$_4$}{QED4}}\label{sec:chiral-defect-revisited}

\subsection{The ABJ Anomaly In \texorpdfstring{Spin$^c$}{Spin-c} \texorpdfstring{QED$_4$}{QED4}}\label{sec:ABJ-spinc}
Classically, the theory has a conserved axial current $j_{\mu}^{A} = \ov{\Psi}\gamma_\mu\gamma^5\Psi$ and a conserved vector current $j_{\mu}^{V} = \ov{\Psi}\gamma_{\mu}\Psi$. However, quantum effects break the axial symmetry
\be
  \Psi \mapsto e^{\imag \frac{\alpha}{2}\gamma^5}\Psi ~,
  \qquad
  \ov\Psi\mapsto\ov\Psi e^{\imag \frac{\alpha}{2}\gamma^5} ~,
\ee
where $\Psi$ and $\ov\Psi$ are independent Euclidean Grassmann fields. This is the Adler--Bell--Jackiw (ABJ) anomaly \cite{Adler:1969gk,Bell:1969ts}.
\tightfootnote{The electromagnetic part of the anomaly arises from the one-loop triangle diagram with one axial current and two vector current insertions. A gauge-invariant regularization preserves the vector Ward identities and places the anomaly in the axial Ward identity. On a curved background, there is also a mixed axial--gravitational contribution, detected by a triangle diagram with one axial current and two stress tensor insertions. One can choose local counterterms to shift the consistent anomaly, but there is no choice that simultaneously preserves vector gauge invariance \emph{and} the classical axial symmetry.}
We regard $j^A:=j_\mu^A dx^\mu$ as the axial current 1-form. Since the rotation parameter is normalized by $\Psi\mapsto e^{\imag\alpha\gamma^5/2}\Psi$, the corresponding normalized current 3-form is
\be\label{eq:axial-current-3form}
  \CJ_A^{(3)}:=\frac{1}{2}\star j^A ~.
\ee
(The conventional current $j^A$ generates $\Psi\mapsto e^{\imag\beta\gamma^5}\Psi$. With $\alpha=2\beta$, the current conjugate to $\alpha$ is $\frac{1}{2} j^A$.)
The anomaly is the failure of this 3-form to be closed:
\be\label{eq:ABJ-anomalous-conservation}
  d\CJ_A^{(3)}
  =\frac{1}{2}\,d\star j^{A}
  =\frac{1}{8\pi^2} F \wedge F - \frac{1}{24}p_1(TM_4)
  =\left[\wh A(TM_4)\exp\left(\frac{F}{2\pi}\right)\right]^{(4)}
  =:\CI_4(\fs,A) ~.
\ee
Here $\wh A(TM_4)=1-\frac{1}{24}p_1(TM_4)+\cdots$ denotes the total $\wh A$-characteristic form of the tangent bundle. For a spin$^c$ structure $\fs$ with determinant line bundle $\CL_{\fs}$, the shifted quantization condition $[F/(2\pi)]=\frac{1}{2}c_1(\CL_{\fs})$ means that $\exp(F/(2\pi))$ represents $\exp\left(\frac{1}{2}c_1(\CL_{\fs})\right)$. Consequently, $\CI_4(\fs,A)$ is the degree-$4$ local index density of the spin$^c$ Dirac operator, and the spin$^c$ Atiyah--Singer index theorem identifies its integral over a closed $M_4$ with $\mathsf{Ind}(\slashed D_{\fs})$ \cite{AtiyahSinger1968}.
The Pontryagin class is related to the signature $\sigma(M_4) \in \IZ$ of $M_4$ by the Hirzebruch signature theorem \cite{Hirzebruch1966}:
\be\label{eq:HirzebruchSignature}
  \int_{M_4} p_1(TM_4) = 3\sigma(M_4) ~.
\ee
Accordingly, in a fixed spin$^c$ sector (see \cite[§36]{Moore:2025tmt}), the fixed sector partition function \eqref{eq:fixed-spinc-partition} transforms under the anomalous axial rotation as
\be
  \sfZ_{M_4}^{\fs}[g;\fo;(e^2,\theta)]
  \mapsto \sfZ_{M_4}^{\fs}[g;\fo;(e^2,\theta)]
  \exp\left(\imag\alpha\,\mathsf{Ind}(\slashed{D}_{\fs})\right) ~.
\ee
Here $\mathsf{Ind}(\slashed{D}_{\fs})$ is constant over the connection space $\CA(\fs)$ of a fixed spin$^c$ structure. Explicitly,
\be\label{eq:spinc-dirac-index}
\begin{aligned}
 \mathsf{Ind}(\slashed{D}_{\fs})
 &= \int_{M_4}\CI_4(\fs,A)
  = \int_{M_4}\left[\wh A(TM_4)\exp\left(\frac{F}{2\pi}\right)\right]^{(4)} \\
 &= \int_{M_4} \left( \frac{1}{8\pi^2} F \wedge F - \frac{1}{24}p_1(TM_4) \right)
  = \int_{M_4} \frac{1}{8\pi^2} F \wedge F - \frac{1}{8}\sigma(M_4) ~.
\end{aligned}
\ee
The fully summed partition function \eqref{eq:full-spinc-partition} includes all spin$^c$ sectors $\fs_0+\varepsilon$, and transforms under the axial rotation above as
\be
  \begin{split}
  \sfZ_{M_4}[g;\fo;(e^2,\theta)] \mapsto \sfZ_{M_4}[g;\fo;(e^2,\theta);\alpha] ={}& \sum_{\varepsilon\in H^2(M_4;\IZ)}\int_{\CA(\fs_0+\varepsilon)/\CG}[\CD A][\CD\Psi][\CD\ov{\Psi}] \\
  &e^{-S_{\text{QED}}[A,\Psi,\ov{\Psi};g;\fo;\fs_0+\varepsilon;(e^2,\theta)]}e^{\imag\alpha \mathsf{Ind}(\slashed{D}_{\fs_0+\varepsilon})} ~.
  \end{split}
\ee
The anomalous phase weights each spin$^c$ sector according to the Dirac index. In the sector $\fs_0+\varepsilon$, this phase is
\be
   e^{\imag\alpha \mathsf{Ind}(\slashed{D}_{\fs_0+\varepsilon})} = \exp\left[\imag\alpha\frac{c_1(\CL_{\fs_0+\varepsilon})^2 - \sigma(M_4)}{8}\right] =: \exp\left[\imag\alpha \frac{c^2-\sigma}{8}\right] ~,
\ee
where, sector by sector, $c=c_1(\CL_{\fs_0+\varepsilon})$ is the first Chern class of the determinant line bundle of $\fs_0+\varepsilon$ (cf. \autoref{sec:spinc}).

For a closed oriented smooth four-manifold $M_4$, $c \equiv w_2(TM_4) \smod 2$, so by the Wu formula $c \smile x \equiv x \smile x \smod 2$ for all $x \in H^2(M_4;\IZ)$. That is, $c$ is a characteristic element. By the van der Blij lemma, $c^2 \equiv \sigma \smod 8$. Therefore, $c^2-\sigma \in 8 \IZ$, and the index of the spin$^c$ Dirac operator is an integer.

Categorically, the ABJ anomaly obstructs the naive chiral wall from defining an absolute object of the symmetry category. The Hall TFT introduced below trivializes its electromagnetic component, while the gravitational component must be canceled before the wall becomes an unframed object.

\subsection{The Chiral Rotation Symmetry Defect Revisited}\label{sec:chiral-rotation-defect}
In this subsection, we give a spin$^c$ version of the argument for the existence of a chiral symmetry rotation defect at the interface of QED$_0$ and QED$_{\theta}$ that implements a chiral rotation, following \cite{Cordova:2022ieu}. Conventions for the action, etc. are spelled out in \appref{app:QEDconventions}. Here $\mathsf{QED}_\theta$ denotes spin$^c$ QED with the gauge--gravitational theta term $-\imag\theta\int_{M_4}\CI_4(\fs,A)$, where $\CI_4(\fs,A)$ is the full spin$^c$ Dirac index density defined in \eqref{eq:ABJ-anomalous-conservation}. This theta term includes the gravitational contribution omitted from \eqref{eq:L-QED-4} (see \appref{app:theta-periodicity}).

\begin{figure}[htbp]
  \centering
  \includegraphics{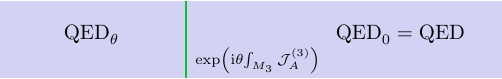}
  \caption{\label{fig:axial-interface}The axial interface $\exp(\imag\theta\int_{M_3}\CJ_A^{(3)})$ (green) separating $\mathsf{QED}_\theta$ from $\mathsf{QED}_0=\mathsf{QED}$.}
\end{figure}

Let $M_4$ be an oriented four-manifold equipped with a spin$^c$ structure, and let $(\omega, A)$ denote the corresponding spin$^c$ connection. On a flat spin manifold, or when the gravitational background is suppressed, the electromagnetic part of the ABJ anomaly is encoded by the descent of the six-form
\be
\frac{F_\chi}{2\pi}\cdot \frac{1}{2}\left(\frac{F}{2\pi}\right)^2 ,
\ee
where $F_{\chi}$ is the fieldstrength of a classical background field for the chiral $\sfU(1)_{\chi}$.
This is the part retained in the gauge sector defect constructions. On a curved manifold it is not the complete anomaly, even when the manifold is spin. The complete anomaly involves the full index density $\CI_4(\fs,A)$ of \eqref{eq:ABJ-anomalous-conservation}. By \eqref{eq:spinc-dirac-index} and the van der Blij lemma, its integral over a closed $M_4$ is
\be\label{eq:spinc-index-integer}
 \int_{M_4} \left[\wh A(TM_4)\exp\left(\frac{F}{2\pi}\right)\right]^{(4)} = \frac{c^2 - \sigma(M_4)}{8} \in \IZ ~.
\ee
To state the relevant mixed part of the anomaly polynomial intrinsically, let $X_6$ be a six-manifold on which the tangent and gauge bundles extend. The term linear in the axial background curvature is
\be
      I_{6} = \frac{F_\chi}{2\pi}\wedge\left[\wh A(TX_6)\exp\left(\frac{F}{2\pi}\right)\right]^{(4)} = \frac{F_\chi}{2\pi} \wedge \left[\frac{F \wedge F}{8\pi^2} - \frac{p_1(TX_6)}{24}\right] ~.
\ee
Here $I_6$ denotes only the mixed terms. The cubic axial anomaly does not contribute below: the wall background is pure gauge, so $F_\chi=0$. The variation follows by descent from $I_6$, whose degree-four factor is the local index density \eqref{eq:ABJ-anomalous-conservation}.
Now insert a codimension-one domain wall at $x=0$ that implements a chiral rotation with angle $\theta$ on the region $x<0$:
\be
  A_{\chi} = d\lambda_{\chi} = -\theta\, \delta(x)\, dx ~, \qquad \lambda_\chi(x) = \theta\, \Theta(-x) ~,
\ee
where $\Theta$ is the Heaviside step function. By descent,
\be
  \delta S = -2\pi\imag\int_{M_4}\frac{\lambda_{\chi}}{2\pi}\left[\wh A(TM_4)\exp\left(\frac{F}{2\pi}\right)\right]^{(4)} = -\imag\theta\int_{x<0}\left[\frac{F \wedge F}{8\pi^2} - \frac{p_1(TM_4)}{24}\right] ~.
\ee
Thus the region $x<0$ acquires the theta term of $\mathsf{QED}_\theta$, while the region $x>0$ remains $\mathsf{QED}_0$: the wall implements a jump of the theta angle by exactly $\theta$. We denote this interface operator $\exp(\imag\theta\int_{M_3}\CJ_A^{(3)})$, where $M_3$ is a closed 3-cycle, oriented as the boundary of the $\mathsf{QED}_\theta$ region, and $\CJ_A^{(3)}=\frac{1}{2}\star j^A$ is the normalized axial current 3-form introduced in \eqref{eq:axial-current-3form}. See \autoref{fig:axial-interface}.
The wall is not yet topological. The gravitational term is independent of the spin$^c$ sector, but its anomaly must still be canceled. Any local wall counterterm used for this purpose must satisfy the appropriate quantization conditions. On a spin manifold, Rokhlin's theorem gives $\sigma(M_4) \in 16\IZ$, so by the Hirzebruch signature theorem \eqref{eq:HirzebruchSignature}, $\int_{M_4}p_1(TM_4) = 3\sigma(M_4) \in 48\IZ$, and hence $\int_{M_4}\frac{p_1(TM_4)}{24} = \frac{\sigma(M_4)}{8} \in 2\IZ$. At a fractional rotation angle $2\pi p/N$, the resulting phase need not be trivial. After the electromagnetic terms cancel, the remaining gravitational anomaly combines this axial contribution with the framing anomaly of the Hall TFT and is governed by $c_- - \frac{p}{N}$, as shown in \remref{rem:gravitational-completion}. A topological wall must cancel this combined anomaly with compatible wall degrees of freedom.

By \eqref{eq:spinc-index-integer}, a theta angle coupling to the full index density, $S \supset -\imag\theta\,\mathsf{Ind}(\slashed{D}_{\fs})$, is exactly $2\pi$-periodic on closed $M_4$. The isolated electromagnetic factor instead transforms by the gravitational phase in \eqref{eq:spinc-em-theta-shift}. \appref{app:theta-periodicity} derives this distinction.

\subsection{Chiral Topological Symmetry Defect Of \texorpdfstring{Spin$^c$}{Spin-c} \texorpdfstring{QED$_4$}{QED4}}\label{sec:coupling-tft-to-spinc-qed4}
Let us begin with the axial rotation wall $\exp(\imag\theta\int_{M_3}\CJ_A^{(3)})$ constructed above between $\mathsf{QED}_{\theta}$ and $\mathsf{QED}_{0}$.  The electromagnetic part of the relative phase between $\mathsf{QED}_{\theta}$ and $\mathsf{QED}_{0}$ in a fixed spin$^c$ sector is
\be
  \exp\left(\frac{\imag\theta}{8\pi^2}\int F \wedge F\right) = \exp\left(\imag \theta \frac{1}{2}\int y \smile y\right) ~.
\ee
Here $y=[F/(2\pi)]\in H^2(M_4;\IR)$. As in \autoref{sec:spinc}, fix a reference spin$^c$ structure with determinant class $\ov w_2$. Twisting it by a line bundle of Chern class $\varepsilon\in H^2(M_4;\IZ)$ gives, in real cohomology,
\be
 y = \varepsilon + \frac{\ov{w}_2}{2} ~, \quad \varepsilon \in H^2(M_4;\IZ) ~.
\ee
Then the relative phase is
\be
 \exp\left[\imag\theta \frac{1}{2}\int\left(\varepsilon + \frac{\ov{w}_2}{2}\right)^2\right] = \exp\left[\imag\theta\left(\frac{1}{2}\int \varepsilon^2 + \frac{1}{2}\int \varepsilon \smile \ov{w}_2 + \frac{1}{8}\int \ov{w}_2^2 \right) \right] ~.
\ee
For $\theta = 2\pi p/N$, this is
\be\label{eq:relphase}
  \exp\left(2\pi\imag \frac{p}{2N}\left[\int \varepsilon^2 + \int \varepsilon \smile \ov{w}_2 + \frac{1}{4}\int \ov{w}_2^2 \right] \right) ~.
\ee
The spin$^c$ mismatch includes the terms $\int \varepsilon\smile\ov{w}_2$ and $\int\ov{w}_2^2$ in addition to the ordinary flux term. We seek a three-dimensional TFT whose coupling to the bulk spin$^c$ connection cancels this mismatch by anomaly inflow.

When $w_2=0$, a candidate is the minimal $\CA^{N,p}$ 3d TFT of \cite{Hsin:2018vcg}, which has a $\IZ_{N}$ 1-form symmetry with anomaly coefficient $p$ \cite{Choi:2022jqy,Cordova:2022ieu} -- up to categorical equivalence \cite{Copetti:2023mcq}. Consider coupling it to a background $\IZ_{N}$ 2-form gauge field $B^{(2)} \in Z^{2}(M_3;\IZ_N)$ for the $\IZ_{N}$ 1-form symmetry. The anomaly implies that the partition function is not a gauge-invariant functional of $B^{(2)}$. To restore gauge invariance, one can use anomaly inflow from a four-manifold $Y_4$ with $\partial Y_4 = M_3$, realizing $\CA^{N,p}$ as an edge mode of the SPT
\be\label{eq:usual-PS}
 \sfZ_{\text{anom}}[B^{(2)}] := \exp\left(\frac{2\pi\imag p}{2N} \int_{Y_4} \CP\big(B^{(2)}\big)\right) ~,
\ee
where $\CP$ denotes the Pontryagin square, with a spin refinement when $pN$ is odd \cite{Hsin:2018vcg}. This four-dimensional invertible theory specifies the anomalous background variation and extension dependence of the three-dimensional partition function. Pairing the relative partition function of a chosen boundary TFT with the appropriate bulk inflow factor cancels that variation. If $y$ were an integral class, we could identify
\be
  B^{(2)} = \rho_{N}(y) = \rho_{N}\left(\varepsilon + \frac{\ov{w}_2}{2}\right) ~, \quad \text{ (if $y$ were integral)}
\ee
where $\rho_{N}$ is the modulo $N$ reduction map. This identification is not generically well defined. Let us define
\be\label{eq:dee}
 c := c_1(\CL_\fs) = 2\varepsilon + \ov{w}_2 \in H^2(M_4;\IZ) ~, \quad c_{\IR}=2y ~, \quad \rho_2(c) = w_2(TM_4) ~.
\ee
Let us consider the anomaly theory defined by the exponentiated action
\be\label{eq:spin-c-refined-PS}
  \sfZ^{\text{spin}^c}_{p,N}(M_4;c) = \exp\left[2\pi\imag\frac{p}{8N}\int_{M_4}c^2\right] ~.
\ee
It depends on $c$ only through $\rho_{4N}(c)$: shifting $c$ by $4N$ times an integral class changes $\int c^2$ by a multiple of $8N$. The spin$^c$ restriction is encoded in the characteristic class $c$ of \eqref{eq:dee}. Substituting $c = 2\varepsilon + \ov{w}_2$ from \eqref{eq:dee}, we find that such an anomaly theory reproduces the electromagnetic part of the relative phase \eqref{eq:relphase} between $\mathsf{QED}_{\theta=\frac{2\pi p}{N}}$ and $\mathsf{QED}_{0}$ in a fixed spin$^c$ sector. This fixes the electromagnetic inflow in the spin$^c$ case.  

Next, we choose a \emph{complete} three-dimensional Hall TFT $\CT_{\mathrm{Hall}}^{N,p}$, coupled to the induced spin$^c$ connection without an independent choice of spin structure. It must obey the local spin-charge relation and have the \emph{exact} Hall coefficient $\sigma_H=p/N$ (not merely this value modulo $\IZ$). 

A concrete sufficient construction is Abelian Chern--Simons theory with compact dynamical connections $a_i$, a nonsingular symmetric integral matrix $K$, and an integral column vector $t$ specifying their electromagnetic coupling. The superscript $\mathsf T$ denotes matrix transpose. We require
\be\label{eq:spinc-Hall-K-condition}
 t_i=K_{ii}\smod 2 ~, \qquad t^{\mathsf T}K^{-1}t=\frac{p}{N} ~.
\ee
These two conditions respectively enforce the spin-charge relation and match the required Hall response. To derive them, consider the local action
\be\label{eq:spinc-Hall-action}
    S_{\mathrm{Hall}}[a,A]
    =\frac{1}{4\pi}\int_{M_3}K_{ij}a_i\wedge da_j
    -\frac{1}{2\pi}\int_{M_3}t_i A\wedge da_i ~,
\ee
where repeated indices are summed. We take the path-integral weight $\exp(\imag S_{\mathrm{Hall}})$ and extensions $Y_4$ with $\partial Y_4=M_3$. These are the conventions in which \eqref{eq:relphase} is written. As shown below, \eqref{eq:spinc-Hall-action} then has the inverse of the electromagnetic response of the axial wall. It therefore describes the Hall theory with the orientation in which it is placed on the wall in \eqref{eq:spinc-defect}. We call the opposite orientation, whose electromagnetic response has the same sign as that of the axial wall, the \emph{standard orientation}, and we always label the theory by its standard-orientation data, in particular $\sigma_H$ and $c_-$. The first condition is the spin$^c$ consistency condition derived by Seiberg and Witten~\cite[Sec.~2.3]{Seiberg:2016rsg}. A local particle has line label $Kn$, where $n$ is an integral column vector, and has charge and topological spin
\be\label{eq:spinc-Hall-local-particle}
    Q=t^{\mathsf T}n ~,\qquad
    h=\frac{1}{2} n^{\mathsf T}Kn\smod 1 ~.
\ee
Taking $n$ with a single unit entry identifies each $t_i$ as a local-particle charge. Its integrality expresses charge quantization in units of the bulk electron charge. The spin-charge relation \eqref{eq:spin-charge-relation} requires $2h=Q\smod2$ for every local particle. Since $K$ is symmetric and $n_i^2=n_i\smod2$,
\be\label{eq:spinc-Hall-characteristic-vector}
    n^{\mathsf T}Kn-t^{\mathsf T}n
    =\sum_i(K_{ii}-t_i)n_i\smod2 ~.
\ee
This vanishes for every integral $n$ precisely when $t_i=K_{ii}\smod2$. Thus odd-charge local particles are fermions and even-charge local particles are bosons. In other words, $t$ is a characteristic covector of the integral lattice defined by $K$.

The same parity condition makes \eqref{eq:spinc-Hall-action} well defined using the spin$^c$ connection $A$, without an independent spin structure. For two four-dimensional extensions, the ambiguity in $S_{\mathrm{Hall}}/(2\pi)$ on the resulting closed spin$^c$ four-manifold is
\tightfootnote{Indeed, $\rho_2(c)=w_2(TX_4)$ and Wu's formula yield $\int z_i^2=\int z_i\smile c\smod2$. The off-diagonal terms in $K_{ij}z_i\smile z_j$ occur in pairs, and the diagonal terms cancel $t_i z_i\smile c$ modulo two. The quantity in \eqref{eq:spinc-Hall-extension} is therefore integral.}
\be\label{eq:spinc-Hall-extension}
 \frac{1}{2}\int_{X_4}\left(K_{ij}z_i\smile z_j-t_i z_i\smile c\right)\in\IZ ~,
 \qquad z_i=c_1(a_i) ~.
\ee

The second condition in \eqref{eq:spinc-Hall-K-condition} follows from the local electromagnetic response. Varying \eqref{eq:spinc-Hall-action} with respect to $a_i$ gives
\be\label{eq:spinc-Hall-field-equation}
    \sum_jK_{ij}\,da_j=t_i\,dA ~.
\ee
Solving this equation locally and substituting into the action gives the electromagnetic response term
\be\label{eq:spinc-Hall-response}
    S_{\mathrm{resp}}[A]
    =-\frac{\sigma_H}{4\pi}\int_{M_3}A\wedge dA ~,\qquad
    \sigma_H=t^{\mathsf T}K^{-1}t ~.
\ee
At $\theta=2\pi p/N$, the electromagnetic phase in \eqref{eq:relphase} has coefficient $\theta/(8\pi^2)=p/(4\pi N)$ multiplying $\int F\wedge F$. The response \eqref{eq:spinc-Hall-response} extends to $\exp(-2\pi\imag\,\sigma_H\int_{Y_4}c^2/8)$, so its cancellation against \eqref{eq:spin-c-refined-PS} requires $\sigma_H=p/N$. Equality only modulo an integer would leave an integer electromagnetic Chern--Simons response. The full Hall TFT retains the global bundle and flat-connection data underlying this local calculation. Its gravitational response will be treated separately below.

For every nonzero $p$, a simple, generally nonminimal example is $K=(4Np)$, $t=(2p)$. This even-charge theory couples through the determinant connection $2A$, and the bulk electron supplies the local charge-one fermion. If an intrinsic electron is desired, append the unimodular block $K=\operatorname{diag}(1,-1)$, $t=(1,1)$, which adds neither Hall conductivity nor chiral central charge.

The vison is the line produced by threading $2\pi$ background flux. Its powers have correlated charge and spin data
\be\label{eq:spinc-vison-pair}
 (Q(v^j),h(v^j))=\left(\frac{pj}{N},\frac{pj^2}{2N}\right) ~,
 \qquad (Q,h)\sim(Q+n,h+n/2) ~,\quad n\in\IZ ~,
\ee
where the equivalence attaches local electrons and spin is also understood modulo one. Independent reductions of $Q$ modulo one and $h$ modulo one half discard this correlation.

The line operators supported on the three-dimensional wall are the worldlines of anyons in its Hall TFT.
\tightfootnote{Henceforth, we refer to these line operators as ``wall lines'' for brevity.}
For an open magnetic surface operator, wall endpoints must be specified together with their junctions. Electric charge matching permits $U_m(2\pi pm/N;\Sigma)$ to meet $v^m$, and, if $pk=1\smod N$, permits $U_m(2\pi/N;\Sigma)$ to meet $v^k$, with the dual line for opposite orientation and possible local electron attachments. These charge-matching conditions identify candidate endpoint lines. They are necessary but do not by themselves construct topological surface--wall junctions. Such junctions must also be compatible with the wall TFT, the relevant spin$^c$ tangential data, and the gravitational surface relation \eqref{eq:Gee-sigma}. The identity $G(\Sigma)=U_m(2\pi;\Sigma)$ holds on closed surfaces. Extending it to open surfaces requires compatible boundary and tangential data. For a charge-one endpoint, the transparent electron on the wall or a dynamical bulk electron is a candidate.

Sweeping the wall with $\Delta\theta=2\pi p/N$ past a magnetic line of charge $m$ gives the Witten-effect transformation \cite{Choi:2022jqy,Cordova:2022ieu}, see also \cite[Sec.~4.1]{Choi:2022fgx},
\be\label{eq:witten-effect-wall}
  H_m(\gamma)\longmapsto H_m(\gamma)\,W^{pm/N}(\gamma;\Sigma) ~,
\ee
with $W^q$ as in \remref{rem:wilson-dressing}. The surface $\Sigma$ ends on $\gamma$ and on the wall, where it terminates on the vison line $v^m$, up to orientation and the attachment of local electrons. The magnetic charge of $H_m$ stays integral, and the fractional quantity supplied by the wall is an attached Wilson surface. Unlike an integral Wilson line, the fractional Wilson line bounding this surface cannot end on dynamical matter. When the wall instead surrounds $H_m$, the line is annihilated unless $N$ divides $m$ \cite[Sec.~4.1]{Choi:2022fgx}. This action on line operators is part of the lower morphism data of the symmetry category. A proposed extension to sheet-bound half-flux lines is formulated in \appref{app:sheet-fusion-transport}.

To choose the \emph{minimal vison sector}, first try to make $v$ have order $N$ modulo local particles. Then $v^N$ has charge $p$ and spin $pN/2$, so locality requires $p(N-1)\in2\IZ$. For odd $N$, this condition holds, and coprimality makes the vison braiding nondegenerate: the sector is the complete minimal electronic TFT $\CV^{N,p}$ of \cite{Cheng:2025ube}. Thus $\CV^{N,p}$ is a possible minimal choice of $\CT_{\mathrm{Hall}}^{N,p}$. The response action alone does not single it out. For $p=1$ and odd $N$, it is the Laughlin $\sfU(1)_N$ theory.
\tightfootnote{\label{foot:complete-spinc}Here \emph{minimal} concerns the charged Hall sector. A spin$^c$ TFT is \emph{complete} if it is fusion-closed and contains the transparent electron. This should be distinguished from the \emph{completion} of a defect, which adjoins wall degrees of freedom that cancel its residual framing anomaly. The neutral gravitational completion need not be minimal.}
For even $N$, coprimality makes $p$ odd and the order-$N$ choice violates the local spin-charge relation. More generally, denote the actual vison order as $s=\ell N$ and its Hall numerator as $r=\ell p$. Consistency requires $r(s-1)\in2\IZ$, which forces $\ell$ even when $N$ is even. The smallest allowed choice is $\ell=2$, giving $\CV^{2N,2p}$. This sector is consistent but incomplete: $v^N$ braids trivially with all visons while its odd charge and integer spin forbid it from being local. Additional lines are therefore necessary. A Hall response alone does not fix $\ell$ or a completion. When selecting minimal vison order we choose a complete TFT containing $\CV^{2N,2p}$. The choice can be Abelian (for example $\sfU(1)_8$ with charge vector $2$ at $p/N=1/2$, which is the example $K=(4Np)$, $t=(2p)$ above, stacked with the electron block) or non-Abelian \cite[Sec.~7.3]{Cheng:2025ube}. These choices are compared in \remref{rem:half-wall}.

\begin{figure}[htbp]
  \centering
  \includegraphics{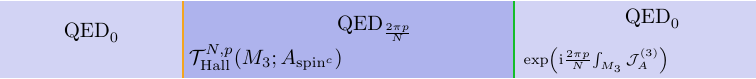}
  \caption{\label{fig:spinc-defect-slab}The Hall TFT $\CT_{\mathrm{Hall}}^{N,p}(M_3;A_{\text{spin}^c})$ (orange) and the axial interface $\exp(\imag\frac{2\pi p}{N}\int_{M_3}\CJ_A^{(3)})$ (green) bound a slab of $\mathsf{QED}_{2\pi p/N}$ inside $\mathsf{QED}_0$. Shrinking the slab gives the defect \eqref{eq:spinc-defect}.}
\end{figure}

The gauge sector of a spin$^c$ analog of the spin QED defect $\wh{\CD}_{p/N}(M_3)$ of \eqref{eq:defect-schematic} is obtained by replacing the 3d TFT $\CA^{N,p}$ by a 3d spin$^c$-coupled TFT -- denoted $\CT_{\mathrm{Hall}}^{N,p}(M_3; A_{\text{spin}^c})$ -- whose anomaly cancels the shifted electromagnetic mismatch. See \autoref{fig:spinc-defect-slab}.

Shrinking the middle region to zero size gives the following \emph{framed spin$^c$ defect}:
\be\label{eq:spinc-defect}
   \boxed{\wh{\CD}^{N,p}_{\text{spin}^c}(M_3) := \CT_{\mathrm{Hall}}^{N,p}(M_3; A_{\text{spin}^c})\otimes \exp\left(\imag \frac{2\pi p}{N}\int_{M_3} \CJ_A^{(3)}\right) ~,}
\ee
where $A_{\text{spin}^c}$ is the restriction of the dynamical bulk spin$^c$ connection to the wall. It is treated as a background only from the viewpoint of the intrinsic three-dimensional TFT. Here $\CT_{\mathrm{Hall}}^{N,p}$ appears with the orientation opposite to its standard one, as explained below \eqref{eq:spinc-Hall-action}, and may be any of the complete choices described above. The Hall theory, its background coupling, and its gravitational response are part of the defect label. This is the gauge sector part of a prospective spin$^c$ analog of the formula in \cite{Choi:2022jqy}. The gravitational completion required for a genuinely topological defect is discussed immediately below.

Equation \eqref{eq:spinc-defect} uses the Hall response of $\CT_{\mathrm{Hall}}^{N,p}$ to cancel the electromagnetic anomaly. A global half-space-gauging derivation, analogous to the spin case of \appref{app:spin-QED4-defect}, requires a differential-cohomological formulation of the shifted flux lattice and the joint magnetic and tangential background data, including torsion. We leave that construction for future work.

Categorically, \eqref{eq:spinc-defect} gives a candidate object only in a framed or relative version of the defect $3$-category. Its residual framing anomaly is the obstruction to regarding it as an object of the unframed symmetry category $\fC_{\chi}^{\mathrm{spin}^c}$.

\begin{remark}[Gravitational completion of the defect]\label{rem:gravitational-completion}
In the terminology introduced in \autoref{sec:contextandmotivation}, the hat in \eqref{eq:spinc-defect} is essential: the operator in \eqref{eq:spinc-defect} cancels the electromagnetic anomaly, but is at most a framed or relative defect until its gravitational anomaly is also canceled. For $\alpha=2\pi p/N$, moving the wall through a region $W_4$ produces the integral of the full index density $F\wedge F/(8\pi^2)-p_1/24$. Comparing two extensions and gluing them to form a closed four-manifold $X_4$ gives
\be
  \exp\left[\frac{2\pi\imag p}{N}\frac{c^2-\sigma(X_4)}{8}\right] ~.
\ee
Equation \eqref{eq:spin-c-refined-PS} captures only the electromagnetic response. Let $c_-^{\mathrm{Hall}}$ denote the chiral central charge of the chosen complete $\CT_{\mathrm{Hall}}^{N,p}$ in the standard orientation, in which $\sigma_H=\frac{p}{N}$. Including its framing anomaly, its four-dimensional response has the form
\begin{equation*}
  \exp\left\{2\pi\imag\left[
    \frac{p}{8N}c^2-
    \frac{c_-^{\mathrm{Hall}}}{8}\sigma(X_4)
  \right]\right\} ~.
\end{equation*}
The orientation-reversed Hall theory is the one that cancels the $c^2$-dependent factor of the axial wall. After this electromagnetic cancellation, the axial wall and the orientation-reversed Hall theory contribute the respective signature phases
\begin{equation*}
  \exp\left[-\frac{2\pi\imag p}{8N}\sigma(X_4)\right]
  \qquad\text{and}\qquad
  \exp\left[\frac{2\pi\imag c_-^{\mathrm{Hall}}}{8}\sigma(X_4)\right] ~.
\end{equation*}
Consequently, its closed-four-manifold signature dependence is
\be\label{eq:residual-gravitational-phase}
  \boxed{\exp\left\{
    \frac{2\pi\imag}{8}
    \left[c_-^{\mathrm{Hall}}-\frac{p}{N}\right]
    \sigma(X_4)
  \right\} ~.}
\ee
Combining the axial Jacobian (see \footrefnew{foot:axial-jacobian}) with the Hall TFT gives the invariant $c_- - \sigma_H$. The Gauss--Milgram formula for a spin$^c$ TFT \cite{Kobayashi:2021jsc,Cheng:2025ube} expresses
\be\label{eq:spinc-Gauss-Milgram}
  \exp\left[\frac{2\pi\imag}{8}(c_- - \sigma_H)\right]
  = \frac{1}{\sqrt{2}\,\CD_{\CT}}
    \sum_{a\in\CT}d_a^2\exp\left(2\pi\imag\,\wh h(a)\right) ~,
\ee
in terms of its quantum dimensions $d_a$ and $\wh h(a)=h(a)+Q(a)/2\smod1$.
\tightfootnote{Here $Q(a)$ is the electric charge of $a$ in electron units, defined jointly with $h(a)$ up to the identification $(Q,h)\sim(Q+1,h+\frac{1}{2})$ of \eqref{eq:spinc-vison-pair}. It is therefore defined modulo $2$. For an Abelian theory \eqref{eq:spinc-Hall-action}, a line $l$ has $Q=t^{\mathsf T}K^{-1}l$ and $h=\frac{1}{2} l^{\mathsf T}K^{-1}l$.}
Here the sum retains both $a$ and $a\psi_{\mathrm e}$, and $\CD_{\CT}^2=\sum_{a\in\CT}d_a^2$. The correlated electron attachment leaves $\wh h$ invariant \cite[Eq.~(C.28)]{Cheng:2025ube}. Hence $c_- - \sigma_H$ is fixed modulo $8$ by the spin$^c$ topological data and cannot be adjusted independently while keeping the same Hall theory.

Triviality of \eqref{eq:residual-gravitational-phase} is a necessary condition for an unframed defect. It tests closed-four-manifold extension dependence. It must be supplemented by cancellation of the local gravitational response and of the full $2$-framing dependence, as specified below. The $p+\imag p$ TFT of \footrefnew{foot:pip-superconductor} contains a neutral fermion. It therefore violates the spin-charge relation and requires a spin structure on the wall, so it is not an admissible spin$^c$ layer. Indeed, at fixed $\sigma_H$, an invertible charge-preserving spin$^c$ phase can shift $c_-$ only by $8\IZ$, whereas an integer quantum Hall layer changes both $c_-$ and $\sigma_H$. It follows that an invertible factor can cancel the residual gravitational anomaly without changing $\sigma_H$ only when
\be\label{eq:invertible-gravitational-completion-criterion}
  c_-^{\mathrm{Hall}}-\frac{p}{N}=0 \smod 8 ~.
\ee
For example, $\CV^{N,1}\simeq\sfU(1)_N$ has $c_-=1$ and $\sigma_H=\frac{1}{N}$, so for $N>1$ (and $N$ odd), the hatted defect retains the nontrivial phase
\be
  \exp\left[\frac{2\pi\imag}{8}
    \left(1-\frac{1}{N}\right)\sigma(X_4)\right] ~.
\ee
\end{remark}

The hatted defect with $\CT_{\mathrm{Hall}}^{N,1}=\CV^{N,1}$ therefore cannot be completed by a neutral invertible factor alone. The same conclusion holds for every Abelian Chern--Simons realization \eqref{eq:spinc-Hall-K-condition} with $N>1$: its chiral central charge is the signature of $K$, an integer, whereas $\frac{p}{N}\notin\IZ$. One must instead add suitable neutral topological order, choose a different charged completion with $c_-=\frac{p}{N} \smod 8$, or retain extra framing or bulk gravitational data. When the Hall sector is Abelian and $N>1$, the added neutral bosonic topological order must be non-Abelian: a bosonic Abelian TFT has integral chiral central charge modulo $8$, because its normalized Gauss sum is an eighth root of unity, whereas the required value $c_-^{\mathrm{Hall}}-\frac{p}{N}$ is not an integer. For the same reason, a charged completion with $c_-=\frac{p}{N}\smod 8$ must be non-Abelian. In the first two cases, a completed defect would have the schematic form
\be\label{eq:completed-spinc-defect}
  \boxed{\CD^{N,p}_{\text{spin}^c}
  = \CT^{N,p}_{\mathrm{comp}}\otimes
    \wh{\CD}^{N,p}_{\text{spin}^c} ~,}
\ee
where $\CT^{N,p}_{\mathrm{comp}}$ cancels the phase \eqref{eq:residual-gravitational-phase} and need not be invertible.

\begin{figure}[htbp]
  \centering
  \includegraphics{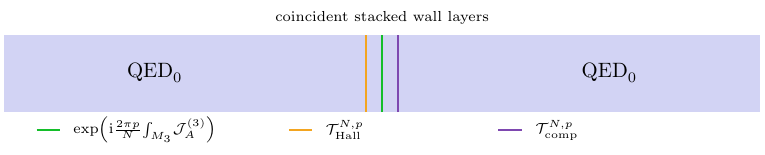}
  \caption{\label{fig:completed-wall}The completed defect \eqref{eq:completed-spinc-defect}. The Hall sector $\CT_{\mathrm{Hall}}^{N,p}$ (orange), the axial interface $\exp(\imag\frac{2\pi p}{N}\int_{M_3}\CJ_A^{(3)})$ (green), and the neutral completion $\CT_{\mathrm{comp}}^{N,p}$ (violet) are coincident layers on one three-dimensional worldvolume, drawn separated for clarity.}
\end{figure}

Schematically, the axial wall, the charged Hall sector, and the neutral framing completion are coincident layers on the same three-dimensional worldvolume. See \autoref{fig:completed-wall}.

For each choice of $\CT^{N,p}_{\mathrm{comp}}$, equation \eqref{eq:completed-spinc-defect} therefore defines an object of $\fC_{\chi}^{\mathrm{spin}^c}$ at the level of anomaly inflow. The object label includes the completion TFT.

In fact, a completion of this kind always exists at the level of anomaly inflow. Define
\be\label{eq:Delta-Np}
  \Delta_{N,p}:=c_-^{\mathrm{Hall}}-\frac{p}{N}\in\IQ ~.
\ee
The construction of \cite{Putrov:2023jqi} associates to any rational framing anomaly a neutral bosonic Witten--Reshetikhin--Turaev TFT with the compensating chiral central charge. In our conventions, we may choose $\CT^{N,p}_{\mathrm{comp}}$ with
\be
  c_-(\CT^{N,p}_{\mathrm{comp}})=\Delta_{N,p} ~,
\ee
so that its four-dimensional signature phase is the inverse
\tightfootnote{The sign convention is that of the four-dimensional response above: a $2{+}1$d topological order of chiral central charge $c_-$, placed on the wall with the induced orientation, contributes the gravitational framing phase $\exp\left(-\tfrac{2\pi\imag}{8}\,c_-\,\sigma(X_4)\right)$ on the closed four-manifold $X_4$ glued from two extensions, i.e., the $-\tfrac{c_-}{8}\sigma$ term of that response. Because this sign is already negative, choosing $c_-(\CT^{N,p}_{\mathrm{comp}})=\Delta_{N,p}$ yields $\exp\left(-\tfrac{2\pi\imag}{8}\,\Delta_{N,p}\,\sigma(X_4)\right)$, the inverse of the residual phase \eqref{eq:residual-gravitational-phase}.}
of \eqref{eq:residual-gravitational-phase}. Because this factor is neutral and bosonic, it preserves $\sigma_H$ and the spin-charge relation of the Hall sector.

Choose a rational lift of the Hall central charge, including its gravitational response, and let $\Delta_{N,p}=a/b$ with coprime integers $a,b$ and $b>0$. For $a\geq0$, a uniform explicit realization is $\CT^{N,p}_{\mathrm{comp}}=\CT_{a/b}$, with
\be\label{eq:explicit-WRT-completion}
  \CT_{a/b}
  = \left[\mathsf{SU}(2)_{-(6b-2)}\right]^{\boxtimes a}
    \boxtimes \left[\sfU(1)_2\right]^{\boxtimes 3a} ~.
\ee
Indeed, each orientation-reversed $\mathsf{SU}(2)_{6b-2}$ factor has $c_-=-3+\frac{1}{b}$, whereas each $\sfU(1)_2$ Chern--Simons factor (the semion theory) has $c_-=1$, and hence $c_-(\CT_{a/b})=a/b$. The trivial theory applies when $a=0$. For $a<0$, use the orientation reversal of $\CT_{|a|/b}$. This is a neutral bosonic topological order, so it may be stacked with either the odd-denominator theory $\CV^{N,p}$ or any complete even-denominator spin$^c$ Hall theory.

The mod-$8$ signature invariant and the full framing anomaly are distinct. A unit change of $2$-framing multiplies a TFT partition function by $\exp(2\pi\imag c_-/24)$ (see the \hyperref[par:framing-data]{framing discussion} in \autoref{sec:contextandmotivation}). Thus the line data fixing $c_-$ modulo $8$ do not fix its framing phase: an $E_8$ layer changes the latter. In \eqref{eq:completed-spinc-defect}, each factor denotes its response-refined partition function, including the gravitational Chern--Simons transgression required to cancel the local Pontryagin density response of the axial wall. Such transgression terms are defined relative to the four-dimensional inflow theory, with coefficients fixed by the combined anomaly. For the combined relative gravitational sector, a change of $2$-framing by $\Delta f\in\IZ$ gives
\begin{equation}\label{eq:full-framing-completion}
  \exp\left[\frac{2\pi\imag\,\Delta f}{24}
  \left(\Delta_{N,p}-c_-(\CT^{N,p}_{\mathrm{comp}})\right)\right] ~.
\end{equation}
The sign is chosen consistently with \eqref{eq:residual-gravitational-phase}. Reversing all wall conventions inverts the phase. Matching the rational lifts exactly cancels both this phase and the local gravitational response. Matching only modulo $8$ cancels the phase associated with changing the four-dimensional extension by a closed manifold and need not cancel the framing phase. If the mismatch lies in $8\IZ$, neutral $E_8$ layers of the appropriate orientation cancel it. The schematic wall formulas include this choice of gravitational response. Bare topological line data are insufficient. We summarize this below:

\begin{proposition}[Unframed completion]\label{prop:unframed-completion}
Let $\gcd(N,p)=1$, let $\CT^{N,p}_{\mathrm{Hall}}$ be a complete spin$^c$ Hall TFT with $\sigma_H=\frac{p}{N}$, and fix a rational lift of its chiral central charge $c_-^{\mathrm{Hall}}$ together with its gravitational response. Then the neutral bosonic TFT $\CT^{N,p}_{\mathrm{comp}}=\CT_{\Delta_{N,p}}$ of \eqref{eq:explicit-WRT-completion} cancels the residual phase \eqref{eq:residual-gravitational-phase}, the framing phase \eqref{eq:full-framing-completion}, and the local gravitational response, while preserving $\sigma_H$ and the spin-charge relation. Hence the completed defect \eqref{eq:completed-spinc-defect} is unframed at the level of anomaly inflow. If $\CT^{N,p}_{\mathrm{Hall}}$ is Abelian and $N>1$, every neutral bosonic completion is non-Abelian.
\end{proposition}

This construction gives a noncanonical, generally noninvertible framing completion. The explicit WRT representative establishes existence. Moreover, as emphasized in \cite{Putrov:2023jqi}, choosing one WRT theory independently for each rational angle need not respect fusion. A fusion-closed formulation naturally retains the completion data and labels the defects by pairs consisting of the rational anomaly and a TFT that realizes it. The hatted operator without a gravitational completion remains a framed or relative gauge sector defect, while \eqref{eq:completed-spinc-defect} with such a choice is unframed at the level of anomaly inflow.

A complementary derivation of this matching condition appears in the monodromy inflow analysis of \cite[Sec.~5.1]{Copetti:2026monodromy}, which transgresses the bulk anomaly along localized axial flux onto the codimension-two termination. After translating the normalization of the axial angle and the orientation of the Hall TFT, its additional sector $\CT_0$ plays the role of the neutral completion $\CT^{N,p}_{\mathrm{comp}}$ used here. In the present closed wall problem the same mismatch instead arises from comparing the two four-dimensional extensions of the wall and combining the signature phase of the spin$^c$ index with the framing anomaly of the Hall TFT.

The equivalence relation used in \cite{Copetti:2023mcq} allows defects to be compared after stacking decoupled TFTs and performing suitable gaugings or condensations. In the present problem, one cannot quotient by the neutral factor before canceling the framing anomaly, because its chiral central charge changes whether the defect is unframed. Two completed objects may be identified only after specifying an anomaly-compatible topological interface or condensation between them. This is why the WRT factor must be retained until the relevant morphism data are constructed.

\begin{remark}[The wall at $p/N=\frac{1}{2}$]\label{rem:half-wall}
The case $N=2$, $p=1$ illustrates the preceding discussion. The wall then interpolates between $\mathsf{QED}_0$ and $\mathsf{QED}_\pi$, the two time-reversal invariant values of the gauge--gravitational theta angle. Since $N$ is even, the vison has order at least $4$, and the Hall sector is a complete spin$^c$ TFT containing $\CV^{4,2}$. Several standard electronic topological orders with $\sigma_H=\frac{1}{2}$ qualify. Their data in the standard orientation are listed in \autoref{tbl:half-wall}, where the non-Abelian entries are the theories of \cite{Moore:1991ks,Levin:2007lpe,Lee:2007zjy,Son:2015xqa}.
\begin{table}[h]
  \centering
  \renewcommand{\arraystretch}{1.15}
  \setlength{\tabcolsep}{10pt}
  \begin{tabular}{|c|c|c|}\hline
    $\CT_{\mathrm{Hall}}^{2,1}$ & $c_-^{\mathrm{Hall}}$ & $\Delta_{2,1}=c_-^{\mathrm{Hall}}-\frac{1}{2}$ \\ \hline
    $\sfU(1)_8$ with $t=2$ & $1$ & $\frac{1}{2}$ \\ \hline
    Moore--Read Pfaffian & $\frac{3}{2}$ & $1$ \\ \hline
    anti-Pfaffian & $-\frac{1}{2}$ & $-1$ \\ \hline
    PH-Pfaffian & $\frac{1}{2}$ & $0$ \\ \hline
  \end{tabular}
  \caption{Spin$^c$ Hall TFTs with $\sigma_H=\frac{1}{2}$ for the wall at $p/N=\frac{1}{2}$, in the standard orientation. Column 3 is the residual framing anomaly \eqref{eq:Delta-Np} that a neutral completion must cancel. The $\sfU(1)_8$ entry is stacked with the electron block $K=\operatorname{diag}(1,-1)$, $t=(1,1)$.}
  \label{tbl:half-wall}
\end{table}

The PH-Pfaffian has $c_-=\sigma_H$. Hence \eqref{eq:invertible-gravitational-completion-criterion} holds without any neutral layer, and since $\Delta_{2,1}=0$ exactly, the phase \eqref{eq:full-framing-completion} and the local gravitational response also cancel with trivial $\CT^{2,1}_{\mathrm{comp}}$. The orientation-reversed PH-Pfaffian therefore completes the $\pi$ wall by itself. This is consistent with the interpretation of the interface between $\theta=0$ and $\theta=\pi$ as the analog of the boundary of a three-dimensional electronic topological insulator: the PH-Pfaffian has the same topological order as the T-Pfaffian surface state of the topological insulator \cite{Chen:2013jha,Metlitski:2013fal,Son:2015xqa}, and a time-reversal breaking surface of the topological insulator has $\sigma_H=c_-=\frac{1}{2}$.

The Pfaffian and anti-Pfaffian have $\Delta_{2,1}=\pm1$. Their completions can be Abelian, for example the semion theory $\sfU(1)_{\pm2}$, which has $c_-=\pm1$. For the Abelian entry, $\Delta_{2,1}=\frac{1}{2}$, and the Ising TFT, with $c_-=\frac{1}{2}$, is a neutral completion. It is much smaller than the WRT representative $\CT_{1/2}=\mathsf{SU}(2)_{-10}\boxtimes[\sfU(1)_2]^{\boxtimes3}$ of \eqref{eq:explicit-WRT-completion}. In the standard orientation of the Hall sector, the resulting stack is
\be
  \sfU(1)_8\boxtimes\overline{\mathsf{Ising}}\boxtimes\{1,\psi_{\mathrm e}\} ~,
  \qquad c_-=\sigma_H=\tfrac{1}{2} ~.
\ee
Let $[4]$ denote the $\sfU(1)_8$ line of label $4$, with charge $1$ and spin $1$, and let $\psi$ be the Ising fermion. Then $\psi[4]$ is a fermion of charge $1$. Condensing the neutral boson $\psi[4]\,\psi_{\mathrm e}$ identifies it with the electron and confines the lines that braid nontrivially with it. The result is the PH-Pfaffian. The two completed walls built from the first and last rows of \autoref{tbl:half-wall} thus carry the same anomaly data and are related by a condensation interface, as anticipated above. This relation alone does not identify them as objects of $\fC_{\chi}^{\mathrm{spin}^c}$.
\end{remark}

\begin{remark}[Comparison with the spin construction]\label{rem:spin-versus-spinc-completion}
Suppose that $X_4$ is spin and that a spin structure has been chosen. The spin$^c$ connection then determines an ordinary twisting line bundle with Chern class $\varepsilon\in H^2(X_4;\IZ)$, so $c=2\varepsilon$ and $[F/(2\pi)]=\varepsilon_{\IR}$. Its full connection data, including possible two-torsion, are not specified by $c$ alone. The electromagnetic response becomes
\be\label{eq:spinc-response-spin-restriction}
 \exp\left(2\pi\imag\frac{p}{8N}\int c^2\right)
 =\exp\left(2\pi\imag\frac{p}{2N}\int \varepsilon^2\right)
 =\sfZ_{\mathrm{anom}}[\rho_N(\varepsilon)] ~.
\ee
This is precisely the 1-form anomaly of \cite{Hsin:2018vcg} pulled back along $B^{(2)}=\rho_N(\varepsilon)$, with the spin convention of \eqref{eq:usual-PS}. Thus the appropriately oriented $\CA^{N,p}$ suffices for the electromagnetic cancellation. This does not mean that every larger spin$^c$ Hall theory becomes $\CA^{N,p}$ upon restriction. The magnetic surface operator at parameter $2\pi$ becomes the identity, and the fermion statistics are defined using the independent spin structure. The wall TFT therefore need not obey the spin-charge relation, and the even-denominator completion problem does not arise.

The mixed gravitational part of the ABJ anomaly nevertheless remains. As shown in \autoref{sec:contextandmotivation}, the spin defect stacked with $\CA^{N,p}$ depends on its four-dimensional extension through the phase \eqref{eq:phase-ratio}, with $c_-(\CA^{N,p})$ the chiral central charge in the standard orientation. This phase is trivial for every closed spin four-manifold precisely when
\be\label{eq:spin-gravitational-completion-criterion}
  c_-(\CA^{N,p})-\frac{p}{N}=0 \smod \frac{1}{2} ~.
\ee
The condition \eqref{eq:spin-gravitational-completion-criterion} tests independence under changing the four-dimensional extension by a closed spin manifold. It does not test dependence on an arbitrary change of $2$-framing. Because $\CA^{N,p}$ is Abelian, its chiral central charge lies in $\frac{1}{2}\IZ$ for every lift compatible with its line data: the normalized Gauss sum of its bosonic factor is an eighth root of unity, and the remaining freedom is the shift by invertible spin layers. Hence \eqref{eq:spin-gravitational-completion-criterion} holds precisely when $\frac{2p}{N}\in\IZ$, that is, since $\gcd(N,p)=1$, precisely when $N\in\{1,2\}$. For example, $\CA^{2,1}\simeq\sfU(1)_2$ has $c_-=1$, and one orientation-reversed $p+\imag p$ layer brings the total to $\frac{1}{2}=\frac{p}{N}$. When this condition holds, we choose a rational central charge lift as above. Stacking a suitable number of the neutral invertible $p+\imag p$ layers (cf. \footrefnew{foot:pip-superconductor}) shifts $c_-$ by the required multiple of $1/2$ and cancels the phase without changing the Hall response. When it does not hold, that is, for every $N>2$, no such invertible spin layer suffices, but the neutral WRT construction above still gives an unframed, generally noninvertible completion.

This also specifies the scope of the existing spin QED constructions \cite{Choi:2022jqy,Cordova:2022ieu}. The abstract construction in \cite{Choi:2022jqy} assumes an anomalous conservation law containing the electromagnetic term $j_m\wedge j_m$. On a curved spin manifold, the QED axial current also has the signature contribution dictated by the Dirac index. Similarly, the QED construction of \cite{Cordova:2022ieu} isolates and cancels the electromagnetic ABJ response. Thus the operators in those analyses give the gauge sector defect and are directly topological on flat backgrounds or with appropriate framing or relative data. Promoting them to unframed defects on arbitrary curved spin manifolds requires compensating wall degrees of freedom and the response refinement described in \eqref{eq:full-framing-completion}. Equation \eqref{eq:spin-gravitational-completion-criterion} gives the criterion for an invertible spin completion. Their electromagnetic-sector calculations remain applicable. A neutral completion leaves electromagnetic charges and the axial action on bulk local operators unchanged, but a noninvertible completion adds wall lines and contributes to the fusion of the full defects, even on flat backgrounds.
\end{remark}

In the defect $4$-category, the gauge--gravitational $\theta$ angle wall is a $1$-endomorphism of QED, or equivalently an object of the associated symmetry $3$-category. Composition of walls adds their rotation angles. For the completed rational walls, fusion also stacks their Hall and gravitational completion TFTs. Identifying the result with a chosen representative at the summed angle requires fusion interfaces and coherent higher morphisms, as discussed in \autoref{sec:conclusion} and developed further in \cite{DelZottoSaxena:symmetry-category}, where we extend the analysis of the higher structure of chiral symmetry studied in \cite{Copetti:2023mcq} to the case of spin$^c$ QED.

\section{The Example Of \texorpdfstring{$\IC\IP^2$}{CP2}}\label{sec:CP2-example}

The first part of this section is a pedagogical interlude: it works the constructions of the preceding sections out explicitly on a single background, the complex projective plane. The final subsection uses the same geometry to formulate a conjectural extension of the fusion and Ward identity analysis of \cite{Copetti:2023mcq}. As the simplest simply connected non-spin four-manifold, $\IC\IP^2$ exhibits all the specifically non-spin features of the construction in one place. Give $\IC\IP^2$ its complex orientation, and let $H\in H^2(\IC\IP^2;\IZ)$ be the positive generator, normalized by
\be
  \int_{\IC\IP^2}H^2=1 ~.
\ee
Then
\be
  w_2(T\IC\IP^2)=\rho_2(H) ~,
  \qquad
  \sigma(\IC\IP^2)=1 ~,
  \qquad
  \int_{\IC\IP^2}p_1(T\IC\IP^2)=3 ~.
\ee
In particular, $\IC\IP^2$ is oriented but non-spin, and its unit signature makes it sensitive to every signature phase that would be invisible on a spin four-manifold by Rokhlin's theorem.

Since $H^2(\IC\IP^2;\IZ)$ has no 2-torsion, its spin$^c$ structures are labeled by the characteristic elements
\be\label{eq:CP2-spinc-classes}
  c=c_1(\CL_\fs)=(2r+1)H ~,
  \qquad r\in\IZ ~.
\ee
The corresponding spin$^c$ gauge flux is
\be\label{eq:CP2-half-flux}
  \left[\frac{F}{2\pi}\right]=\frac{c}{2}
  =\left(r+\frac{1}{2}\right)H ~.
\ee
Thus, on the standard embedded projective line $\IC\IP^1\subset\IC\IP^2$,
\be
  \int_{\IC\IP^1}\frac{F}{2\pi}=r+\frac{1}{2} ~.
\ee
This immediately exhibits the $w_2$-twisted magnetic 1-form symmetry. The magnetic surface operator at parameter $2\pi$ obeys
\be
  U_m(2\pi;\IC\IP^1)
  =\exp\left(2\pi\imag\int_{\IC\IP^1}\frac{F}{2\pi}\right)
  =-1
  =(-1)^{\langle w_2(T\IC\IP^2),[\IC\IP^1]\rangle} ~,
\ee
whereas $U_m(4\pi;\IC\IP^1)=1$. Hence the failure of $2\pi$ periodicity of the magnetic parameter is already detected by the generator of $H_2(\IC\IP^2;\IZ)$.

This detection uses the half-integral flux through a nontrivial global 2-cycle and requires no magnetically charged line: a small sphere linking a genuine magnetic line still has integral flux (see \autoref{sec:fractionalization}).

The spin$^c$ Dirac index in the sector \eqref{eq:CP2-spinc-classes} is
\be\label{eq:CP2-index}
  \mathsf{Ind}(\slashed D_\fs)
  =\frac{c^2-\sigma(\IC\IP^2)}{8}
  =\frac{(2r+1)^2-1}{8}
  =\frac{r(r+1)}{2} ~.
\ee
The integrality is manifest. An axial rotation through $\alpha=2\pi p/N$ therefore contributes
\be\label{eq:CP2-axial-phase}
  \exp\left[\frac{2\pi\imag p}{N}\frac{r(r+1)}{2}\right] ~.
\ee
For the minimal characteristic element $c=H$, corresponding to $r=0$, the full index and hence \eqref{eq:CP2-axial-phase} are trivial. Nevertheless, its electromagnetic and gravitational pieces are separately nontrivial: they contribute $\exp(2\pi\imag p/(8N))$ and $\exp(-2\pi\imag p/(8N))$, respectively. This simple sector makes clear why retaining the full index density is essential even when its integral happens to vanish.

There is also a natural geometry for the defect. Let $Y_4:=\nu(\IC\IP^1)$, where $\nu(\IC\IP^1)$ denotes a closed tubular neighborhood of the standard $\IC\IP^1\subset\IC\IP^2$. The normal bundle of this embedding is $\CO_{\IC\IP^1}(1)$, the complex line bundle of degree one over $\IC\IP^1$. Thus $Y_4$ is its closed unit disk bundle, and its boundary is the unit circle bundle of $\CO_{\IC\IP^1}(1)$, namely the Hopf fibration over $\IC\IP^1$:
\be
  M_3:=\partial Y_4\cong S^3\subset\IC\IP^2 ~.
\ee
In \autoref{sec:magnetic-spinc}, $\CO$ instead denotes the trivial real line bundle.

\begin{figure}[htbp]
  \centering
  \includegraphics{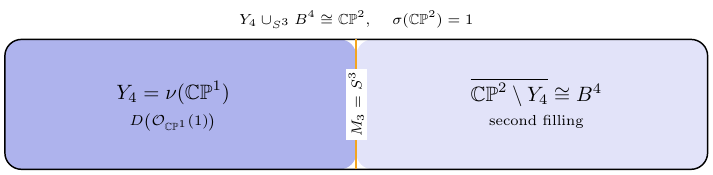}
  \caption{\label{fig:cp2-fillings}The decomposition $\IC\IP^2=Y_4\cup_{S^3}B^4$. The disk bundle $Y_4=\nu(\IC\IP^1)$ of $\CO_{\IC\IP^1}(1)$ and the complementary four-ball are the two fillings of $M_3=S^3$ (orange). Comparing wall extensions across them measures the anomaly on a closed four-manifold of signature one.}
\end{figure}

The closure of the complement is a four-ball. Hence $M_3$ has two natural fillings inside this decomposition, and gluing them with the appropriate orientation recovers $\IC\IP^2$ (see \autoref{fig:cp2-fillings}). Comparing the two extensions of a wall theory across $M_3$ therefore measures its anomaly on a closed four-manifold of signature one. This comparison tests whether the wall partition function is independent of the chosen four-dimensional extension and determines the coefficient of its signature response modulo $8$.

For a rational rotation $\alpha=2\pi p/N$, the gauge sector defect on this $S^3$ is
\be\label{eq:CP2-hatted-defect}
  \wh{\CD}^{N,p}_{\mathrm{spin}^c}(S^3)
  =\CT_{\mathrm{Hall}}^{N,p}(S^3;\iota^*A)
  \otimes
  \exp\left(\frac{2\pi\imag p}{N}
  \int_{S^3}\CJ_A^{(3)}\right) ~.
\ee
Here $\iota:S^3\hookrightarrow\IC\IP^2$ is the inclusion, and $\CT_{\mathrm{Hall}}^{N,p}=\CV^{N,p}$ for odd $N$, while for even $N$ it is a chosen complete spin$^c$ Hall TFT containing the required vison sector. It is placed in the orientation that reverses its standard electromagnetic response, so its $c^2$-dependent phase cancels the electromagnetic part of \eqref{eq:CP2-axial-phase}. If $c_-^{\mathrm{Hall}}$ is its chiral central charge in the standard orientation, the ratio obtained from the two fillings of $S^3$ is then
\be\label{eq:CP2-residual-framing}
  \exp\left[\frac{2\pi\imag}{8}
  \left(c_-^{\mathrm{Hall}}-\frac{p}{N}\right)\right] ~.
\ee
Because $\sigma(\IC\IP^2)=1$, no divisibility condition weakens this phase. Thus $\IC\IP^2$ directly detects the residual phase relating the two extensions: the hatted operator \eqref{eq:CP2-hatted-defect} gives the same result for both fillings only if $c_-^{\mathrm{Hall}}-p/N=0\smod8$, or if additional wall degrees of freedom cancel \eqref{eq:CP2-residual-framing}. Full framing independence further requires the response refinement and lift matching of \eqref{eq:full-framing-completion}. Reference \cite[App.~D]{Copetti:2026monodromy} also uses the partition function on $\IC\IP^2$ with half-integral flux to compute the invariant $c_- - \sigma_H$.
\tightfootnote{Partition functions on $\IC\IP^2$ with half-integral flux were used earlier to extract the chiral central charge of spin$^c$ topological orders \cite{Kobayashi:2021jsc}.}
 The role here is more geometric: the decomposition $\IC\IP^2=Y_4\cup_{S^3}B^4$ places the closed wall on the common $S^3$ and turns the same invariant into the ratio between its two four-dimensional extensions.

For the unframed completion, choose the gravitational response and rational central charge lifts as in \eqref{eq:full-framing-completion}. Then \eqref{eq:completed-spinc-defect} becomes
\be\label{eq:CP2-completed-defect}
  \CD^{N,p}_{\mathrm{spin}^c}(S^3)
  =\CT^{N,p}_{\mathrm{comp}}(S^3)
  \otimes\wh{\CD}^{N,p}_{\mathrm{spin}^c}(S^3) ~,
  \qquad
  c_-(\CT^{N,p}_{\mathrm{comp}})=c_-^{\mathrm{Hall}}-\frac{p}{N} ~.
\ee
This exact matching cancels the local gravitational response and full framing phase, as well as the extension-dependence phase \eqref{eq:CP2-residual-framing}. Matching only modulo $8$ would establish the latter cancellation alone. For example, when $p=1$ and $N>1$ is odd, one may take $\CT_{\mathrm{Hall}}^{N,1}\simeq\sfU(1)_N$, for which $c_-^{\mathrm{Hall}}=1$. The uncompleted defect then has the nontrivial $\IC\IP^2$ phase
\be
  \exp\left[\frac{2\pi\imag}{8}\left(1-\frac{1}{N}\right)\right] ~.
\ee
For the chosen Hall lift $c_-^{\mathrm{Hall}}=1$, take the neutral completion to have central charge lift $1-1/N$, with the associated gravitational response. In this way, $\IC\IP^2$ exhibits the shifted magnetic flux, the full spin$^c$ index, and the framing completion of the rational chiral defect.


We now describe a conjectural extension of the fusion and Ward--Takahashi analysis of \cite{Copetti:2023mcq} to the spin$^c$ defects constructed here. The fusion and coherence analysis planned in \cite{DelZottoSaxena:symmetry-category} is intended to provide a categorical framework in which this proposal can be tested.

There is nevertheless a fusion statement that already follows at the level of anomaly inflow. Label a decorated wall by its rotation angle, its charged Hall theory, and its neutral framing completion as
\be
  \CD_{\alpha}[\CT_{\mathrm{Hall}},\CT_{\mathrm{comp}}] ~.
\ee
Before reducing the angle modulo its physical period, and before condensing any common topological sectors, fusion is the raw stacking operation
\be\label{eq:raw-completed-defect-fusion}
  \begin{aligned}
    \CD_{\alpha}[\CT_{\mathrm{Hall}},\CT_{\mathrm{comp}}]
    \otimes
    \CD_{\beta}[\CT'_{\mathrm{Hall}},\CT'_{\mathrm{comp}}]
    &:=
    \CD_{\alpha+\beta}
    \bigl[\CT_{\mathrm{Hall}}\boxtimes\CT'_{\mathrm{Hall}},
    \CT_{\mathrm{comp}}\boxtimes\CT'_{\mathrm{comp}}\bigr] ~.
  \end{aligned}
\ee
For such a wall, define its residual extension-dependence class by
\be\label{eq:defect-extension-class}
  \delta(\CD_{\alpha})
  :=c_-(\CT_{\mathrm{Hall}})
  -\sigma_H(\CT_{\mathrm{Hall}})
  -c_-(\CT_{\mathrm{comp}})
  \in\IQ/8\IZ ~.
\ee
Our completed walls have $\delta(\CD_{\alpha})=0$ and, with the chosen rational lifts, vanishing full gravitational mismatch. The class $\delta$ alone encodes only the signature-dependent extension phase modulo $8$. Since both Hall conductivity and chiral central charge are additive under stacking, \eqref{eq:raw-completed-defect-fusion} implies
\be
  \delta(\CD_{\alpha}\otimes\CD_{\beta})
  =\delta(\CD_{\alpha})+\delta(\CD_{\beta}) ~.
\ee
On $\IC\IP^2$, where the signature is one, this becomes the multiplicative relation
\be\label{eq:CP2-fusion-extension-phase}
  \Phi_{\IC\IP^2}(\CD_{\alpha}\otimes\CD_{\beta})
  =\Phi_{\IC\IP^2}(\CD_{\alpha})
   \Phi_{\IC\IP^2}(\CD_{\beta}) ~,
  \qquad
  \Phi_{\IC\IP^2}(\CD):=
  \exp\left(\frac{2\pi\imag}{8}\delta(\CD)\right) ~.
\ee
Thus the completed walls are closed under raw stacking at the level of their electromagnetic and gravitational responses.

This does not yet identify the right-hand side of \eqref{eq:raw-completed-defect-fusion} with a chosen representative of the defect at angle $\alpha+\beta$. Such an identification requires a two-dimensional topological fusion interface
\be\label{eq:prospective-fusion-interface}
  \mathfrak J_{\alpha,\beta}:
  \CD_{\alpha}[\CT_{\mathrm{Hall}}(\alpha),\CT_{\mathrm{comp}}(\alpha)]
  \otimes
  \CD_{\beta}[\CT_{\mathrm{Hall}}(\beta),\CT_{\mathrm{comp}}(\beta)]
  \longrightarrow
  \CD_{\alpha+\beta}
  [\CT_{\mathrm{Hall}}(\alpha+\beta),
   \CT_{\mathrm{comp}}(\alpha+\beta)] ~,
\ee
together with associator interfaces and their higher coherence data. These interfaces must implement the necessary condensations in the charged Hall sectors and equivalences between the neutral WRT completion factors. For even denominators they must moreover be defined using a chosen complete spin$^c$ Hall TFT containing the required vison sector. Constructing these interfaces remains open.

The handle presentation of $\IC\IP^2$ makes it a natural first test of the corresponding Ward identity. It has one $+1$-framed two-handle and
\be
  H^2(\IC\IP^2;\IZ_N)\cong\IZ_N ~.
\ee
Let $c=(2r+1)H$ be a characteristic element and write $b=j\rho_N(H)$, with $j\in\IZ_N$, for a prospective discrete magnetic background. By analogy with \cite[Sec.~4]{Copetti:2023mcq}, the expected identity with local insertions has the schematic form
\be\label{eq:conjectural-CP2-Ward-identity}
  \left\langle\prod_i\CO_i\right\rangle_{\IC\IP^2,c}
  \stackrel{?}{=}
  \frac{1}{\sqrt N}
  \exp\left(\frac{2\pi\imag p}{N}\sum_i q_i\right)
  \sum_{j\in\IZ_N}
  \CK^{\mathrm{spin}^c}_{N,p}(j;c)
  \left\langle\prod_i\CO_i
  \right\rangle_{\IC\IP^2,c;j} ~.
\ee
Here $q_i$ is the normalized chiral charge of $\CO_i$, and the last correlator denotes coupling to the background $b=j\rho_N(H)$. The kernel $\CK^{\mathrm{spin}^c}_{N,p}(j;c)$ should be the $+1$-framed link invariant of the complete wall, stacked with its Hall TFT and WRT completion. It should replace the ordinary Pontryagin square kernel of the spin construction by a quadratic function compatible with the characteristic class $c$, the spin$^c$-refined response \eqref{eq:spin-c-refined-PS}, and the $w_2$-twisted magnetic composition law. Formula \eqref{eq:conjectural-CP2-Ward-identity} is deliberately marked with a question mark because these requirements do not yet determine the kernel uniquely.

A derivation should address four concrete problems:
\begin{enumerate}
  \item Define the $w_2$-twisted discrete magnetic background and determine whether the Ward identity is sectorwise at fixed $c$ or instead acts on the compact QED sum over all characteristic classes.

  \item Choose a complete spin$^c$ Hall theory and a neutral WRT completion for every rational angle, then evaluate their framed link invariant for the $+1$-framed handle of $\IC\IP^2$.

  \item Construct the interfaces \eqref{eq:prospective-fusion-interface}, their condensation data, and the TFT-valued associators that enter the surgery argument.

  \item Include fermion zeromodes. Since \eqref{eq:CP2-index} is nonzero in generic spin$^c$ sectors, the partition function without insertions can vanish, and a nontrivial Ward identity may require operators that saturate the zeromodes.
\end{enumerate}
Completing these steps would turn the anomaly-level relation \eqref{eq:CP2-fusion-extension-phase} into a genuine fusion rule and would promote \eqref{eq:conjectural-CP2-Ward-identity} from a proposed spin$^c$ extension to a derived Ward--Takahashi identity. We hope to return to this and related issues in \cite{DelZottoSaxena:symmetry-category}.

\section{Preliminary Remarks About Constraints On RG Flows}\label{sec:dynamics}

The mixed anomaly of \propref{prop:bordism-anomaly} and the rational chiral defects of \autoref{sec:chiral-defect-revisited} constrain the RG flow of theories on oriented four-manifolds with an exact magnetic 1-form symmetry carrying this anomaly. Below, we list few conditional constraints on possible infrared phases and examples of their non-Abelian realizations (we will return to these applications in more detail in \cite{DelZottoSaxena:symmetry-category}). They apply to any theory with this symmetry structure. Massless spin$^c$ QED$_4$ itself is infrared free and realizes only the Coulomb phase of item~\ref{item:dyn-coulomb}, with $a=A$. The other phases require additional matter or interactions.

\begin{enumerate}[leftmargin=*]
  \item \emph{Absence of trivially gapped phases.} A theory with this anomaly cannot flow to a trivially gapped phase that preserves the magnetic 1-form symmetry: a symmetry-preserving standalone invertible infrared theory cannot realize the nontrivial deformation class of $\alpha_5$. The infrared must therefore have gapless degrees of freedom, as in a Coulomb phase where the magnetic symmetry is spontaneously broken, or nontrivial topological order that matches the anomaly \cite[Secs.~3.3 and 4.1]{BergCappelliVilla2026}. At the cohomological level, the anomaly becomes trivial on backgrounds pulled back along the $4\pi$-periodic extension \eqref{eq:extension-sequence} (see \autoref{sec:magnetic-gravitational-anomaly}). This is the starting point of the symmetry-extension mechanism of \cite{Wang:2017loc}. A concrete symmetry-preserving gapped phase is the charge-two Higgs phase of item~\ref{item:dyn-higgs}.

  \item \emph{Shifted photon flux in Coulomb phases.}\label{item:dyn-coulomb} Suppose an infrared photon $a$, with curvature $f_a$, carries the ultraviolet magnetic current with unit normalization, $F/2\pi\to f_a/2\pi$, and that $-\frac{\imag}{2\pi}\int_{M_4}B\wedge f_a$ is the only infrared coupling to the magnetic background. Matching the large gauge variation of \autoref{sec:magnetic-gravitational-anomaly} then requires
  \be\label{eq:coulomb-flux-quantization}
    \int_{\Sigma_2}\frac{f_a}{2\pi}
    \in\IZ+\frac{1}{2}\int_{\Sigma_2}w_2(TM_4) ~,
    \qquad [\Sigma_2]\in H_2(M_4;\IZ) ~.
  \ee
  This fixes the free flux lattice. Torsion and flat data require a global refinement, as in \footrefnew{foot:spinc-surface-torsion}. The usual Maxwell realization is a spin$^c$ connection, whose unit Wilson line is fermionic by the spin-charge relation. If the ultraviolet theory and all remaining infrared sectors are separately free of purely gravitational anomalies, the $w_2w_3$ anomaly of all-fermion electrodynamics \cite[Sec.~3.2]{Wang:2018qoy} (see also \cite[Sec.~4.2]{DavighiLohitsiriDebray2023}) excludes $\mathsf{Spin}^c_{\mathrm f}$ and selects $\mathsf{Spin}^c_{\mathrm b}$.
   \tightfootnote{For an $\mathsf{SU}(2)$ gauge theory with the new $\mathsf{SU}(2)$ anomaly that is Higgsed to $\sfU(1)$, with the $w_2w_3$ anomaly matched in the Abelian theory, see \cite[Sec.~4.2]{Brennan:2023vsa} and item~\ref{item:dyn-nonabelian}.}
  A neutral anomalous sector could otherwise compensate that gravitational anomaly.

  \item \emph{Fermionic $\IZ_2$ gauge order and vortices on non-spin backgrounds.}\label{item:dyn-higgs} Add to spin$^c$ QED$_4$ a charge-two scalar $\phi$, either elementary or a composite of two fermions. It is bosonic by the spin-charge relation. Its condensate gaps the photon and leaves a $\IZ_2$ gauge theory with fermionic electric charge \cite{Thorngren:2014pza}. Assuming the fermions and other matter modes are also gapped, this sector with its inherited magnetic-background coupling matches $\alpha_5$ \cite[Sec.~4.1.1]{BergCappelliVilla2026}.

  The Higgs field $\phi$ is a section of $\CL_\fs$, which is nontrivial on a non-spin manifold. It must therefore have zeros, since an everywhere nonzero section would trivialize the bundle. For a transverse section, these zeros form the vortex worldvolume $V$, representing the Poincar\'e dual of $c_1(\CL_\fs)$ and carrying minimal vortex flux $\pi$. Since $\rho_2\big(c_1(\CL_\fs)\big)=w_2(TM_4)$,
  \be\label{eq:Higgs-vortex-twist}
    U_m(2\pi;\Sigma_2)=(-1)^{\Sigma_2\cdot V}=(-1)^{\langle w_2(TM_4),[\Sigma_2]\rangle}=G(\Sigma_2) ~.
  \ee
  These vortices are required by the non-spin background and realize the $w_2$-twist. A nowhere-vanishing condensate requires $c_1(\CL_\fs)=0$ and hence a spin background.
  \tightfootnote{\label{foot:higgs-topological-limit}The minimal strict infrared TFT sums over dynamical spin structures and has vanishing partition function on non-spin manifolds. Its anomaly is carried by the magnetic coupling $\exp\big(\imag\pi\int_{M_4}\eta\smile\rho_2[dB/2\pi]\big)$ of the $\IZ_2$ cochain $\eta$ with $\delta\eta=w_2(TM_4)$. Under a change of cochain representative $w_2\mapsto w_2+\delta f$, $\eta\mapsto\eta+f$, it shifts by $(-1)^{\int_{M_4}f\smile\rho_2[dB/2\pi]}$, which is the boundary variation of $\alpha_5$. The full Higgs theory includes the finite-tension vortex worldvolumes required by the non-spin background, which are omitted in this limit.}

  \item \emph{Non-Abelian realizations of spin$^c$ photons.}\label{item:dyn-nonabelian} There are two useful constructions with generalized spin structures \cite{Wang:2018qoy,Moore:2024vsd}. Confinement can leave an existing Abelian photon, while Higgsing can select a photon in the Cartan. For $\mathsf{SU}(N)$ gauge theories on $\IC\IP^2$ with background baryon, color, and flavor fluxes, see \cite{Anber:2020gig}.

  First, take $\sfU(N)$ gauge theory with $N_f$ massive fundamental Dirac fermions and a $(\mathsf{Spin}(4)\times\sfU(N))/\IZ_2$ structure. If the $\mathsf{SU}(N)$ sector confines and gaps, with no additional gapless or topological sector, the surviving photon is normalized by the determinant map:
  \tightfootnote{The map $\det: \sfU(N) \to \sfU(1)$ is surjective, with kernel $\mathsf{SU}(N)$. So it induces an isomorphism \eqref{eq:un-determinant-photon}. The derivative of $\det$ at the identity is the trace, and $\Tr(g^{-1}dg) = d\,\log\,\det\,g$. Under a $\sfU(N)$ gauge transformation, $A \mapsto g^{-1}Ag + g^{-1}dg$, therefore $\Tr\,A \mapsto \Tr\,A + d\,\log\,\det\,g$. Thus $\Tr\,A$ is an Abelian connection whose transition functions are $\det\,g_{\alpha\beta}$. Equivalently, $a=\det_*A_{\sfU(N)}$ is obtained by the transfer of structure group along $\det$, as for the determinant line bundle of a spin$^c$ structure (see \autoref{subsec:spinc-connections} and \cite[App.~B]{Moore:2024vsd}). With a $(\mathsf{Spin}(4)\times\sfU(N))/\IZ_2$ structure there is no $\sfU(N)$ bundle, and the transfer is along the homomorphism $[(g,u)]\mapsto[(g,\det\,u)]$ to $\mathsf{Spin}^c(4)$. It is well defined for odd $N$, since $\det\,(-\mathbbm{1}_N)=-1$. For even $N$, the well-defined homomorphism is $[(g,u)]\mapsto([g],\det\,u)$ to $\sfS\sfO(4)\times\sfU(1)$, and $a$ is an ordinary connection.}
  \be\label{eq:un-determinant-photon}
    \det:\sfU(N)/\mathsf{SU}(N)\xrightarrow{\ \cong\ }\sfU(1) ~,
  \qquad a=\Tr A_{\sfU(N)} ~,
  \qquad f_a=\Tr F_{\sfU(N)} ~.
  \ee
  The trace extracts the central $\fu(1)$ curvature, which need not vanish. Thus confinement leaves an Abelian field already present in the ultraviolet. The remaining photon couples to baryon charge, normalized so that a baryon has charge one. Baryons have fermion parity $(-1)^N$. For odd $N$, $\det\,(-\mathbbm{1}_N)=-1$, so $a$ inherits spin$^c$ structure and matches $\alpha_5$. Its determinant-line connection is $2a$. For even $N$, $a$ is ordinary, the baryons are bosons, and this mixed anomaly is absent. For $N=2$, this is the spin-$\sfU(2)$ setting of \cite{Davighi:2020bvi}, in which fermions have odd charge under the central $\sfU(1)$. The determinant photon $a$ couples instead to bosonic baryons and is an ordinary $\sfU(1)$ connection.

  Second, take $\mathsf{SU}(2)$ gauge theory with one Dirac doublet and an adjoint Higgs field \cite[Sec.~3.1]{Wang:2018qoy}. A $(\mathsf{Spin}(4)\times\mathsf{SU}(2))/\IZ_2$ structure induces spin$^c$ structure on the surviving Cartan photon: doublet fermions have charges $\pm1$ and $W$ bosons have charges $\pm2$. With massive doublet fermions, the minimum-charge monopole is bosonic, giving $\mathsf{Spin}^c_{\mathrm b}$. Since the photon is selected by Higgsing, its magnetic symmetry is only emergent below the mass of the dynamical monopoles, which explicitly break it in the full theory.
  \tightfootnote{A single Weyl multiplet of isospin $\frac{3}{2}$ instead gives all-fermion electrodynamics \cite[Sec.~3.1]{Wang:2018qoy}. Its ultraviolet theory already has the new $\mathsf{SU}(2)$ anomaly and requires additional spin or background spin$^c$ data, or five-dimensional inflow \cite[Secs.~2.5, 2.6 and 4.7]{Wang:2018qoy}. On embedding in $\sfU(2)$ with a spin-$\sfU(2)$ structure, this anomaly becomes a perturbative mixed gauge and gauge--gravitational anomaly \cite{Davighi:2020bvi}.}
  %


  \item \emph{Chiral defects along RG flows.} The rational chiral defects of \autoref{sec:chiral-defect-revisited} also constrain RG flows. Their Hall TFT couples to the bulk through the magnetic 1-form symmetry, so they are exact only when that symmetry is exact. When dynamical monopoles break it, as in the $\mathsf{SU}(2)$ example of item~\ref{item:dyn-nonabelian}, their selection rules are violated non-perturbatively, which leads to exponentially small fermion masses \cite{Cordova:2022ieu}.

  Suppose that an axial rotation by $2\pi p/N$ remains a topological defect along a flow with an exact magnetic 1-form symmetry. For Dirac fermions with electric charges $q_i$ and integer axial charges $\chi_i$, normalized as in \eqref{eq:ABJ-anomalous-conservation},
  \be\label{eq:general-ABJ-coefficients}
    d\CJ_A^{(3)}=k_{\mathrm{EM}}\,\frac{F\wedge F}{8\pi^2}-k_{\mathrm{G}}\,\frac{p_1(TM_4)}{24} ~,
    \qquad k_{\mathrm{EM}}=\sum_i\chi_i q_i^2 ~,
    \qquad k_{\mathrm{G}}=\sum_i\chi_i ~.
  \ee
  The defect then requires a Hall TFT with $\sigma_H=k_{\mathrm{EM}}\,p/N$ and leaves the residual framing anomaly $c_-^{\mathrm{Hall}}-k_{\mathrm{G}}\,p/N$. The coefficients $(k_{\mathrm{EM}},k_{\mathrm{G}})$ fix the anomaly inflow of the defect and therefore match between the ultraviolet and the infrared. The difference $c_-^{\mathrm{Hall}}-\sigma_H$ is not itself invariant along the flow, since it depends on the chosen Hall TFT. The invariant combination is that of the completed wall, the Hall sector together with its neutral completion $\CT_{\mathrm{comp}}$, whose total chiral central charge is $k_{\mathrm{G}}\,p/N$:
  \be\label{eq:completed-wall-matching}
    \big(c_--\sigma_H\big)_{\mathrm{wall}}
    =c_-^{\mathrm{Hall}}+c_-(\CT_{\mathrm{comp}})-\sigma_H
    =\big(k_{\mathrm{G}}-k_{\mathrm{EM}}\big)\frac{p}{N} ~,
  \ee
  at both ends of the flow. Both ends therefore face the same framing completion problem. This holds exactly, as required by the framing phase \eqref{eq:full-framing-completion} and the local gravitational response. Matching only the signature phase \eqref{eq:residual-gravitational-phase} would fix the total chiral central charge only modulo $8$. For a single unit-charge Dirac fermion, $k_{\mathrm{EM}}=k_{\mathrm{G}}=1$ and \eqref{eq:completed-wall-matching} vanishes.

  On spin$^c$ backgrounds, the two coefficients are not independent. On a closed $M_4$ with characteristic class $c$, \eqref{eq:spinc-instanton-number} and \eqref{eq:HirzebruchSignature} give $\int F\wedge F/8\pi^2=c^2/8$ and $\int p_1(TM_4)/24=\sigma(M_4)/8$, so the integrated anomaly $\int_{M_4}d\CJ_A^{(3)}$ of \eqref{eq:general-ABJ-coefficients} is
  \be\label{eq:ABJ-integral-decomposition}
    \frac{k_{\mathrm{EM}}\,c^2-k_{\mathrm{G}}\,\sigma(M_4)}{8}
    =k_{\mathrm{G}}\,\frac{c^2-\sigma(M_4)}{8}+\frac{k_{\mathrm{EM}}-k_{\mathrm{G}}}{8}\,c^2 ~.
  \ee
  The first term is $k_{\mathrm{G}}$ times the Dirac index $\mathsf{Ind}(\slashed D_\fs)$, and the second is $\frac{1}{8}(k_{\mathrm{EM}}-k_{\mathrm{G}})$ times the integer $c^2$. The index and $c^2$ generate the $2\pi$-periodic topological terms on spin$^c$ backgrounds \cite[App.~A.2]{Seiberg:2016rsg}, and they vary independently. In particular, there are spin$^c$ backgrounds with $(\mathsf{Ind}(\slashed D_\fs),c^2)=(1,0)$ and $(0,1)$.
  \tightfootnote{Both $\mathsf{Ind}(\slashed D_\fs)$ and $c^2$ are additive under connected sum, so it suffices to exhibit backgrounds whose values form a basis of $\IZ^2$. On $\IC\IP^2$, the spin$^c$ structures are $c=(2r+1)H$ as in \eqref{eq:CP2-spinc-classes}, with $c^2=(2r+1)^2$ and index $r(r+1)/2$ by \eqref{eq:CP2-index}. The choices $r=0$ and $r=1$ give $(0,1)$ for $c=H$ and $(1,9)$ for $c=3H$, and these two vectors form a basis of $\IZ^2$. Reversing the orientation of $\IC\IP^2$ with $c=H$ flips the signs of $c^2$ and $\sigma$, so $\ov{\IC\IP}{}^2$ with $c=H$ has $(0,-1)$. The connected sum of $\IC\IP^2$ with $c=3H$ and nine copies of $\ov{\IC\IP}{}^2$ with $c=H$ therefore realizes $(1,0)$, while $\IC\IP^2$ with $c=H$ realizes $(0,1)$. In fact, the pair $(\mathsf{Ind}(\slashed D_\fs),c^2)$ gives an isomorphism $\Omega_4^{\mathsf{Spin}^c}\to\IZ\oplus\IZ$ \cite{Stong1968}.}
  On these backgrounds, $\int_{M_4}d\CJ_A^{(3)}$ equals $k_{\mathrm{G}}$ and $\frac{1}{8}(k_{\mathrm{EM}}-k_{\mathrm{G}})$, respectively. The integrated anomaly is therefore an integer on every spin$^c$ background precisely when both coefficients are integers, that is,
  \be\label{eq:kEM-kG-congruence}
    k_{\mathrm{G}}\in\IZ ~,
    \qquad k_{\mathrm{EM}}\equiv k_{\mathrm{G}}\smod8 ~.
  \ee
  For a fermion spectrum, this holds automatically, since the charges $q_i$ are odd and $q_i^2\equiv1\smod8$. In an infrared phase in which the axial symmetry is realized by a compact scalar $\vartheta\sim\vartheta+2\pi$, coupled through $\imag\vartheta$ times the right-hand side of \eqref{eq:general-ABJ-coefficients}, it is instead a consistency condition, namely the periodicity of $\vartheta$. This is the dynamical counterpart of the exact $2\pi$-periodicity of the gauge--gravitational theta angle (see \appref{app:theta-periodicity}). On spin manifolds, $c$ is even and the intersection form is even, so $c^2/8\in\IZ$, while $\sigma(M_4)/8\in2\IZ$ by Rokhlin's theorem. The two pieces $k_{\mathrm{EM}}\,c^2/8$ and $k_{\mathrm{G}}\,\sigma(M_4)/8$ of \eqref{eq:ABJ-integral-decomposition} are then separately integers, and $k_{\mathrm{EM}}$ and $k_{\mathrm{G}}$ are unrelated. A discrete analog, which ties baryon and gravitational anomaly coefficients on $\IC\IP^2$, appears in \cite{Anber:2020gig}.

  These constraints do not apply to the $\sfU(N)$ example of item~\ref{item:dyn-nonabelian}. With massive quarks, the infrared has no chiral symmetry. With massless quarks, $\mathsf{SU}(N)$ instantons break the flavor-singlet axial symmetry to an invertible $\IZ_{2N_f}$, so no rational defects exist in the ultraviolet. More generally, in gauge theories with a non-Abelian factor and an Abelian factor, only a discrete set of chiral defects survives \cite{Anber:2023pny}.
\end{enumerate}

\section{Conclusions}\label{sec:conclusion}

Spin$^c$ geometry changes both the electromagnetic flux lattice and the global interpretation of the magnetic 1-form symmetry. The gravitational sign is detected on global surfaces, where it equals $(-1)^{\Sigma\cdot\Sigma}$, fixed by the normal bundle, while the genuine local magnetic charge lattice remains integral. A half-flux electromagnetic sheet by itself does not realize the gravitational kernel. The relation $U_m(\alpha+2\pi)=(-1)^{\int w_2}U_m(\alpha)$ is most naturally regarded as a mixing of magnetic and tangential symmetry data, and its oriented bordism anomaly is that of the fully summed compact theory. As discussed in \autoref{sec:dynamics}, this anomaly forbids a symmetry-preserving trivial gap and constrains Coulomb and Higgs phases under the assumptions stated there. Along such flows, the rational chiral defects also match their electromagnetic and gravitational ABJ coefficients, which on non-spin backgrounds ties the framing completion problem of the ultraviolet to that of the infrared. Rational chiral rotations are controlled by the full spin$^c$ Dirac index, so canceling their electromagnetic response with a Hall TFT leaves a signature-dependent framing anomaly governed by $c_- - \sigma_H$.

This last point also sharpens the status of the previous spin manifold constructions \cite{Choi:2022jqy,Cordova:2022ieu}. Those works determine the electromagnetic gauge sector defect and its flat-space consequences, but do not combine the gravitational contribution to the spin Dirac index governing the curved-space ABJ anomaly with the framing anomaly of the Hall TFT. Electromagnetic anomaly cancellation therefore establishes a framed or relative defect. An unframed defect on an arbitrary curved spin manifold requires cancellation of the gravitational response and full framing anomaly. The test of extension independence in \remref{rem:spin-versus-spinc-completion} is not sufficient by itself. This additional condition leaves their flat-space gauge sector fusion rules and operator actions unchanged, although the neutral completion contributes additional data to the fusion of the completed defects.

The present construction concerns the global closed wall problem. We formulate compact spin$^c$ QED as a sum over dynamical spin$^c$ sectors, distinguish that theory from a fixed spin$^c$ sector, and derive the $4\pi$-periodic, $w_2$-twisted magnetic operator algebra. We also interpret its bordism anomaly \cite{DavighiLohitsiriDebray2023} for the summed theory. For the chiral wall, the rationality of the residual framing anomaly gives a uniform anomaly-level unframed completion: following \cite{Putrov:2023jqi}, one may stack \eqref{eq:spinc-defect} with a neutral bosonic WRT TFT of compensating rational chiral central charge. This factor preserves the Hall response and the spin-charge relation. For even denominator, it must be stacked with a complete spin$^c$ Hall TFT containing the required vison sector. The $\IC\IP^2$ example realizes the extension dependence of this closed wall as the comparison of its two fillings. On spin backgrounds the signature criterion weakens by Rokhlin's theorem, and with the minimal Abelian TFT $\CA^{N,p}$ an invertible completion exists precisely for $N\in\{1,2\}$.

The recent monodromy defect analysis of \cite{Copetti:2026monodromy} arrives at the same response matching condition. Its monodromy defects give a complementary, dynamical view of the codimension-two termination, whereas the index-theoretic and bordism methods used here fix the global definition and framing independence of the closed wall. Since a gapless termination need not be a topological $1$-morphism, joining the two pictures -- terminations compatible with the $w_2$-twisted magnetic 1-form symmetry and with the framing completion -- is a concrete target for the ambient defect category and its topological symmetry subcategory.

The individual WRT-completed walls constructed above give candidate objects of $\fC_{\chi}^{\mathrm{spin}^c}$, or equivalently $1$-endomorphisms of QED in the full defect $4$-category. The conjectural $\IC\IP^2$ program of \autoref{sec:CP2-example} gives a concrete setting in which to test their fusion and the resulting Ward identities. What remains is to organize them into a fusion-closed symmetry category. Choosing one completion $\CT_{\mathrm{comp}}(\alpha)$ for each rational angle amounts to choosing a section of the larger family of pairs $(\alpha,\CT_{\mathrm{comp}})$. Under fusion the completion factors stack, whereas the WRT existence construction does not provide coherent equivalences
\be
  \CT_{\mathrm{comp}}(\alpha)\boxtimes
  \CT_{\mathrm{comp}}(\beta)
  \simeq \CT_{\mathrm{comp}}(\alpha+\beta) ~.
\ee
A fusion-closed formulation must therefore either retain the completion TFT as part of the defect label, as in the commutative monoid construction of \cite{Putrov:2023jqi}, or construct two-dimensional topological interfaces or condensations that realize these equivalences. These interfaces are $1$-morphisms in the symmetry $3$-category. Their junction lines and point junctions provide the higher morphisms needed for associativity and its coherence conditions, as in \cite{Copetti:2023mcq,Carqueville:2023jhb}. The construction of these fusion interfaces and their higher coherence data for the chiral defects of spin$^c$ QED remains an open problem. Determining a natural minimal choice of completion, particularly one compatible with the even-denominator spin$^c$ Hall completions, is a separate categorical problem. It does not obstruct the existence of any individual completed defect. Extending this organization to manifolds with torsion further requires a differential cohomology refinement.

More broadly, the interplay between generalized symmetries, anomalies, and tangential structures suggests connections with generalized cohomology and the spectrum of invertible field theories. This perspective is motivated in part by themes discussed in \cite{Moore:2025tmt}. These are natural directions for further investigation 

\medskip

Further directions include:
\begin{itemize}
  \item Work out the anomaly matching for the $\sfU(N)$ and generalized-spin ultraviolet realizations of \autoref{sec:dynamics}, including the flavor dependence and the case of massless quarks, whose pions would add a gapless sector.

  \item Construct open versions of the gravitational surface $G$ with the relative tangential data required at their boundaries, and determine their junctions with magnetic surface operators. Such a construction could realize the gravitational kernel on decorated boundary lines, but would not by itself assign them half-integral magnetic charge. The independent quantum completion and fusion questions for electromagnetic sheets are summarized in \appref{app:sheet-fusion-transport}.

  \item Determine the action of the completed walls on the full spectrum of genuine and non-genuine Wilson--'t Hooft lines, magnetic surface operators, and their junctions on non-spin backgrounds. This should turn the $w_2$-twisted magnetic composition law into concrete fusion and selection rules.

  \item Formulate the completed defects in the SymTFT and quiche approaches to topological symmetry \cite{Freed:2022qnc,Apruzzi:2021nmk,Kaidi:2023maf,Bhardwaj:2023kri,Bhardwaj:2023ayw,Brennan:2024fgj}. In particular, identify a five-dimensional topological symmetry theory and boundary conditions that encode the spin$^c$ tangential data, the $w_2$-twisted magnetic 1-form symmetry, and the framing completion. It would also be interesting to relate this to Putrov's discussion of $\IQ/\IZ$ symmetry and its relation to the noninvertible defects of four-dimensional QED \cite{Putrov:2022qz}, together with the proposed continuous symmetry SymTFT of \cite{Brennan:2024fgj}.

  \item Extend the framework to multiflavor QED and to the generalized spin$^c$ structures introduced in \cite[App.~C]{Moore:2024vsd}, in which the $\sfU(1)$ factor of $\mathsf{Spin}^c$ is replaced by a torus and the central quotient is generalized. A further extension is to theories defined using $\mathsf{Spin}^G$ structures \cite{Teleman:1996spg,Wang:2018qoy,Brennan:2023vsa}. Their charge lattices and gauge--gravitational index densities should lead to richer mixing between tangential and internal symmetries.
  
\end{itemize}

\appendix
\numberwithin{table}{section}
\section{Notation List}\label{app:symbol-list}

\begingroup
\footnotesize
\renewcommand*{\arraystretch}{0.96}
\setlength{\extrarowheight}{3pt}
\setlength{\LTpre}{0.5\baselineskip}
\setlength{\LTpost}{0pt}
\begin{longtable}[c]{|>{\centering\arraybackslash}m{0.20\textwidth}|m{0.52\textwidth}|>{\centering\arraybackslash}m{0.16\textwidth}|}
  \caption{\label{tbl:notation-list}Notation used in the paper.}\\
  \hline
  \textbf{Symbol(s)} & \multicolumn{1}{c|}{\textbf{Description}} & \textbf{Reference} \\
  \hline\hline
  \endfirsthead
  \caption[]{(Continued) Notation used in the paper.}\\
  \hline
  \textbf{Symbol(s)} & \multicolumn{1}{c|}{\textbf{Description}} & \textbf{Reference} \\
  \hline\hline
  \endhead
  \hline
  \multicolumn{3}{r}{\emph{Continued on next page.}}\\
  \endfoot
  \hline
  \endlastfoot
  $\star$ & Hodge star. & \eqref{eq:appendix-form-conventions} \\ \hline
  $\boxtimes$ & Decoupled stacking of TFTs. & \eqref{eq:HLS-factorization} \\ \hline
  $\otimes$ & Stacking and fusion of defects. & \eqref{eq:spinc-defect} \\ \hline
  $a$ (Coulomb) & Infrared photon connection. & \eqref{eq:coulomb-flux-quantization}, \eqref{eq:un-determinant-photon} \\ \hline
  $A$ & Real electromagnetic potential. & \autoref{subsec:fixed-spinc-sector} \\ \hline
  $A_{\mathrm{det}}$ & Determinant-line connection. & \autoref{sec:spinc} \\ \hline
  $A_{\sfU(N)}$ & Ultraviolet $\sfU(N)$ gauge potential. & \eqref{eq:un-determinant-photon} \\ \hline
  $\CA^{N,p}$ & Minimal Abelian TFT. & \appref{app:minimal-TFT} \\ \hline
  $\alpha$ (axial) & Axial rotation angle. & \eqref{eq:axial-rotation} \\ \hline
  $\alpha$ (magnetic) & Magnetic surface parameter. & \eqref{eq:Um}, \eqref{eq:twisted-identification} \\ \hline
  $\alpha_5[B;X_5]$ & Mixed magnetic--gravitational anomaly phase. & \eqref{eq:anomaly-theory-5d} \\ \hline
  $B$ & Magnetic two-form background. & \eqref{eq:anomaly-theory-5d} \\ \hline
  $b$ (gauging) & Dynamical two-form gauging field. & \eqref{eq:action-QED-coupling-notorsion} \\ \hline
  $b'$ & Discrete gauging class. & \eqref{eq:GaussSum-1-notorsion} \\ \hline
  $B^{(2)}$ & Hall-sector 1-form symmetry background. & \eqref{eq:usual-PS} \\ \hline
  $B$, $L$ (unification) & Baryon and lepton number. & \autoref{sec:dynamics} \\ \hline
  $c$ (determinant) & Determinant Chern class. & \eqref{eq:dee} \\ \hline
  $c_-(\CT)$ & Chiral central charge. & \eqref{eq:metric-framed-TFT}, \eqref{eq:full-framing-completion} \\ \hline
  $c_-^{\mathrm{Hall}}$ & Hall-sector chiral central charge. & \eqref{eq:residual-gravitational-phase} \\ \hline
  $\chi_i$ & Fermion axial charges. & \eqref{eq:general-ABJ-coefficients} \\ \hline
  $\fC_{\chi}^{\mathrm{spin}^c}$ & Proposed symmetry $3$-category. & \autoref{sec:contextandmotivation} \\ \hline
  $\mathsf{CS}_{\mathrm{grav}}$ & Gravitational Chern--Simons functional. & \eqref{eq:gravitational-CS-functional}, \eqref{eq:framing-conventions} \\ \hline
  $D$ (sheet) & External $\pi$-monodromy sheet. & \appref{app:magnetic-sheet} \\ \hline
  $\begin{gathered}\CD_\alpha[\CT_{\mathrm{Hall}},\\[-2pt]\CT_{\mathrm{comp}}]\end{gathered}$ & Decorated chiral wall. & \eqref{eq:raw-completed-defect-fusion} \\ \hline
  $\CD^{N,p}_{\mathrm{spin}^c}$ & Completed spin$^c$ chiral wall. & \eqref{eq:completed-spinc-defect} \\ \hline
  $\wh{\CD}^{N,p}_{\mathrm{spin}^c}$ & Framed spin$^c$ chiral wall. & \eqref{eq:spinc-defect} \\ \hline
  $\delta_D$ & Two-form current dual to the sheet. & \eqref{eq:distributional-sheet-curvature} \\ \hline
  $\Delta_{N,p}$ & Gravitational mismatch before neutral completion. & \eqref{eq:Delta-Np} \\ \hline
  $\delta_\gamma$ & Current dual to the boundary line. & \eqref{eq:distributional-sheet-curvature} \\ \hline
  $\delta(\CD_\alpha)$ & Residual gravitational extension class. & \eqref{eq:defect-extension-class} \\ \hline
  $e$ & Maxwell gauge coupling. & \eqref{eq:Maxwell} \\ \hline
  $\varepsilon$ & Twist of the reference spin$^c$ structure. & \autoref{subsec:summed-spinc-theory}, \eqref{eq:dee} \\ \hline
  $\eta$ & Emergent spin structure, Higgs phase. & \footrefnew{foot:higgs-topological-limit} \\ \hline
  $f$ (framing) & Label of a $2$-framing. & \eqref{eq:relative-Pontryagin-framing-shift}, \eqref{eq:framing-conventions} \\ \hline
  $F$ (electromagnetic) & Electromagnetic curvature. & \eqref{eq:spin-c-physics-v3} \\ \hline
  $f_a$ & Infrared photon curvature. & \eqref{eq:coulomb-flux-quantization}, \eqref{eq:un-determinant-photon} \\ \hline
  $F_{\mathrm{reg}}$ & Regular curvature in a sheet background. & \eqref{eq:half-flux-profile} \\ \hline
  $F_{\mathrm{sing}}$ & Distributional curvature including the sheet. & \eqref{eq:distributional-sheet-curvature} \\ \hline
  $F_{\sfU(N)}$ & Ultraviolet $\sfU(N)$ curvature. & \eqref{eq:un-determinant-photon} \\ \hline
  $G(\Sigma_2)$ & Gravitational magnetic surface factor. & \eqref{eq:Gee-sigma} \\ \hline
  $\mathsf{GCS}(\omega)$ & Local gravitational Chern--Simons form. & \eqref{eq:gravitational-CS-form}, \eqref{eq:Pontryagin-CS-conventions} \\ \hline
  $H$ (projective plane) & Positive hyperplane class on $\IC\IP^2$. & \eqref{eq:CP2-spinc-classes} \\ \hline
  $H_m(\gamma)$ & 't Hooft line of charge $m$. & \autoref{sec:fractionalization} \\ \hline
  $h(a)$ & Anyon topological spin. & \eqref{eq:ANp-spins}, \eqref{eq:spinc-vison-pair} \\ \hline
  $\wh h(a)$ & Spin combination invariant under electron attachment. & \eqref{eq:spinc-Gauss-Milgram} \\ \hline
  $H_{1/2}(\gamma;D)$ & Classical half-flux sheet prescription. & \appref{app:magnetic-sheet} \\ \hline
  $\CI_4(\fs,A)$ & Local spin$^c$ index density. & \eqref{eq:ABJ-anomalous-conservation} \\ \hline
  $\mathsf{Ind}(\slashed D_\fs)$ & Spin$^c$ Dirac index. & \eqref{eq:spinc-dirac-index}, \eqref{eq:spinc-index-integer} \\ \hline
  $\iota_3$ & Fundamental class of $K(\IZ,3)$. & \appref{app:bordism-anomaly-proof} \\ \hline
  $\CJ_A^{(3)}$ & Axial current three-form. & \eqref{eq:axial-current-3form} \\ \hline
  $j_m$ & Magnetic 1-form symmetry current. & \appref{app:spin-QED4-defect} \\ \hline
  $K$ (Hall matrix) & Abelian Hall Chern--Simons matrix. & \eqref{eq:spinc-Hall-action} \\ \hline
  $k_{\mathrm{EM}}$, $k_{\mathrm{G}}$ & ABJ coefficients. & \eqref{eq:general-ABJ-coefficients}, \eqref{eq:kEM-kG-congruence} \\ \hline
  $\CK^{\mathrm{spin}^c}_{N,p}(j;c)$ & Conjectural Ward identity kernel on $\IC\IP^2$. & \eqref{eq:conjectural-CP2-Ward-identity} \\ \hline
  $\CL_\fs$ & Determinant line bundle. & \eqref{eq:sbluorg} \\ \hline
  $\Lambda$ & Torsion-free integral flux lattice. & \eqref{eq:spin-c-moore-witten} \\ \hline
  $\lambda$ (electromagnetic gauge) & Electromagnetic gauge parameter. & \autoref{subsec:fixed-spinc-sector} \\ \hline
  $\lambda$ (magnetic gauge) & Large magnetic gauge parameter. & \autoref{sec:magnetic-gravitational-anomaly} \\ \hline
  $\mathsf{Lk}(\Sigma_2,\gamma)$ & Linking number. & \autoref{sec:fractionalization} \\ \hline
  $M_3$ & Worldvolume of a chiral wall. & \autoref{sec:contextandmotivation} \\ \hline
  $M_4$ & Spacetime four-manifold. & \autoref{sec:spinc} \\ \hline
  $N$ (color) & Rank of $\sfU(N)$. & \autoref{sec:dynamics} \\ \hline
  $N$ (denominator) & Denominator of the rational axial angle. & \autoref{sec:contextandmotivation} \\ \hline
  $N_f$ & Number of flavors. & \autoref{sec:dynamics} \\ \hline
  $N_{\Sigma_2/M_4}$ & Normal bundle of an embedded surface. & \eqref{eq:gravitational-surface-self-intersection} \\ \hline
  $\wt\Omega_5^{\mathsf{SO}}(K(\IZ,3))$ & Reduced oriented bordism group. & \propref{prop:bordism-anomaly} \\ \hline
  $\wt\Omega_5^{\mathsf{Spin}^c}(K(\IZ,3))$ & Reduced spin$^c$ bordism group. & \eqref{eq:spinc-bordism-extension} \\ \hline
  $p$ & Numerator of the rational axial angle. & \autoref{sec:contextandmotivation} \\ \hline
  $\CP$ & Pontryagin square. & \eqref{eq:usual-PS} \\ \hline
  $p_1(R)$ & First Pontryagin form. & \eqref{eq:Pontryagin-CS-conventions}, \eqref{eq:phase-ratio} \\ \hline
  $p_1(2TY_4,f)$ & Relative Pontryagin number. & \eqref{eq:gravitational-CS-functional}, \eqref{eq:relative-Pontryagin-framing-shift} \\ \hline
  $\phi$ (Higgs) & Charge-two Higgs field. & \autoref{sec:dynamics} \\ \hline
  $\psi_{\mathrm e}$ & Transparent electron line. & \appref{app:minimal-spinc-Hall-TFT} \\ \hline
  $Q$ & Electric charge in electron units. & \eqref{eq:spinc-Hall-local-particle}, \eqref{eq:spinc-vison-pair} \\ \hline
  $q$ (sheet profile) & Sheet endpoint flux. & \appref{app:magnetic-sheet} \\ \hline
  $q_i$ & Fermion electric charges. & \eqref{eq:general-ABJ-coefficients} \\ \hline
  $Q_e$ & Physical electric charge. & \eqref{eq:dyon-charges} \\ \hline
  $Q_m$ & Physical magnetic charge. & \eqref{eq:dyon-charges} \\ \hline
  $q_N$ & Quadratic function for discrete magnetic gauging. & \eqref{eq:spin-gauging-quadratic-function} \\ \hline
  $\mathsf{QED}_\theta$ & QED with gauge--gravitational theta coupling. & \autoref{sec:chiral-rotation-defect} \\ \hline
  $\rho_N$ & Reduction modulo $N$. & \autoref{sec:coupling-tft-to-spinc-qed4} \\ \hline
  $s$ (vison order) & Vison order modulo local particles. & \appref{app:minimal-spinc-Hall-TFT} \\ \hline
  $\mathsf{Spin}^c_{\mathrm b}$, $\mathsf{Spin}^c_{\mathrm f}$ & Spin$^c$ Maxwell variants. & \autoref{tbl:u1-variants} \\ \hline
  $\fs$ & Spin$^c$ structure. & \autoref{sec:spinc} \\ \hline
  $\fs_0$ & Reference spin$^c$ structure. & \autoref{subsec:summed-spinc-theory} \\ \hline
  $\Sigma$ (Wilson surface) & Support of a Wilson surface. & \remref{rem:wilson-dressing} \\ \hline
  $\Sigma_2$ & Closed 2-cycle for measuring flux. & \eqref{eq:spin-c-physics-v3} \\ \hline
  $\sigma_H$ & Hall conductivity. & \eqref{eq:spinc-Hall-response}, \eqref{eq:spinc-Hall-K-condition} \\ \hline
  $\sigma(M_4)$ & Four-manifold signature. & \eqref{eq:HirzebruchSignature}, \eqref{eq:phase-ratio} \\ \hline
  $t$ (Hall charge) & Hall charge vector. & \eqref{eq:spinc-Hall-K-condition}, \eqref{eq:spinc-Hall-characteristic-vector} \\ \hline
  $\CT_{a/b}$ & Neutral WRT TFT, $c_-=a/b$. & \eqref{eq:explicit-WRT-completion} \\ \hline
  $\CT_{\mathrm{comp}}^{N,p}$ & Neutral gravitational completion TFT. & \eqref{eq:completed-spinc-defect}, \eqref{eq:full-framing-completion} \\ \hline
  $\CT_{\mathrm{Hall}}^{N,p}$ & Chosen complete spin$^c$ Hall TFT. & \eqref{eq:spinc-defect} \\ \hline
  $\tau$ & Complex Maxwell coupling. & \autoref{sec:u1-variants} \\ \hline
  $\theta$ (gauge--gravitational) & Gauge--gravitational theta angle. & \eqref{eq:spinc-full-theta-factor} \\ \hline
  $\theta$ (ordinary electromagnetic) & Electromagnetic theta angle for integral flux. & \eqref{eq:Maxwell}, \appref{app:theta-periodicity} \\ \hline
  $\theta$ (spin$^c$, electromagnetic) & Electromagnetic theta angle for spin$^c$ flux. & \eqref{eq:spinc-em-theta-shift}, \appref{app:theta-periodicity} \\ \hline
  $\vartheta$ & Axion-like compact scalar. & \eqref{eq:kEM-kG-congruence} \\ \hline
  $\sfU(1)_{\mathrm b}$, $\sfU(1)_{\mathrm f}$ & Ordinary Maxwell variants. & \autoref{tbl:u1-variants} \\ \hline
  $\sfU(1)_m^{(1)}$ & Magnetic quotient group. & \eqref{eq:extension-sequence} \\ \hline
  $\wt{\sfU(1)}_m^{(1)}$ & Covering magnetic parameter group. & \eqref{eq:extension-sequence} \\ \hline
  $U_m(\alpha;\Sigma_2)$ & Magnetic symmetry surface. & \eqref{eq:Um}, \eqref{eq:twisted-identification} \\ \hline
  $U_m^{\mathrm{reg}}(2\pi;\Sigma_2\!\mid\!D)$ & Surface factor measuring regular flux. & \eqref{eq:Um-with-sheet} \\ \hline
  $V$ (Higgs vortex) & Higgs vortex sheet. & \eqref{eq:Higgs-vortex-twist} \\ \hline
  $v$ (vison) & Vison generated by background electromagnetic flux. & \eqref{eq:spinc-vison-pair}, \eqref{eq:VNp-def} \\ \hline
  $\CV^{N,p}$ & Minimal spin$^c$ Hall TFT, odd $N$. & \eqref{eq:VNp-def} \\ \hline
  $\CV^{2N,2p}$ & Incomplete vison sector for even denominators. & \appref{app:minimal-spinc-Hall-TFT} \\ \hline
  $W^nT^m$ & Wilson--'t Hooft line. & \eqref{eq:dyon-charges} \\ \hline
  $\ov w_2(TM_4)$ & Chosen integral lift of $w_2(TM_4)$. & \autoref{sec:spinc}, \eqref{eq:dee} \\ \hline
  $w_2(TM_4)$ & Second Stiefel--Whitney class. & \eqref{eq:spin-c-math} \\ \hline
  $w_3$ & Third Stiefel--Whitney class. & \eqref{eq:anomaly-Wu-manifold}, \autoref{sec:dynamics} \\ \hline
  $W^q(\gamma;\Sigma)$ & Charge-$q$ Wilson surface. & \eqref{eq:wilson-dressing-surface-change} \\ \hline
  $W_3$ & Third integral Stiefel--Whitney class. & \propref{prop:bordism-anomaly} \\ \hline
  $W_5$ & Wu manifold $\mathsf{SU}(3)/\mathsf{SO}(3)$. & \propref{prop:bordism-anomaly} \\ \hline
  $X_4$ & Glued closed four-manifold. & \eqref{eq:phase-ratio}, \eqref{eq:residual-gravitational-phase} \\ \hline
  $y$ & Real electromagnetic flux class. & \autoref{sec:coupling-tft-to-spinc-qed4}, \eqref{eq:dee} \\ \hline
  $Y_4$ & Four-dimensional extension (filling) of the wall. & \eqref{eq:usual-PS}, \footrefnew{foot:spin-wall-filling} \\ \hline
  $\IZ_2^{\mathrm{grav}}$ & Tangential kernel of the magnetic extension. & \eqref{eq:extension-sequence} \\ \hline
  $\IZ_N^{(1)}$ & Finite magnetic 1-form symmetry subgroup. & \eqref{eq:defect-schematic} \\ \hline
  $\sfZ_{\mathrm{anom}}[B^{(2)}]$ & Discrete 1-form anomaly inflow factor. & \eqref{eq:usual-PS} \\ \hline
  $\sfZ_{M_4}$ & Summed QED partition function. & \eqref{eq:full-spinc-partition} \\ \hline
  $\sfZ_{M_4}^{\fs}$ & Fixed-sector partition function. & \eqref{eq:fixed-spinc-partition} \\ \hline
  $\sfZ^{\mathrm{spin}^c}_{p,N}(M_4;c)$ & Spin$^c$ electromagnetic anomaly inflow factor. & \eqref{eq:spin-c-refined-PS} \\ \hline
  $\sfZ_{\CT}^{\mathrm{fr}}$ & Framed TFT partition function. & \eqref{eq:framing-conventions}, \eqref{eq:metric-framed-TFT} \\ \hline
  $\sfZ_{\CT}^{\mathrm{met}}$ & Metric-dependent TFT partition function. & \eqref{eq:metric-framed-TFT} \\ \hline
\end{longtable}
\endgroup

\section{Differential Form And \texorpdfstring{QED$_4$}{QED4} Conventions}\label{app:QEDconventions}
All formulas in the body of the paper use Euclidean signature. Let $(M_4,g,\fo)$ be an oriented Riemannian four-manifold, and choose positively oriented local coordinates $x^\mu$. We write
\be
  \vol(g)=\sqrt{g}\,d^4x ~, \qquad \varepsilon^{1234}=+1 ~,
\ee
where $\varepsilon^{\mu\nu\rho\sigma}$ is the metric-independent alternating symbol. A $p$-form is expanded as
\be
  \alpha=\frac{1}{p!}\alpha_{\mu_1\cdots\mu_p}
  dx^{\mu_1}\wedge\cdots\wedge dx^{\mu_p} ~,
  \qquad
  (d\alpha)_{\mu_0\cdots\mu_p}
  =(p+1)\nabla_{[\mu_0}\alpha_{\mu_1\cdots\mu_p]} ~,
\ee
with antisymmetrization normalized by, for example, $T_{[\mu\nu]}=\frac{1}{2}(T_{\mu\nu}-T_{\nu\mu})$. These satisfy
\be
  \alpha\wedge\beta=(-1)^{pq}\beta\wedge\alpha ~,
  \qquad
  d(\alpha\wedge\beta)=d\alpha\wedge\beta+(-1)^p\alpha\wedge d\beta ~,
  \qquad
  \int_{M_4}d\eta=\int_{\partial M_4}\eta ~,
\ee
for $\alpha\in\Omega^p(M_4)$, $\beta\in\Omega^q(M_4)$, and $\eta\in\Omega^3(M_4)$. The Hodge star is defined by
\be
  \alpha\wedge\star\beta
  =\frac{1}{p!}\alpha_{\mu_1\cdots\mu_p}\beta^{\mu_1\cdots\mu_p}\vol(g) ~,
  \qquad
  \star^2\alpha=(-1)^{p(4-p)}\alpha ~,
\ee
for $\alpha,\beta\in\Omega^p(M_4)$. The identities used repeatedly in the text are
\be\label{eq:appendix-form-conventions}
  F\wedge\star F
  =\frac{1}{2}F_{\mu\nu}F^{\mu\nu}\sqrt{g}\,d^4x ~,
  \qquad
  F\wedge F
  =\frac{1}{4}\varepsilon^{\mu\nu\rho\sigma}F_{\mu\nu}F_{\rho\sigma}\,d^4x ~,
  \qquad
  d\star j=(\nabla_\mu j^\mu)\vol(g) ~.
\ee
We use real gauge potentials. Thus $A=A_\mu dx^\mu$ has $A_\mu\in\IR$, the corresponding anti-Hermitian connection form is $\imag A$, and the real fieldstrength used throughout the paper is
\be
  F:=dA ~,
  \qquad
  F_{\mu\nu}=\partial_\mu A_\nu-\partial_\nu A_\mu ~.
\ee
The Euclidean QED action is therefore
\begin{align}
  S_{\mathrm{QED}}
  &=\int_{M_4}\left(
    \frac{1}{2e^2}F\wedge\star F
    -\frac{\imag\theta}{8\pi^2}F\wedge F
    +\imag\ov\Psi\slashed D_A\Psi\,\vol(g)
  \right) \nonumber\\
  &=\int_{M_4}d^4x\left(
    \frac{\sqrt g}{4e^2}F_{\mu\nu}F^{\mu\nu}
    -\frac{\imag\theta}{32\pi^2}\varepsilon^{\mu\nu\rho\sigma}F_{\mu\nu}F_{\rho\sigma}
    +\imag\sqrt g\,\ov\Psi\gamma^\mu D_\mu\Psi
  \right) ~.
\end{align}
Here $D_\mu=\nabla_\mu+\imag A_\mu$ and $\slashed D_A=\gamma^\mu D_\mu$, with $\{\gamma^\mu,\gamma^\nu\}=2g^{\mu\nu}$. We take the Euclidean gamma matrices to be Hermitian. On a closed manifold, $\slashed D_A$ is then formally anti-Hermitian and $\imag\slashed D_A$ is Hermitian. In an oriented orthonormal frame we take $\gamma^5=\gamma^1\gamma^2\gamma^3\gamma^4$, so $(\gamma^5)^2=1$ and the chiral projectors are $P_\pm=\frac{1}{2}(1\pm\gamma^5)$. In the Euclidean functional integral, $\Psi$ and $\ov\Psi$ are treated as independent Grassmann fields. Under $A\mapsto A+d\lambda$, $\Psi\mapsto e^{-\imag\lambda}\Psi$ and $\ov\Psi\mapsto\ov\Psi e^{\imag\lambda}$. Accordingly, the Wilson line associated with this fermion is $\exp(-\imag\oint A)$. Choosing its inverse as the electric generator simply relabels $n\mapsto-n$.

As reviewed in \autoref{sec:spinc}, on a spin four-manifold the bundles $S_\pm$ exist globally: a neutral Dirac spinor is a section of $S_+\oplus S_-$, while a unit-charge QED spinor coupled to an electromagnetic line bundle $L$ is a section of $(S_+\oplus S_-)\otimes L$. On a non-spin four-manifold, only the spin$^c$ bundles $\Sigma_\pm=S_\pm\otimes\CW$ are globally defined. The spin$^c$ QED field is a section of $\Sigma_+\oplus\Sigma_-$, with $\imag A$ already incorporated as the local $\fu(1)$ component of the spin$^c$ connection.

\section{Theta Angle Periodicities}\label{app:theta-periodicity}

The integrality of the electromagnetic and gravitational topological densities on spin$^c$ manifolds, and the resulting pair of $2\pi$-periodic theta angles, are worked out in \cite[Apps.~A.1 and A.2]{Seiberg:2016rsg}. For the normalization used in this paper, let $y=[F/(2\pi)]$ and isolate the electromagnetic topological factor
\be\label{eq:em-theta-factor}
 \sfZ_{\mathrm{em}}(\theta)
 =\exp\left(\frac{\imag\theta}{8\pi^2}\int_{M_4}F\wedge F\right)
 =\exp\left(\frac{\imag\theta}{2}\int_{M_4}y\smile y\right) ~.
\ee
For an ordinary $\sfU(1)$ connection, $y\in H^2(M_4;\IZ)$. If $M_4$ is spin, its intersection form is even, so $\frac{1}{2}\int y^2\in\IZ$ and \eqref{eq:em-theta-factor} is $2\pi$-periodic. On a general oriented four-manifold, $\int y^2$ can be odd, so the universal period is $4\pi$. Equivalently, $T$ exchanges $\sfU(1)_{\text{b}}$ and $\sfU(1)_{\text{f}}$ (see \autoref{sec:u1-variants}). A particular non-spin manifold whose free intersection form is even can still have period $2\pi$.

For a spin$^c$ gauge potential, $y=c/2$, where $c=c_1(\CL_\fs)$ is characteristic. The electromagnetic instanton number is therefore
\be\label{eq:spinc-instanton-number}
 \frac{1}{8\pi^2}\int_{M_4}F\wedge F=\frac{c^2}{8} ~.
\ee
The characteristic property gives $c^2\equiv\sigma(M_4)\smod 8$. Consequently,
\be\label{eq:spinc-em-theta-shift}
 \sfZ_{\mathrm{em}}(\theta+2\pi)
 =\sfZ_{\mathrm{em}}(\theta)
 \exp\left(\frac{\imag\pi}{4}\sigma(M_4)\right) ~.
\ee
Unlike the ordinary case, this phase is the same in every spin$^c$ sector, so it is removed by a shift of the gravitational theta term. This is the statement that $T$ fixes $\mathsf{Spin}^{c}_{\text{b}}$. The natural spin$^c$ theta coupling includes the full Dirac index,
\be\label{eq:spinc-full-theta-factor}
 \sfZ_{\mathrm{gauge\text{-}grav}}(\theta)
 =\exp\left[\imag\theta\,\frac{c^2-\sigma(M_4)}{8}\right] ~,
\ee
which is exactly $2\pi$-periodic because $(c^2-\sigma(M_4))/8\in\IZ$ (see \cite[App.~A.2]{Seiberg:2016rsg}). Equivalently, a $2\pi$ shift of the electromagnetic term must be accompanied by the corresponding shift of the gravitational counterterm. On a spin manifold, $\rho_2(c)=w_2(TM_4)=0$, so $c$ is an even integral class, and the intersection form is even. Hence $c^2\in8\IZ$, and the isolated electromagnetic factor is already $2\pi$-periodic.

\section{Bordism Computation For \texorpdfstring{\propref{prop:bordism-anomaly}}{the Bordism Proposition}}\label{app:bordism-anomaly-proof}

This appendix derives \propref{prop:bordism-anomaly} and the torsion-free spin$^c$ group used in \footrefnew{foot:fixed-spinc-trivial}. See also \cite[App.~B.1]{DavighiLohitsiriDebray2023}. Deformation classes of reflection-positive invertible field theories are classified by the Anderson dual of bordism \cite{Freed:2016rqq}, extended to include background maps. When the degree-six group has no free part, as below, they are the characters of the torsion subgroup of the degree-five group. The background is a map $f:X_5\to K(\IZ,3)$ with $f^*\iota_3=[dB/2\pi]$, where $\iota_3\in H^3(K(\IZ,3);\IZ)$ is the fundamental class. Reduced bordism discards the purely gravitational theories.

\medskip
\medskip
\noindent \emph{Homology of $K(\IZ,3)$.}

Let $\wt H_n:=\wt H_n(K(\IZ,3);\IZ)$. Since $K(\IZ,3)$ is $2$-connected with $\pi_3=\IZ$ and $\pi_4=0$, the Hurewicz theorem gives $\wt H_n=0$ for $n\le2$, $\wt H_3\cong\IZ$ and $\wt H_4=0$. The integral cohomology in degrees $5,6,7$ is $0,\IZ_2,0$ \cite{CartanSeminar1955}, so the universal coefficient theorem gives $\wt H_5\cong\IZ_2$ and $\wt H_6=0$. The mod-$2$ cohomology $H^5(K(\IZ,3);\IZ_2)\cong\IZ_2$ is generated by $\ssq^2\rho_2\iota_3$ \cite{Serre1953}, which therefore detects $\wt H_5$.

\medskip
\medskip
\noindent \emph{Oriented Bordism.}

\begin{figure}[h]
\centering
\includegraphics[width=\textwidth]{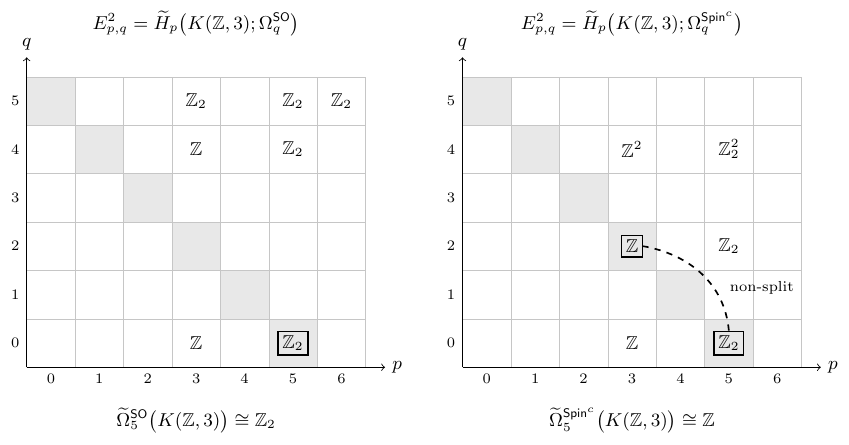}
\caption{\label{fig:AHSS-E2-pages}The $E^2$ pages of the reduced Atiyah--Hirzebruch spectral sequences for oriented (left) and spin$^c$ (right) bordism of $K(\IZ,3)$. Blank entries vanish, and the diagonal $p+q=5$ is shaded. No differential can be nonzero on the boxed entries. In the oriented case, the boxed $\IZ_2$ is the anomaly group. In the spin$^c$ case, the two boxed entries form a non-split extension, which gives a torsion-free group.}
\end{figure}

The reduced Atiyah--Hirzebruch spectral sequence has $E^2_{p,q}=\wt H_p(K(\IZ,3);\Omega_q^{\mathsf{SO}})$. Since $\Omega_1^{\mathsf{SO}}=\Omega_2^{\mathsf{SO}}=\Omega_3^{\mathsf{SO}}=0$ \cite{Stong1968} and $\wt H_p=0$ for $p\le2$, the only nonzero entry with $p+q=5$ is $E^2_{5,0}\cong\IZ_2$ (see the left panel of \figref{fig:AHSS-E2-pages}).
No differential enters $E_{5,0}$, and those leaving it land on $E_{3,1}$, $E_{2,2}$, $E_{1,3}$, and $E_{0,4}$, which all vanish.
Hence
\be
  \wt\Omega_5^{\mathsf{SO}}\big(K(\IZ,3)\big)\cong\IZ_2 ~,
\ee
with no extension problem. Rationally $K(\IZ,3)\simeq_{\IQ}S^3$, so $\wt\Omega_6^{\mathsf{SO}}(K(\IZ,3))\otimes\IQ=0$, and the Pontryagin dual of $\IZ_2$ is the full anomaly group.

The edge map sends $[X_5,f]$ to $f_*[X_5]\in\wt H_5$, which is detected by pairing with $\ssq^2\rho_2\iota_3$. By the Wu formula, $\langle\ssq^2 \rho_2f^*\iota_3,[X_5]\rangle=\langle w_2\smile \rho_2f^*\iota_3,[X_5]\rangle$, so $\alpha_5$ is $(-1)$ raised to this mod-$2$ pairing. On $(W_5,W_3)$, $\rho_2f^*\iota_3=\ssq^1w_2=w_3$, and $\langle w_2w_3,[W_5]\rangle=1$ by the proof of \eqref{eq:anomaly-Wu-manifold}. Thus $[W_5,W_3]-[W_5,0]$ generates $\wt\Omega_5^{\mathsf{SO}}(K(\IZ,3))$, and $\alpha_5$ is the nontrivial character. Adding $\Omega_5^{\mathsf{SO}}\cong\IZ_2$, generated by $W_5$ and detected by $w_2w_3$ \cite{Wall1960}, gives the unreduced $\Omega_5^{\mathsf{SO}}(K(\IZ,3))\cong\IZ_2\oplus\IZ_2$, in agreement with \cite{DavighiLohitsiriDebray2023,JiaWangZhang2026}.

That spin$^c$ QED$_4$ realizes the nontrivial character follows from its large gauge variation in \autoref{sec:magnetic-gravitational-anomaly}, whose mapping torus $\IC\IP^2\times S^1$ has $\alpha_5=-1$ (see the discussion following \propref{prop:bordism-anomaly}).

\medskip
\medskip
\noindent \emph{Spin$^c$ Bordism.}

A fixed spin$^c$ sector is governed by $\wt\Omega_5^{\mathsf{Spin}^c}(K(\IZ,3))$. In degrees $q=0,\dots,4$, $\Omega_q^{\mathsf{Spin}^c}$ is $\IZ,0,\IZ,0,\IZ\oplus\IZ$ \cite{Stong1968}. All these groups are free, so $E^2_{p,q}=\wt H_p\otimes\Omega_q^{\mathsf{Spin}^c}$, and the nonzero entries with $p+q=5$ are (see the right panel of \figref{fig:AHSS-E2-pages})
\be
  E^2_{5,0}\cong\wt H_5\cong\IZ_2 ~,
  \qquad
  E^2_{3,2}\cong\wt H_3\otimes\Omega_2^{\mathsf{Spin}^c}\cong\IZ ~.
\ee
Every differential entering or leaving these entries starts or ends in an odd row, in a column $p\le2$, or at $E_{6,0}=\wt H_6=0$, so it vanishes. The spectral sequence therefore gives an extension
\be\label{eq:spinc-bordism-extension}
  0\to\IZ\to\wt\Omega_5^{\mathsf{Spin}^c}\big(K(\IZ,3)\big)\to\IZ_2\to0 ~.
\ee
The subgroup is generated by $S^3\times S^2$, with $f^*\iota_3$ the generator of $H^3(S^3;\IZ)$ and the spin$^c$ structure on $S^2$ with $\int_{S^2}c_1(\fs)=2$, which generates $\Omega_2^{\mathsf{Spin}^c}$. The quotient map is the edge map, detected by pairing with $\ssq^2\rho_2\iota_3$.

To resolve the extension, consider the characteristic number
\be
  n[X_5,\fs,f]=\int_{X_5}c_1(\fs)\smile f^*\iota_3\in\IZ ~,
\ee
which is a homomorphism on the reduced group. It equals $2$ on the generator of the subgroup. By the Wu formula and $w_2=\rho_2(c_1(\fs))$, its reduction modulo $2$ is the pairing with $\ssq^2\rho_2\iota_3$, so $n$ is odd exactly off the subgroup. If $n(x)=0$, then $x$ lies in the subgroup, on which $n$ is injective, so $x=0$. On $\IC\IP^2\times S^1$ with $c_1(\fs)=H$ and $f^*\iota_3=H\smile u$, $n=1$. Hence $n$ is an isomorphism,
\be
  \wt\Omega_5^{\mathsf{Spin}^c}\big(K(\IZ,3)\big)\cong\IZ ~,
\ee
and \eqref{eq:spinc-bordism-extension} does not split. The group has no torsion, so the restriction $(-1)^n$ of $\alpha_5$ to spin$^c$ backgrounds (see \remref{rem:anomaly-vs-extension}) lies in the family $\exp(\imag\theta n)$ and is deformation-trivial.

\section{Review Of The Minimal TFTs \texorpdfstring{$\CA^{N,p}$}{A(N,p)} And \texorpdfstring{$\CV^{N,p}$}{V(N,p)}}\label{app:minimal-TFT}

In this appendix, we briefly review the minimal 3d TFTs entering the construction of noninvertible chiral symmetry defects: the minimal Abelian TFT $\CA^{N,p}$ of \cite{Hsin:2018vcg}, which is stacked on the defect in spin QED$_4$ (see \appref{app:spin-QED4-defect}), and its spin$^c$/electronic counterpart $\CV^{N,p}$ of \cite{Cheng:2025ube}, which provides a minimal charged Hall TFT for odd $N$ in spin$^c$ QED$_4$. Even $N$ requires a larger complete Hall theory (see \autoref{sec:coupling-tft-to-spinc-qed4}).

\subsection{The Minimal Abelian TFT \texorpdfstring{$\CA^{N,p}$}{A(N,p)}}

Consider a 3d TFT with a $\IZ_N$ 1-form symmetry generated by an Abelian anyon $a$, with fusion rules $a^{s} \times a^{s'} = a^{[s+s']_N}$ (where $[s+s']_N$ denotes addition modulo $N$) and $a^{N} = 1$. Consistency of fusion and braiding fixes the topological spins of the symmetry lines in terms of a single integer $p$ defined modulo $2N$ \cite{Hsin:2018vcg}:
\be\label{eq:ANp-spins}
  h[a^s] = \frac{p s^2}{2N} \smod 1 ~.
\ee
The condition $a^{N} = 1$ requires $h[a^N] = \frac{pN}{2} \equiv 0 \smod 1$, i.e., $pN \in 2\IZ$, in a bosonic TFT. The positive additive monodromy determined by the ribbon identity is
\be
  M(a^{s}, a^{s'}) = h[a^{s+s'}] - h[a^{s}] - h[a^{s'}] = \frac{p\, s s'}{N} \smod 1 ~.
\ee
Following the orientation convention of \cite{Hsin:2018vcg}, the symmetry action and modular $S$-matrix use the oppositely oriented braid, whose additive phase is $-M$. Thus the generating line carries $\IZ_N$ charge $q(a)=-p\smod N$. Since a 1-form symmetry can be gauged only if its symmetry lines are neutral, $p$ is the 't Hooft anomaly coefficient of the $\IZ_N$ 1-form symmetry. The associated 4d anomaly inflow action is the Pontryagin square term \eqref{eq:usual-PS}.

When $\gcd(N,p) = 1$ and $pN \in 2\IZ$, the $N$ symmetry lines by themselves form a consistent bosonic TFT: the \emph{minimal Abelian TFT} $\CA^{N,p}$, with modular $S$-matrix
\be
  S_{ss'} = \frac{1}{\sqrt{N}}\exp\left(-\frac{2\pi\imag\, p\, s s'}{N}\right) ~,
\ee
which is unitary precisely when $\gcd(N,p) = 1$. Its chiral central charge modulo 8 is computed by the Gauss sum $e^{2\pi\imag c/8} = \frac{1}{\sqrt{N}}\sum_{s=1}^{N}e^{2\pi\imag\, h[a^s]}$ and admits a closed-form expression in terms of Jacobi symbols \cite[§2.2]{Hsin:2018vcg}. The Gauss sum fixes the chiral central charge modulo $8$ and hence the signature-dependent phase under a change of four-dimensional extension. A choice of central charge lift and gravitational Chern--Simons response is additionally required to fix the full framing anomaly of $\CA^{N,p}$, as explained in \eqref{eq:full-framing-completion}. This chiral central charge is distinct from the 1-form anomaly coefficient $p$ and need not equal the Hall coefficient $p/N$ used in the defect coupling. Familiar realizations include $\CA^{N,1} \leftrightarrow \sfU(1)_N$ and $\CA^{N,N-1} \leftrightarrow \mathsf{SU}(N)_1$. Relabeling the generator $a \mapsto a^{r}$ with $\gcd(r,N) = 1$ yields the identification $\CA^{N,p r^2} \cong \CA^{N,p}$.

If $pN \in 2\IZ + 1$, the theory is instead a spin TFT (i.e., it relies on a choice of spin structure). It contains a transparent fermion line $\psi$ with topological spin $\frac{1}{2}$ and a $\IZ_2$ 1-form symmetry generated by $\psi$. This symmetry does not act faithfully on the remaining lines and therefore does \emph{not} extend the 1-form symmetry from $\IZ_N$ to $\IZ_{2N}$. In summary,
\be
 \CA^{N,p} = \left\{
  \begin{array}{ll}
    \text{a bosonic TFT} \quad & \text{if } pN \in 2\IZ \\
    \CA^{N,p+N} \boxtimes \{1,\psi\} = \text{a spin TFT} \quad & \text{if } pN \in 2\IZ+1
  \end{array}
 \right.
\ee
where $\{1,\psi\}$ is the almost trivial invertible spin TFT generated by the transparent fermion line $\psi$. For example, for odd $N$, the spin TFT $\sfU(1)_N$ factorizes as $\sfU(1)_N \leftrightarrow \CA^{N,N+1}\boxtimes\{1,\psi\}$, reproducing the level-rank duality $\sfU(1)_N \leftrightarrow \mathsf{SU}(N)_{-1}$, valid as spin TFTs \cite{Hsin:2018vcg}.

In the spin case, the line data determine the chiral central charge only modulo $1/2$ \cite{Kitaev:2005hzj}, because stacking a neutral invertible $p+\imag p$ theory shifts $c_-$ by $1/2$ without changing those lines. When $\CA^{N,p}$ is stacked on the spin QED wall, its framing anomaly combines with the gravitational part of the axial Jacobian through the difference $c_-(\CA^{N,p})-p/N$, as in \eqref{eq:phase-ratio}.

The central significance of $\CA^{N,p}$ is the factorization theorem of \cite{Hsin:2018vcg}: any 3d TFT $\CT$ with a $\IZ_N$ 1-form symmetry of anomaly coefficient $p$ such that $\gcd(N,p) = 1$ factorizes as
\be\label{eq:HLS-factorization}
  \CT = \CA^{N,p} \boxtimes \CT' ~,
\ee
where $\CT'$ is neutral under the symmetry. In this sense, $\CA^{N,p}$ is the unique minimal carrier of a $\IZ_N$ 1-form symmetry with anomaly $p$: it saturates the anomaly with no superfluous degrees of freedom, which is why it is the canonical choice of TFT in the noninvertible defect \eqref{eq:defect-schematic}.

\subsection{The Minimal \texorpdfstring{Spin$^c$}{Spin-c} Hall TFT \texorpdfstring{$\CV^{N,p}$}{V(N,p)}}\label{app:minimal-spinc-Hall-TFT}

We now review the electronic/spin$^c$ counterpart, following \cite{Cheng:2025ube}, whose $V^{s,r}$ has actual vison order $s$ and Hall coefficient $r/s$. The minimal odd-denominator choice has $(s,r)=(N,p)$. In general, $s=\ell N$, $r=\ell p$.

The setting is a gapped $2{+}1$d system with a global $\sfU(1)$ symmetry -- e.g., the topological subsector of a fractional quantum Hall state -- described by a spin TFT containing a transparent fermion $\psi_{\mathrm e}$, the electron, with unit $\sfU(1)$ charge and topological spin $\frac{1}{2}$, subject to the spin-charge relation: local operators of odd (even) charge are fermions (bosons). Electron attachment makes the joint identification $(Q,h)\sim(Q+n,h+n/2)$, $n\in\IZ$, with spin also taken modulo one. Reducing the two quantities independently loses the spin-charge correlation.

A fractionally quantized Hall conductivity $\sigma_H$ implies the existence of a distinguished Abelian anyon: threading $2\pi$ units of background $\sfU(1)$ flux nucleates the \emph{vison} $v$, with
\be
  Q(v) = \sigma_H \smod 1 ~, \qquad h(v) = \frac{\sigma_H}{2} \smod \frac{1}{2} ~,
\ee
and the positive monodromy with $v$ measures the fractional $\sfU(1)$ charge of every anyon. If we additionally choose vison order $N$ and fix the exact response $\sigma_H = \frac{p}{N}$, the response theory defines a $\IZ_N^{(1)}$ 1-form symmetry with anomaly coefficient $p$, generated by the vison lines
\be\label{eq:VNp-def}
  \CV^{N,p} := \left\{1, v, v^2, \ldots, v^{N-1}\right\} ~, \qquad Q(v^j) = \frac{p j}{N} \smod 1 ~, \quad h(v^j) = \frac{p j^2}{2N} \smod \frac{1}{2} ~,
\ee
with the correlated electron identification of \eqref{eq:spinc-vison-pair}. The identification $\CV^{N,p}\cong\CV^{N,p+2N}$ concerns the line sector. It does not identify fixed Hall responses, which differ by $2$, or specify their gravitational response. As a set of anyons, $\CV^{N,p}$ consists of the lines of $\CA^{N,p}$ (together with $\psi_{\mathrm e}$), refined by a $\sfU(1)$ charge assignment. The assignment in \eqref{eq:VNp-def} is the one for which the symmetry lines are generated by the vison itself, corresponding to the generator of $H^{2}(B\sfU(1);\IZ_N) \cong \IZ_N$, the group of possible charge classes \cite{Cheng:2025ube}.

The spin-charge relation imposes a consistency condition absent in the bosonic case: $v^{N}$ is a local particle of integer charge $p$, so its topological spin must satisfy $h(v^N) = \frac{pN}{2} \equiv \frac{p}{2} \smod 1$. Subtracting the two sides gives
\be\label{eq:spin-charge-consistency}
  p(N-1) \in 2\IZ
  \qquad\Longleftrightarrow\qquad
  p(N+1) \in 2\IZ ~,
\ee
because the two expressions differ by the even integer $2p$. We retain the $p(N+1)$ form when comparing with \cite{Cheng:2025ube}. Either form replaces the bosonic condition $pN \in 2\IZ$. For odd $N$, the condition \eqref{eq:spin-charge-consistency} always holds: $\CV^{N,p}$ exists as a minimal Abelian spin$^c$ TFT, and for $p = 1$ it is the topological order of the Laughlin state at filling $\nu = \frac{1}{N}$, i.e., the spin TFT $\sfU(1)_N$. For even $N$ -- whence $p$ is odd by coprimality -- an order-$N$ vison would violate locality. The actual order is $s=\ell N$ with even $\ell$, since $r(s-1)=\ell p(\ell N-1)$ must be even. The smallest possible order gives the consistent but incomplete sector $\CV^{2N,2p}$. The Hall coefficient alone does not impose this smallest choice. Complete minimal theories containing this sector can be Abelian or non-Abelian. Pfaffian/Moore--Read-type theories are distinguished non-Abelian examples \cite{Cheng:2025ube}.

A framing anomaly is defined for a complete spin$^c$ TFT. For odd $N$, $\CV^{N,p}$ is itself complete and its gravitational Chern--Simons response is controlled by its chiral central charge $c_-$. The spin$^c$ Gauss--Milgram relation \eqref{eq:spinc-Gauss-Milgram} fixes the combination $c_- - \sigma_H$ modulo $8$ from the refined topological spins. For even $N$, the order-$N$ construction is inconsistent with locality, while the smallest consistent sector $\CV^{2N,2p}$ is incomplete. Its framing anomaly is specified only after choosing a complete Hall theory, and different completions can have different chiral central charges.

Finally, the analog of the factorization \eqref{eq:HLS-factorization} continues to hold for odd $N$ \cite{Cheng:2025ube}: a spin$^c$ TFT $\CT$ whose vison generates $\CV^{N,p}$ with $\gcd(N,p) = 1$ factorizes as $\CT = \CV^{N,p} \boxtimes \CT'$, where all anyons of $\CT'$ may be taken $\sfU(1)$-neutral. For even $N$, the Hall TFT is instead a chosen complete theory, and the minimal-vison-order choice contains $\CV^{2N,2p}$. For odd $N$, the factorization singles out $\CV^{N,p}$ as a natural minimal \emph{gauge sector} choice in \eqref{eq:spinc-defect}: its response theory reproduces the spin$^c$ inflow action \eqref{eq:spin-c-refined-PS} with $\sigma_H = \frac{p}{N}$. The gauge sector factorization alone does not cancel the residual framing anomaly, but its rational value can always be canceled by the neutral WRT completion \eqref{eq:explicit-WRT-completion}.

\section{The Spin QED Defect From A Finite Gauss Sum}\label{app:spin-QED4-defect}

Starting from the 1-form symmetry gauging construction of $\wh{\CD}_{p/N}$ in \cite{Choi:2022jqy}, we evaluate the gauging path integral as a finite sum over cohomology classes. For $\gcd(p,N)=1$ and spin QED$_4$ on a manifold with torsion-free (co)homology, the even intersection form defines a quadratic function $q_N$ on the finite group $H^2(M_4;\IZ_N)$. After the continuum fields imposing the discrete gauging are integrated out, the remaining path integral is the Gauss sum \eqref{eq:GaussSum-1-notorsion} of $q_N$ coupled linearly to the electromagnetic flux class. Completing the square separates this Gauss sum into a flux-dependent prefactor, which produces the theta angle shift, and a flux-independent normalization. The gravitational part of the anomaly, which leaves a signature-dependent framing anomaly on a curved spin background, is treated in \remref{rem:spin-versus-spinc-completion}. Relaxing the torsion-free assumption requires a differential cohomology refinement.

We begin with the action of spin QED$_4$. Here $A$ is the real gauge potential representative of a connection on a globally defined principal $\sfU(1)$ bundle. The theory has a magnetic 1-form symmetry $\sfU(1)^{(1)}$ with conserved current $j_{m} = F/2\pi$.
\tightfootnote{We represent conserved currents by closed differential forms, so the conservation law is expressed using the exterior derivative.}
In order to gauge a discrete subgroup $\IZ_{N}^{(1)}$ of the $\sfU(1)^{(1)}$ 1-form symmetry, one adopts a four-step process. First, couple the theory to a two-form $\sfU(1)^{(1)}$ background field $B$ by adding $-\frac{\imag}{2\pi}B \wedge F$ to the Euclidean action. Second, in anticipation of making $B$ dynamical, relabel $B \to b$.
\tightfootnote{We use lowercase symbols for all dynamical fields unless stated otherwise, except for the QED $\sfU(1)$ gauge field $A$ and its curvature $F$, for which we use uppercase symbols.}
Third, in order to ensure that the subgroup gauged is $\IZ_{N}^{(1)}$, we make $b$ flat and require its periods to lie in $\frac{2\pi}{N}\IZ$. This is achieved by coupling $b$ to a dynamical 1-form Lagrange multiplier $c$ through $-\frac{\imag N}{2\pi} b \wedge dc$. The fourth and final step is to add a topological discrete torsion
\tightfootnote{Discrete torsion is to be distinguished from torsion in (co)homology.}
term of the form $-\frac{\imag N k}{4\pi} b\wedge b$ for an integer $k$ satisfying
\tightfootnote{By B\'ezout's lemma \cite{Bezout1779}, there exist (non-unique) integers $k, k' \in \IZ$ such that $p k  + N k' = \gcd(p,N)$. Since $p$ and $N$ are coprime, it follows that $pk + Nk' = 1$. This implies that $pk \equiv 1 \smod N$.}
$p k \equiv 1 \smod N$. This is like a discrete theta angle for a discrete gauge field. The advantage of this formulation is that discrete gauge fields can be written in terms of $\sfU(1)$ gauge fields with suitably constrained holonomies, allowing the use of de Rham cohomology.
\tightfootnote{It is important to stress that the connection is globally defined as a differential form on $M_4$ \emph{only} when the principal bundle is topologically trivial. Keeping track of bundle nontriviality and torsion requires a differential cohomology refinement \cite{Moore:2025tmt}.}
Therefore, one considers the action
\be\label{eq:action-QED-coupling-notorsion}
  S_{\text{QED}}[\Psi,\ov{\Psi},A] - \int_{M_4}\frac{\imag}{2\pi} b\wedge F - \int_{M_4}\frac{\imag N}{2\pi} b \wedge dc - \int_{M_4}\frac{\imag N k}{4\pi} b \wedge b ~,
\ee
with $(\Psi, \ov{\Psi}, A, b, c)$ treated as fluctuating fields that are path integrated over.

In this formulation, it is instructive to carry out the path integral over $b$ and $c$ explicitly to see the effect on the $\theta$ angle. Let us assume as above that $M_4$ is spin and has torsion-free (co)homology. The path integral over $c$ makes $b$ flat and gives it $\frac{2\pi}{N}\IZ$-valued periods. Every class $b'\in H^2(M_4;\IZ_N)$ then admits an integral lift $\wt b'\in H^2(M_4;\IZ)$. Because the intersection form of a spin four-manifold is even, the quadratic function
\be\label{eq:spin-gauging-quadratic-function}
 q_N(b'):=\frac{1}{2}\int_{M_4}\wt b'\smile\wt b'\smod N
\ee
is independent of the lift. This is the spin, torsion-free realization of the relevant quadratic refinement. Without these assumptions it must be replaced by the appropriate Pontryagin square in differential cohomology. Set $\varepsilon=[F/(2\pi)]\in H^2(M_4;\IZ)$. Writing $b=\frac{2\pi}{N}b'$, the path integral leaves the finite Gauss sum
\be\label{eq:GaussSum-1-notorsion}
 \sfG := \sum_{b' \in H^{2}(M_4;\IZ_N)}\exp\left\{\frac{2\pi\imag}{N}\left(\int_{M_4}b'\smile\rho_N(\varepsilon)+kq_N(b')\right)\right\} ~.
\ee
We define a $\IZ_N$-valued bilinear form
\be
  \langle\!\langle b'_1, b'_2 \rangle\!\rangle = \int_{M_4}b_1' \smile  b_2' = \left\langle b_1'\smile  b_2', [M_4]\right\rangle \in \IZ_{N} ~,
\ee
 where $\langle \cdot, \cdot\rangle$ is the canonical pairing between cohomology and homology, and $[M_4] \in H_{4}(M_4;\IZ)$ denotes the fundamental class.
\tightfootnote{The fundamental class is associated with the orientation, and makes sense even if $M_4$ is not connected.}
Then \eqref{eq:GaussSum-1-notorsion} is
\be\label{eq:GaussSum-2-notorsion}
  \sfG = \sum_{b' \in H^{2}(M_4;\IZ_N)}\exp\left\{ \frac{2\pi\imag}{N}\left(kq_N(b') + \langle\!\langle b', \rho_N(\varepsilon) \rangle\!\rangle \right) \right\} ~.
\ee
One can ``complete the square'' by shifting $b' \mapsto b' + p\rho_N(\varepsilon)$. This shift is valid in cohomology with $\IZ_N$ coefficients because $pk \equiv 1 \smod N$ implies that $p$ is the multiplicative inverse of $k$ modulo $N$. This yields
\be\label{eq:GaussSum-3-notorsion}
 \sfG = \exp\left[-\frac{2\pi\imag p}{2N}\int_{M_4}\varepsilon\smile \varepsilon\right]\sum_{b' \in H^{2}(M_4;\IZ_N)}\exp\left[\frac{2\pi\imag k}{N}q_N\big(b' + p\rho_N(\varepsilon)\big)\right] ~.
\ee
Since $H^{2}(M_{4};\IZ_N)$ is an Abelian group, translation by the fixed element $p\rho_N(\varepsilon)$ yields a bijection. Therefore,
\be
  \sum_{b' \in H^{2}(M_4;\IZ_N)}\exp\left[\frac{2\pi\imag k}{N}q_N\big(b' + p\rho_N(\varepsilon)\big)\right] = \sum_{b' \in H^{2}(M_4;\IZ_N)}\exp\left[\frac{2\pi\imag k}{N}q_N(b')\right] ~,
\ee
and hence the summation over $b'$ in \eqref{eq:GaussSum-3-notorsion} is, in fact, independent of $\varepsilon$, or equivalently, independent of $F$. So the entire $F$-dependence is contained in the prefactor of the sum in \eqref{eq:GaussSum-3-notorsion}. Identifying $\int_{M_4} \varepsilon \smile  \varepsilon = \frac{1}{4\pi^2}\int_{M_4} F \wedge F$, it is clear that this prefactor induces a term
\be
  \exp\left[-\frac{2\pi\imag p}{N}\cdot\frac{1}{2}\int_{M_4}\frac{F}{2\pi} \wedge \frac{F}{2\pi} \right] ~.
\ee
Comparing with the theta term, this discrete gauging effectively induces a shift in the $\theta$ angle given by
\be\label{eq:theta-shift-gauging}
  \theta \mapsto \theta - \frac{2\pi p}{N} ~.
\ee
In spin QED$_4$, this shift can be compensated by an axial rotation of the fermions
\be
  \Psi \mapsto \exp\left(\frac{\imag \alpha \gamma^5}{2}\right)\Psi ~,
  \qquad
  \ov\Psi\mapsto\ov\Psi\exp\left(\frac{\imag \alpha \gamma^5}{2}\right) ~,
  \qquad
  \alpha = \frac{2\pi p}{N} ~.
\ee
At the level of the electromagnetic response, it follows that the discrete gauging can be compensated by the axial rotation. The gravitational part of the axial Jacobian is addressed separately in \remref{rem:spin-versus-spinc-completion}.

It was argued by \cite{Choi:2022jqy} that an abstract $d=1+3$ QFT with (1) a $\sfU(1)^{(1)}$ 1-form global symmetry with conserved 2-form current $j_m$ (which is closed: $dj_m = 0$), and (2) a spin-one operator $j^A$ obeying the anomalous conservation equation $d\star j^{A} = j_m \wedge j_m$, has a codimension-one noninvertible symmetry defect. This assumption isolates the electromagnetic part of the QED axial anomaly. In the convention $\CJ_A^{(3)}=\frac{1}{2}\star j^A$ used throughout this paper, the corresponding gauge sector operator is
\be
   \wh{\CD}_{\frac{p}{N}}(M_3) = \exp\left(\frac{2\pi\imag p}{N}\oint_{M_3}\CJ_A^{(3)}\right)\,\CA^{N,p}\left[\frac{2\pi}{N}j_m\right] ~,
\ee
where $\CA^{N,p}$ is the minimal 3d Abelian TFT of \cite{Hsin:2018vcg}, coupled to the 1-form symmetry current of the bulk 4d theory.
\tightfootnote{Equivalently, the exponent of the first factor is $\frac{2\pi\imag p}{2N}\oint_{M_3}\star j^A$. Since $d\CJ_A^{(3)}=\frac{1}{2} j_m\wedge j_m$, sweeping $\exp(\imag\alpha\oint_{M_3}\CJ_A^{(3)})$ across a spacetime region shifts the theta angle by $\alpha$. Thus $\alpha=\frac{2\pi p}{N}$ compensates \eqref{eq:theta-shift-gauging}, in agreement with \eqref{eq:defect-schematic}.}

For the $\sfU(1)^{(1)}$ magnetic 1-form symmetry of spin QED$_4$, the coupling to the minimal TFT is
\be\label{eq:minTFT-coupling-spin}
  \CA^{N,p}\left[\frac{2\pi}{N}j_m\right] := \CA^{N,p}\left[\frac{2\pi}{N}\iota^{*}\frac{F}{2\pi}\right] ~,
\ee
where $\int_{\Sigma_2}\frac{F}{2\pi} \in \IZ$ for any 2-cycle $\Sigma_2$ in $M_4$. Here $\iota^*$ denotes pullback by the embedding $\iota: M_3 \hookrightarrow M_4$. In particular, for spin QED$_4$ on a torsion-free manifold, the gauge sector defect is
\be
   \left.\wh{\CD}_{\frac{p}{N}}(M_3)\right|_{\text{spin QED$_4$}} = \exp\left(\frac{2\pi \imag p}{N}\oint_{M_3}\CJ_A^{(3)}\right)\,\CA^{N,p}\left[\iota^{*}\frac{F}{N}\right] ~.
\ee
This is realized as a composition of an axial rotation and gauging a $\IZ_{N}^{(1)} \subset \sfU(1)^{(1)}$ magnetic 1-form symmetry of the bulk theory in half of the spacetime. It is an unframed topological defect on a general curved spin manifold only after its local gravitational response and full framing anomaly have also been canceled. Equation \eqref{eq:phase-ratio} tests only its dependence on changing the four-dimensional extension by a closed spin manifold. The authors of \cite{Choi:2022jqy} also show that this discrete gauging is associated with a sequence of operations for a group $\mathsf{SL}_2(\IZ)^{(k)}$ (parametrized by $k$), the composite effect of which is to shift the theta angle. 

Here ``half of spacetime'' means a region $M_4^+$ with boundary $\partial M_4^+=M_3$. Gauging $\IZ_N^{(1)}$ only on $M_4^+$ first produces an interface between QED and the theory obtained by this discrete gauging. At the gauge sector level, composing that interface with the compensating axial rotation identifies the theory on the gauged side with QED at the original coupling, producing an endodefect of QED. The defect is noninvertible because gauging sums over discrete two-form gauge fields: fusion with the oppositely oriented interface gives the condensation operator associated with this discrete gauging.

\section{Electromagnetic Sheets And Fractional Regular Flux}\label{app:magnetic-sheet}

This appendix distinguishes intrinsic half-integral spin$^c$ flux from a classical configuration with an externally imposed electromagnetic sheet. The notation $H_{1/2}(\gamma;D)$ below denotes a classical boundary prescription with regular endpoint flux $+\frac{1}{2}$. A quantum definition requires the additional data described below. Its sheet sign is present already on spin spacetime and does not detect the gravitational extension \eqref{eq:extension-sequence}.

\subsection{The Classical Sheet Prescription}

\begin{figure}[htbp]
  \centering
  \includegraphics[width=0.5\textwidth]{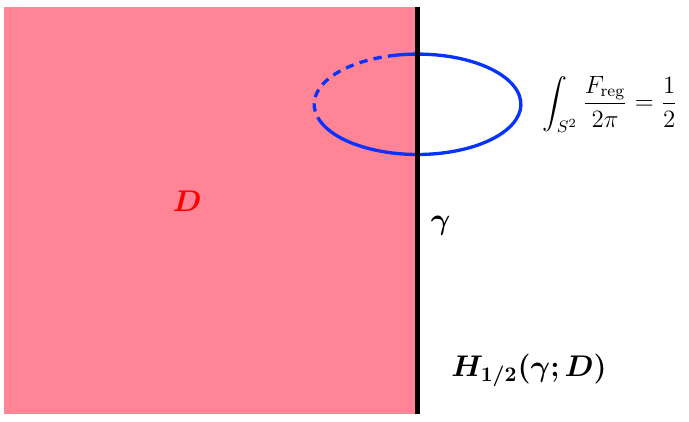}
  \caption{\label{fig:halftoofta}The classical surface--line prescription $H_{1/2}(\gamma;D)$. The pink sheet ends on $\gamma$. The separate linking sphere intersects $D$ once and encloses regular flux $\frac{1}{2}$.}
\end{figure}

On $\IR_t\times\IR^3$, take $\gamma=\IR_t\times\{0\}$ and $D=\IR_t\times\{(0,0,z):z\leq0\}$, with $\partial D=\gamma$.
\tightfootnote{\label{foot:compact-local-model}Here $t$ is a Euclidean coordinate. The same model applies in a normal neighborhood $I_t\times B^3_\epsilon\subset M_4$. A global sheet with $\partial D=\gamma$ requires $[\gamma]=0$ in $H_1(M_4;\IZ)$ and compatible bundle data. A small circle can bound a disk inside a coordinate ball of either $S^4$ or $\IC\IP^2$.}
In spherical coordinates $(r,\theta,\phi)$, with $\phi$ of period $2\pi$, choose on the complement of $D$
\begin{equation}\label{eq:half-flux-profile}
  A=\frac{1}{4}(1-\cos\theta)\,d\phi ~,
  \qquad F_{\mathrm{reg}}=dA=\frac{1}{4}\sin\theta\,d\theta\wedge d\phi ~.
\end{equation}
Here $F_{\mathrm{reg}}$ is obtained by differentiating off the sheet and extending the resulting two-form smoothly across it away from $\gamma$. For this profile,
\begin{equation}
  \int_{S^2_{\mathrm{link}}}\frac{F_{\mathrm{reg}}}{2\pi}=\frac{1}{2} ~,
  \qquad
  \lim_{\mathfrak m\to D}\exp\left(\imag\oint_{\mathfrak m} A\right)=-1 ~,
\end{equation}
where $\mathfrak m$ is a shrinking meridian of $D$ away from its boundary. The second equality follows from $A\to\frac{1}{2}d\phi$ near the negative $z$-ray. Thus $D$ is a physical $\pi$-monodromy sheet, locally of Gukov--Witten type \cite{GukovWitten2006}, detected by a unit-charge electron (see \figref{fig:halftoofta}).

Removing the singular term would require a half-winding transformation $\lambda=-\phi/2$, whose $e^{\imag\lambda}$ is not single-valued on the electron with the ambient spin$^c$ data held fixed. The sheet therefore cannot be removed by an allowed gauge transformation. Its monodromy is gauge invariant: an allowed transformation changes $\oint_{\mathfrak m} A$ by $2\pi n$, with $n\in\IZ$. Smooth fluctuations $a$ about the profile preserve both its limiting holonomy and the smooth extension of $F_{\mathrm{reg}}$ away from $\gamma$.

\begin{figure}[htbp]
  \centering
  \includegraphics[width=0.95\textwidth]{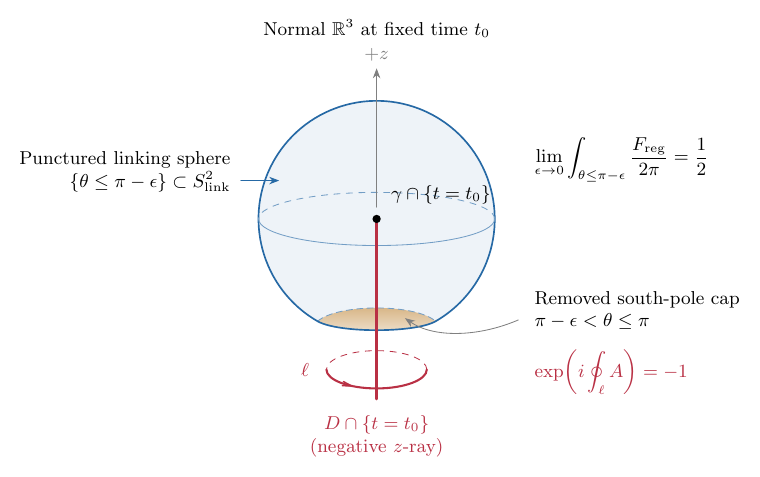}
  \caption{\label{fig:half-magnetic-punctured-sphere}Regular flux measured on a punctured linking sphere. In a spatial slice, $\gamma$ is the central point and $D$ is the negative $z$-ray. The excluded south-pole cap avoids $D$, and its boundary meridian has limiting electron holonomy $-1$. The dashed arc marks the back of the puncture rim.}
\end{figure}

One can measure the regular flux without touching the singularity by removing the south-pole cap $\pi-\epsilon<\theta\leq\pi$ and taking $\epsilon\to0$. The flux tends to $\frac{1}{2}$, while the holonomy around the shrinking puncture stays $-1$, as in \figref{fig:half-magnetic-punctured-sphere}.

A quantum defect requires compatible fermion boundary conditions, a regularization of the singular action, and sheet and endpoint counterterms. In pure Maxwell theory an electric 1-form symmetry surface can implement the corresponding monodromy topologically. In QED the dynamical unit-charge electron breaks that symmetry, so this construction gives no topological protection for $D$. A decorated topological sheet would require an explicit defect theory and its coupling to QED.

\subsection{Separating The Sheet Sign From The Gravitational Sign}

\begin{figure}[htbp]
  \centering
  \includegraphics[width=0.95\textwidth]{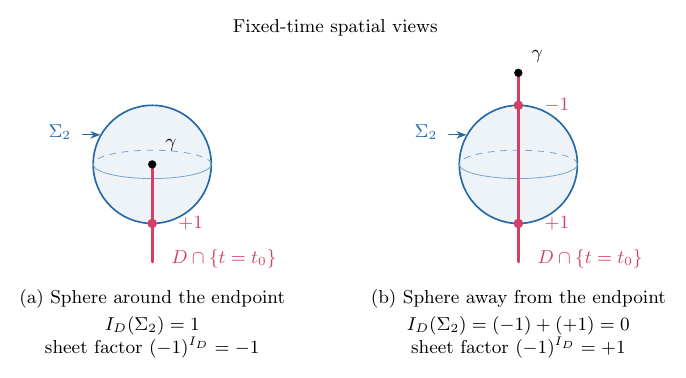}
  \caption{\label{fig:magnetic-sheet-intersections}Measuring surfaces intersecting the sheet. In spatial slices, the blue sphere avoids the endpoint (black point), while the pink ray is the sheet. Panel (a) has one positive intersection, and panel (b) has two of opposite signs. The resulting sheet factors are $-1$ and $+1$, independently of the gravitational factor $G$.}
\end{figure}

Let $\Sigma_2$ be a closed measuring surface disjoint from $\gamma$ and transverse to $D$, as in \figref{fig:magnetic-sheet-intersections}. Define
\begin{equation}\label{eq:relative-regular-flux}
  \Phi_D(\Sigma_2):=\int_{\Sigma_2}\frac{F_{\mathrm{reg}}}{2\pi} ~,
  \qquad I_D(\Sigma_2):=\int_{\Sigma_2}\delta_D ~,
\end{equation}
where $\delta_D$ is the two-form current dual to $D$ and $I_D$ is the signed intersection number. Distributionally, differentiating the singular potential includes the sheet flux:
\begin{equation}\label{eq:distributional-sheet-curvature}
  F_{\mathrm{sing}}=F_{\mathrm{reg}}-\pi\delta_D ~,
  \qquad dF_{\mathrm{reg}}=\pi\delta_\gamma ~,
  \qquad d\delta_D=\delta_\gamma ~.
\end{equation}
Here $\delta_\gamma$ is the current dual to the oriented boundary line.
\tightfootnote{\label{foot:distributional-sheet-flux}Near the negative $z$-ray, $A\sim\frac{1}{2}d\phi$ and $d(d\phi)=2\pi\delta(x)\delta(y)\,dx\wedge dy$ distributionally \cite[Sec.~2.1]{Tan2009SurfaceOperators}. With the orientation for which the outward linking sphere has $I_D=+1$, $\delta_D=-\delta(x)\delta(y)\,dx\wedge dy$, giving the sign in \eqref{eq:distributional-sheet-curvature}.}
Consequently $dF_{\mathrm{sing}}=0$: the sheet compensates the apparent half-unit source of the regular field. In the local model, the sheet contribution cancels the regular flux through the linking sphere, giving zero total flux.

For the global comparison, assume an underlying spin$^c$ bundle whose total distributional curvature obeys \eqref{eq:spin-c-physics-v1}. This global flux prescription must be specified in addition to the local monodromy. Then
\begin{equation}\label{eq:twisted-spin-c}
  2\Phi_D(\Sigma_2)-I_D(\Sigma_2)
  \equiv\int_{\Sigma_2}w_2(TM_4)\smod 2 ~.
\end{equation}
Writing $U_m^{\mathrm{reg}}$ for the exponential evaluated with the regular field gives the useful separation
\begin{equation}\label{eq:Um-with-sheet}
  U_m^{\mathrm{reg}}(2\pi;\Sigma_2\!\mid\! D)
  :=e^{2\pi\imag\Phi_D(\Sigma_2)}
  =G(\Sigma_2)(-1)^{I_D(\Sigma_2)} ~.
\end{equation}
Including the sheet restores the intrinsic identity:
\begin{equation*}
  \exp\left(2\pi\imag\int_{\Sigma_2}\frac{F_{\mathrm{sing}}}{2\pi}\right)
  =U_m^{\mathrm{reg}}(2\pi;\Sigma_2\!\mid\! D)(-1)^{I_D(\Sigma_2)}
  =G(\Sigma_2) ~.
\end{equation*}
On a spin background $G=1$, and the regular-flux exponential reduces to the intersection sign. On the local linking sphere, $I_D=1$ and the entire regular-flux sign comes from the sheet. This sheet sign occurs on both spin and non-spin backgrounds. The gravitational extension is detected by $G$ on global surfaces.

The local construction also allows other regular fluxes: the profile $A_q=\frac{q}{2}(1-\cos\theta)\,d\phi$ has regular flux $q$ and meridian holonomy $e^{2\pi\imag q}$ for any real $q$. Global embeddings impose further consistency conditions \cite[Sec.~2.1]{Tan2009SurfaceOperators}. Such imposed sheets do not enlarge the genuine magnetic charge lattice.

\subsection{Fusion And Wall Transport}\label{app:sheet-fusion-transport}

Superposing two half-flux profiles on the same sheet gives unit regular flux and trivial electron monodromy, but retains the distributional sheet term $2F_{\mathrm{sing}}=2F_{\mathrm{reg}}-2\pi\delta_D$. In particular, the regular flux near the boundary of $D$ is not by itself the intrinsic magnetic charge: in the local prescription, the sheet contribution cancels the regular flux in the total distributional curvature. Doubling the regular flux therefore does not establish that the fused configuration is a genuine unit magnetic line. Accordingly,
\begin{equation}\label{eq:half-magnetic-fusion}
  H_{1/2}(\gamma;D)\otimes H_{1/2}(\gamma;D)
  \stackrel{?}{\simeq}H_1(\gamma) ~,
\end{equation}
would require a quantum completion that removes or suitably identifies the doubled sheet, supplies the necessary bundle and endpoint data to obtain a bare unit magnetic line, and preserves the action of magnetic surface operators through specified junctions. Fusion for different sheets $D_1,D_2$ would moreover have to account for their relative surface data. Constructing such fusion junctions remains open.

Likewise, inserting $m=\frac{1}{2}$ into the Witten-effect transformation \eqref{eq:witten-effect-wall} would only suggest
\begin{equation*}
  H_{1/2}(\gamma;D)
  \stackrel{?}{\longmapsto}
  H_{1/2}(\gamma;D)W^{p/(2N)}(\gamma;\Sigma) ~.
\end{equation*}
The theta term involves sheet singularities and counterterms, so regular flux does not fix the junction charge. If a completion realized both this action of the wall and \eqref{eq:half-magnetic-fusion}, a wall endpoint line $\eta_{\mathrm{wall}}$ would need $\eta_{\mathrm{wall}}^2\simeq v$ and $Q(\eta_{\mathrm{wall}})=p/(2N)\smod1$, together with compatible spin, braiding, and surface junctions. A fusion-algebra square root alone is insufficient.

For example, in the odd-denominator minimal Hall theory $\CV^{3,1}\simeq\sfU(1)_3$, the fractional electric charges of the vison sectors are $0,\frac{1}{3},\frac{2}{3}\smod1$. Although $v^2$ is a square root of $v$ at the level of the vison fusion algebra modulo local electrons, it has charge $\frac{2}{3}\smod1$, whereas the proposed action of the wall on the half-magnetic line requires $\frac{1}{6}\smod1$. No line in this minimal charged sector has that required charge. A neutral gravitational framing completion cannot supply the missing electric charge. This elementary test does not exclude a larger charged wall theory or an intrinsically surface-dependent junction, but shows that fusion-algebra compatibility alone is insufficient.

Finally, a gravitational decoration is a different boundary problem. For a mod-two cocycle representative $\omega_2$ of $w_2(TM_4)$ on an open surface $D$ and a mod-two one-cochain $\xi_1$, changing $\omega_2$ by $\delta\xi_1$ gives
\begin{equation}\label{eq:open-gravitational-surface-variation}
  (-1)^{\langle\omega_2+\delta\xi_1,D\rangle}
  =(-1)^{\langle\omega_2,D\rangle}
   (-1)^{\langle\xi_1,\partial D\rangle} ~.
\end{equation}
Equivalently, in the normal-bundle form $G=(-1)^{\int e(N)}$ of \autoref{subsubsec:w2-twist}, the normal Euler number of an open surface is defined only relative to a trivialization of $N_{D/M_4}$ along $\partial D$. Changing that trivialization by a loop of degree $d$ in $\mathsf{SO}(2)$ shifts it by $d$, and modulo two this is the variation above. An open $G$ requires a boundary trivialization or line data that cancel this variation. A fermionic boundary line is a natural candidate for such data because of its projective Lorentz transformation \cite[Sec.~2.5]{Brennan:2022tyl}. This does not establish a compatible boundary junction in the coupled spin$^c$ QED theory. Its fermionic statistics alone do not assign it a magnetic charge. Defining its junctions with $U_m$ and their coupling to the bulk spin$^c$ field is an open problem. Any magnetic charge must obey \eqref{eq:local-spinc-magnetic-quantization}.

\phantomsection
\interlinepenalty=10000
\addcontentsline{toc}{section}{References}
\bibliographystyle{ytamsalpha}
\bibliography{refshigherstructure}

\end{document}